\documentclass[12pt,british]{article}
\usepackage[utf8]{inputenc}
\usepackage{geometry}
\usepackage[dvipsnames]{xcolor}
\usepackage{babel}
\usepackage{mathtools}
\usepackage{amsmath}
\usepackage{amsthm}
\usepackage{amssymb}
\usepackage{cancel}
\usepackage{setspace}
\usepackage{microtype}
\usepackage[unicode=true,pdfusetitle,
 bookmarks=true,bookmarksnumbered=false,bookmarksopen=false,
 breaklinks=false,pdfborder={0 0 0},pdfborderstyle={},backref=false,colorlinks=true]
 {hyperref}
\hypersetup{
colorlinks=true,urlcolor=darkblue,anchorcolor=darkblue,citecolor=darkblue,filecolor=darkblue,linkcolor=darkblue,menucolor=darkblue,linktocpage=true,pdftitle=Geometric Flows and Supersymmetry}

\makeatletter

\usepackage{slashed}	 	%
\usepackage{amsfonts} 		%
\usepackage{cite}	%
\SetTracking{encoding={*}, shape=sc}{0} 	%
\allowdisplaybreaks		%
\date{\today} 		%
\numberwithin{equation}{section}	%
\usepackage[small]{titlesec} 		%

\makeatletter
\g@addto@macro\bfseries{\boldmath}
\makeatother

\definecolor{darkblue}{rgb}{0.1,0.0,0.5}

\makeatletter
\g@addto@macro\@floatboxreset\centering
\makeatother

\let\originalleft\left
\let\originalright\right
\renewcommand{\left}{\mathopen{}\mathclose\bgroup\originalleft}
\renewcommand{\right}{\aftergroup\egroup\originalright}

\renewenvironment{thebibliography}[1]{%
\begin{oldthebibliography}{#1}%
\small%
\raggedright%
\setlength{\itemsep}{0pt}%
}%
{%
\end{oldthebibliography}%
}

\makeatother

\begin{document}

\global\long\def\rep#1{\boldsymbol{#1}}%
\global\long\def\repb#1{\overline{\boldsymbol{#1}}}%
\global\long\def\dd{\text{d}}%
\global\long\def\ii{\text{i}}%
\global\long\def\ee{\text{e}}%
\global\long\def\Dorf{L}%
\global\long\def\tDorf{\hat{L}}%
\global\long\def\GL#1{\text{GL}(#1)}%
\global\long\def\Orth#1{\text{O}(#1)}%
\global\long\def\SO#1{\text{SO}(#1)}%
\global\long\def\Spin#1{\text{Spin}(#1)}%
\global\long\def\Symp#1{\text{Sp}(#1)}%
\global\long\def\Uni#1{\text{U}(#1)}%
\global\long\def\SU#1{\text{SU}(#1)}%
\global\long\def\Gx#1{\text{G}_{#1}}%
\global\long\def\Fx#1{\text{F}_{#1}}%
\global\long\def\Ex#1{\text{E}_{#1}}%
\global\long\def\ExR#1{\text{E}_{#1}\times\mathbb{R}^{+}}%
\global\long\def\ex#1{\mathfrak{e}_{#1}}%
\global\long\def\gl#1{\mathfrak{gl}_{#1}}%
\global\long\def\SL#1{\text{SL}(#1)}%
\global\long\def\Stab{\operatorname{Stab}}%
\global\long\def\vol{\operatorname{vol}}%
\global\long\def\tr{\operatorname{tr}}%
\global\long\def\ad{\operatorname{ad}}%
\global\long\def\ext{\Lambda}%
\global\long\def\AdS#1{\text{AdS}_{#1}}%
\global\long\def\op#1{\operatorname{#1}}%
\global\long\def\im{\operatorname{im}}%
\global\long\def\re{\operatorname{re}}%
\global\long\def\eqspace{\mathrel{\phantom{{=}}{}}}%
\global\long\def\bZ{\mathbb{Z}}%
\global\long\def\bC{\mathbb{C}}%
\global\long\def\bP{\mathbb{P}}%
\global\long\def\bR{\mathbb{R}}%
\global\long\def\feyn#1{\slashed{#1}}%
\global\long\def\id{\operatorname{id}}%
\global\long\def\ap{\alpha'}%
\global\long\def\oap{\mathcal{O}(\alpha')}%
\global\long\def\del{\partial}%
\global\long\def\bdel{\bar{\partial}}%
\global\long\def\transpose{{\scriptscriptstyle \mathsf{T}}}%
\global\long\def\charge{\text{c}}%
\let\oldstar\star\renewcommand{\star}{\mathop{}\mathopen{}{\oldstar}}

\begin{titlepage}
\begin{flushright} %
\end{flushright}
\vfill
\begin{center}
{\setstretch{1.3}\Large\bf Geometric Flows and Supersymmetry\par} 
\vskip 1cm 
Anthony Ashmore,\textsuperscript{a} Ruben Minasian\textsuperscript{b} and Yann Proto\textsuperscript{c}
\vskip 1cm
\textit{\small{}\textsuperscript{a}Enrico Fermi Institute \& Kadanoff Center for Theoretical Physics,\\ University of Chicago, Chicago, IL 60637, USA}
\\[.2cm]
\textit{\small{}\textsuperscript{b}Institut de Physique Th\'eorique, Universit\'e Paris-Saclay, CNRS,\\ CEA, F-9119, Gif-sur-Yvette, France}
\\[.2cm] 
\textit{\small\textsuperscript{c}Sorbonne Universit\'{e}, CNRS, Laboratoire de Physique Th\'{e}orique\\ et Hautes \'{E}nergies, LPTHE, F-75005 Paris, France}
\end{center}
\vfill
\begin{center} \textbf{Abstract} \end{center}
\begin{quote}
We study the relation between supersymmetry and geometric flows driven by the Bianchi identity for the three-form flux $H$ in heterotic supergravity. We describe how the flow equations can be derived from a functional that appears in a rewriting of the bosonic action in terms of squares of supersymmetry operators. On a complex threefold, the resulting equations match what is known in the mathematics literature as ``anomaly flow''. We generalise this to seven- and eight-manifolds with G$_2$ or Spin(7) structures and discuss examples where the manifold is a torus fibration over a K3 surface. In the latter cases, the flow simplifies to a single scalar equation, with the existence of the supergravity solution implied by the long-time existence and convergence of the flow. We also comment on the $\alpha'$ expansion and highlight the importance of using the proper connection in the Bianchi identity to ensure that the flow's fixed points satisfy the supergravity equations of motion.
\end{quote}
\vfill
{\begin{NoHyper}\let\thefootnote\relax\footnotetext{\tt \!\!\!\!\!\!\!\!\!\!\!\!\! ashmore@uchicago.edu, ruben.minasian@ipht.fr, yproto@lpthe.jussieu.fr}\end{NoHyper}}
\end{titlepage}

\microtypesetup{protrusion=false} %
\tableofcontents %
\microtypesetup{protrusion=true} %

\clearpage{}

\section{Introduction}

The relationship between theoretical physics and mathematics has a rich and productive history. This is particularly evident in string theory, where the exploration of supersymmetric vacua has been both influenced by and driven developments in complex, symplectic, and K\"ahler geometry. A noteworthy example of this in mathematics is Perelman's proof of Thurston's geometrisation conjecture using Ricci flow~\cite{math/0211159}. Key to this was his introduction of a functional whose gradient flow is gauge-equivalent to Ricci flow. Somewhat fascinatingly, this functional is actually the string-frame effective action for a metric and dilaton~\cite{Fradkin:1984pq,Fradkin:1985ys,Callan:1985ia,Callan:1986jb}.

Ricci flow is only one example of a geometric flow -- a partial differential equation that describes the evolution of a geometric structure with time. There are now a wide variety of geometric flows in both physics and mathematics. These include examples that assume extra structure on the underlying manifold, such as K\"ahler--Ricci flow~\cite{Cao1985} or $\Gx2$ flows~\cite{2003math......5124B,2019arXiv190410068D,2011arXiv1108.2192K,2007math......2077K}, or higher-order variants of Ricci flow, such as Calabi flow~\cite{calabi_flow}. When writing down a geometric flow, there are at least two properties that one is seeking. First, the fixed points of the flow should correspond to solutions of a meaningful geometric equation; second, the flow itself should be (weakly) parabolic so that solutions to the evolution equations are unique and exist at least for some short time.\footnote{A flow is strongly/strictly parabolic if the partial differential operator driving the flow has a positive-definite principal symbol. If the relevant operator admits a gauge symmetry, such as diffeomorphisms, the principal symbol will have null directions and the flow is said to be weakly parabolic. The gauge redundancies can be fixed using ``DeTurck's trick'' to give a strongly parabolic flow, with short-time existence then following. (See, for example, \cite[Section 6]{Hamilton1}.)} The first of these conditions leads one to wonder whether a given geometric flow has any physical interpretation, or if a flow that appears in the description of a physical system has any geometric interpretation, such as interpreting renormalisation group flow for a sigma model as Ricci flow on the target space~\cite{Oliynyk:2005ak,Oliynyk:2004ey,Tseytlin:2006ak}.

The aim of the present work is to analyse a geometric flow known as ``anomaly flow'' from a physics perspective. Anomaly flow is a coupled flow for a conformally balanced hermitian metric and a Yang--Mills gauge field on a complex threefold, first introduced in \cite{1508.03315} and further studied in \cite{1610.02740,1610.02739,1705.09763,1711.08186}. The leading term in the flow for the metric is simply the Ricci tensor, with higher-order corrections constructed from curvatures and torsions, so that anomaly flow is a kind of modified Ricci flow, specifically a non-K\"ahler generalisation of K\"ahler--Ricci flow.\footnote{See \cite{10.1093/imrn/rnp237,Fu2022} for a different example of a flow for a hermitian metric with torsion.} In fact, the flow is intimately related to solutions of the Hull--Strominger system~\cite{Strominger:1986uh,Hull:1986kz}, a set of PDEs which follow from requiring that the threefold with its metric, torsion and gauge field preserves $N=1$ supersymmetry as a Minkowski background in heterotic supergravity. Indeed, the name ``anomaly flow'' comes from the observation that the flow is driven by the anomaly cancellation condition in heterotic string theory~\cite{Green:1984sg}, which appears as a non-trivial Bianchi identity for the torsion/three-form flux $H$.

Anomaly flow provides one of very few tools for analysing non-K\"ahler geometries. To date, there are still very few examples of non-K\"ahler vacua in heterotic string theory: these include what are known in the mathematics literature as the Fu--Yau solutions~\cite{hep-th/0604063}, which were first found by Dasgupta, Rajesh and Sethi~\cite{Dasgupta:1999ss} via a chain of T- and S-dualities, variations of this in the physics literature~\cite{Becker:2002sx,LopesCardoso:2002vpf,Adams:2006kb,Becker:2008rc,Becker:2009df,Andriot:2009fp,Becker:2009zx,Israel:2011urb,Israel:2023itj}, some examples on parallelisable manifolds~\cite{Fernandez:2008wa,Grantcharov:2009qv,Fernandez:2014kwa}, and solutions on generalised Calabi--Gray manifolds~\cite{Fei:2017ctw}.\footnote{It seems likely that these do not actually solve the equations of motion due to the choice of connection in the anomaly cancellation condition.} At a more fundamental level, unlike Calabi--Yau metrics, given some choice of topological data there are no guarantees that a non-K\"ahler solution will exist. Without existence proofs of this form or explicit metrics, it is difficult to use these backgrounds for phenomenology or model building.

For any elliptic system of equations, such as the Hull--Strominger system, there are infinitely many parabolic flows whose fixed points solve the elliptic system. Due to this, short-time existence of a flow is not greatly constraining. Instead, a better measure of whether a parabolic flow is worth exploring further is its convergence and long-time existence properties. Although long-time existence for anomaly flow has not been shown in general, it has been proven for geometries known as Goldstein--Prokushkin fibrations~\cite{1610.02740}. These are threefolds given as two-torus fibrations over K3 surfaces~\cite{Goldstein:2002pg}, with the fibration structure leading to an enhanced $N=2$ supersymmetry. In this case, long-time existence of the flow provides a simpler version of Fu and Yau's proof for the existence of Hull--Strominger solutions~\cite{hep-th/0411136,hep-th/0604063}. These results suggest that anomaly flow stands out as special and deserving of further investigation.

Two questions come to mind when considering anomaly flow. First, as originally defined, fixed points of the flow do not actually give solutions to the equations of motion of heterotic supergravity due to a subtlety with the choice of connection in the anomaly cancellation condition. Second, there is no apparent physical reason for why this particular flow has such desirable properties, making it difficult to generalise to higher-dimensional manifolds. We will argue that the key to both of these questions is supersymmetry. As we will review, heterotic supergravity admits a Lichnerowicz-like identity which relates the bosonic action to squares of supersymmetry operators and the Bianchi identity~\cite{1702.01156}, with a corresponding decomposition of the equations of motion. Using this, we present a direct link between anomaly flow and supersymmetry. Moreover, by deriving the anomaly flow directly from supergravity, it becomes clear that it is crucial to use the correct connection in the anomaly cancellation condition and to consider carefully the $\ap$ expansion of both the equations of motion and the anomaly flow itself.

Essential for understanding anomaly flow as a modified Ricci flow is the observation that the heterotic equations of motion can be written in terms of certain supersymmetry operators acting on a nowhere-vanishing spinor. This holds in both the full ten-dimensional theory and when reduced on geometries of the form $\bR^{1,9-n}\times X_n$. For the latter, the Einstein equation -- the equation of motion for the metric on $X_n$ -- can be expressed as~\cite{0908.2927,Gauntlett:2002sc,Kunitomo:2009mx,Martelli:2010jx}
\begin{equation}\begin{split}
		&R_{mn}+2\nabla_{m}\nabla_{n}\varphi-\tfrac{1}{4}H_{m}{}^{p_{1}p_{2}}H_{np_{1}p_{2}}\\
		&=\left(\epsilon^{\dagger}\gamma_{(m}\feyn{D}D_{n)}\epsilon-\epsilon^{\dagger}\gamma_{(m}D_{n)}\feyn{D}\epsilon+\tfrac{1}{2}H_{(m}{}^{pq}\,\epsilon^{\dagger}\gamma_{n)}\gamma_{p}D_{q}\epsilon+\text{c.c.}\right)\\
		&\eqspace+\tfrac{1}{12}\dd H_{(m}{}^{p_{1}\dots p_{3}}\,\epsilon^{\dagger}\gamma_{n)p_{1}\dots p_{3}}\epsilon,
	\end{split}
\end{equation}
where $\epsilon$ is an invariant spinor characterising a $G$-structure on $X_n$ and we have included only the zeroth-order terms in $\ap$ for simplicity. The final $\dd H$-dependent term is exactly what appears in the evolution equation for the metric in anomaly flow, and so the flow of the metric is driven by the Einstein equation plus terms that depend on the supersymmetry operators $D_m$ and $\feyn{D}$. This also explains why anomaly flow is particularly interesting on $\SU3$ structure manifolds -- it turns out that if one initially preserves supersymmetry, $D_m \epsilon=\feyn{D}\epsilon=0$, the flow continues to preserve these conditions. The flow driven by $\dd H$ then reduces to a flow by the Einstein equation without any correction terms. A similar result holds when the leading $\ap$ corrections are included. 

We also analyse a natural extension of anomaly flow to seven- or eight-manifolds with $\Gx2$ or $\Spin7$ structures. These flows generically do not preserve supersymmetry, with the flow equations driven by the equations of motion plus corrections that depend at most quadratically on the torsion $H$. These flow equations may be mathematically interesting in their own right.\footnote{See \cite{2022arXiv221105197F} for recent work on flows of $G$-structures.} However, there are backgrounds where the $G$-structures reduce further to $\SU3$ or $\SU2$ and the flow \emph{does} preserve supersymmetry. This is similar in spirit to \cite{2022arXiv220903411P} where the flow of a $G$-structure simplifies to a flow for a reduced $G$-structure given an appropriate choice of initial data. In particular, upon reducing to $\SU 2$, the flow simplifies to a scalar evolution equation for the dilaton. For example, on a seven-manifold, this gives solutions which are T$^{3}$ fibrations over K3 preserving $N=2$ or $N=4$ supersymmetry in three dimensions.\footnote{There also seems to be more freedom in these equations which might allow for $N=1$ supersymmetry.} This example can be generalised to give AdS solutions on $\text{K3}\times\text{S}^{3}$, where again the anomaly flow reduces to a single equation.

This paper is organized as follows. In Section \ref{sec:heterotic_supergravity}, we recall some facts about spacetime heterotic string theory to order $\ap$ and the conditions for $N=1$ supersymmetry on backgrounds of the form $\bR^{1,3}$ times a six-manifold. In Section \ref{sec:anomaly_flow}, we review anomaly flow as originally defined and explain why the choice of connection in the Bianchi identity is crucial. Following this, we reformulate anomaly flow using the Hull connection and show that this enables a direct link between the flow equations and the supergravity equations of motion. We also comment on the importance of the $\ap$ expansion for understanding heterotic flux solutions. In Section \ref{sec:functional}, we introduce a functional from which one can derive anomaly flow and show how this functional is related to the bosonic supergravity action via a heterotic generalisation of the Lichnerowicz identity without imposing the Bianchi identity. In Section \ref{sec:diverse}, we use this functional to generalise anomaly flow to $\Gx2$ and $\Spin7$ structure manifolds. We give a number of examples, including two where the flow equations reduce to a single scalar equation for the dilaton, resulting in three- and four-torus analogues of Fu--Yau solutions. In Section \ref{sec:gradient}, we describe how anomaly flow can be interpreted as the gradient flow of a functional after dualising the three-form flux $H$. We finish in Section \ref{sec:discussion} with a discussion of open questions and future work. Appendices \ref {app:conventions} and \ref{app:flow_equation} contain our conventions for heterotic supergravity, gamma matrices and $G$-structures, and a detailed derivation via supersymmetry of the link between the flow equations and the equations of motion.

\section{Heterotic supergravity and supersymmetry}\label{sec:heterotic_supergravity}

Heterotic supergravity is the spacetime theory which follows from a truncation of heterotic string theory to first order in $\ap$, the square of the string length. The bosonic fields of this theory are a metric $g$, a scalar dilaton field $\varphi$, a three-form $H$ with local two-form potential $B$, and a gauge field $A$. The fermionic fields are a gravitino $\psi$, a dilatino $\lambda$ and a gaugino $\chi$. One has a bosonic solution to heterotic supergravity if the supergravity equations of motion are satisfied, $H$ obeys a non-trivial Bianchi identity and the fermionic fields vanish. In detail, the bosonic equations of motion are\footnote{See Appendix \ref{app:eom} for the action and variations. Our conventions match those of \cite{1702.01156}. Alternatively, they match \cite{Kunitomo:2009mx} after sending $H$ to $-H$.}
\begin{subequations}\label{eq:bosonic_eom}
	\begin{align}
		\begin{split}\text{eom}[g]_{MN} & \equiv R_{MN}+2\nabla_{M}\nabla_{N}\varphi-\frac{1}{4}H_{MPQ}H_{N}{}^{PQ}\\
			& \eqspace-\frac{\ap}{4}(\tr F_{MP}F_{N}{}^{P}-\tr R_{MP}^{+}R_{N}^{+}{}^{P})+\mathcal{O}(\ap^{2}),
		\end{split}
		\label{eq:metric_eom}\\
		\begin{split}\text{eom}[\varphi] & \equiv\frac{1}{4}R+\nabla^{2}\varphi-(\nabla\varphi)^{2}-\frac{1}{48}H_{MNP}H^{MNP}\\
			& \eqspace-\frac{\ap}{32}(\tr F_{MN}F^{MN}-\tr R_{MN}^{+}R^{+MN})+\mathcal{O}(\ap^{2}),
		\end{split}
		\label{eq:dilaton_eom}\\
		\text{eom}[B]_{MN} & \equiv\nabla^{P}(\ee^{-2\varphi}H_{PMN})+\mathcal{O}(\ap^{2}),\label{eq:B_eom}\\
		\text{eom}[A]_{M} & \equiv\mathcal{D}^{-N}(\ee^{-2\varphi}F_{NM})+\mathcal{O}(\ap),\label{eq:A_eom}
	\end{align}
\end{subequations}
where $R^{+}$ is the curvature two-form of the Hull connection $\nabla^{+}$, defined in \eqref{eq:Hull}, $\nabla^{-}$ is the Bismut connection, defined in \eqref{eq:Bismut}, $F$ is the field strength for $A$, and $\mathcal{D}^-$ is the gauge-covariant Bismut connection.\footnote{Care should be taken when naively comparing orders of $\ap$ in these equations since the gauge connection is already down by a factor of $\ap$ in the action and everywhere else.} The Bianchi identity or ``anomaly cancellation condition'' for $H$ is
\begin{equation}
\dd H-\frac{\ap}{4}(\tr F\wedge F-\tr R^{+}\wedge R^{+})=0.\label{eq:bianchi}
\end{equation}
The theory itself is supersymmetric, with the supersymmetry transformations of the fermionic fields given by
\begin{subequations}\label{eq:10_susy}
\begin{align}
\delta\psi_{M} & =\nabla_{M}\varepsilon+\tfrac{1}{8}H_{MNP}\Gamma^{NP}\varepsilon+\mathcal{O}(\ap^{2}),\\
\delta\lambda & =\left(\Gamma^{M}\partial_{M}\phi+\tfrac{1}{12}H_{MNP}\Gamma^{MNP}\right)\varepsilon+\mathcal{O}(\ap^{2}),\\
\delta\chi & =-\tfrac{1}{2}F_{MN}\Gamma^{MN}\varepsilon+\mathcal{O}(\ap^2),
\end{align}
\end{subequations}
where the $\mathcal{O}(\ap)$ corrections to these supersymmetry variations are entirely captured by the modification of $H$ that leads to \eqref{eq:bianchi}. A bosonic solution of the theory is said to be supersymmetric if the supersymmetry variations of the fermionic fields vanish. Of particular interest for us are supersymmetric solutions that are products of four-dimensional Minkowski space with a six-manifold $X$. The supersymmetry parameter can then be decomposed in terms of spinors on Minkowski space and spinors on $X$. 
Supersymmetry of the solution then requires the existence of (at least) one nowhere-vanishing spinor $\epsilon$ on $X$ which sets the above supersymmetry variations to zero. Together with the Bianchi identity, the resulting system of equations has come to be known as the Hull--Strominger system~\cite{Hull:1986kz,Strominger:1986uh}.

In more detail, the ten-dimensional string-frame metric takes the form
\begin{equation}\label{eq:10d_ansatz}
\dd s_{10}^{2}=\dd s^{2}(\mathbb{R}^{1,3})+\dd s^{2}(X).
\end{equation}
Solutions of this form preserve Poincar\'e invariance in four dimensions only if the metric on $\bR^{1,3}$ is the Minkowski metric, the dilaton $\varphi$ is constant on $\bR^{1,3}$, and the three-form $H$ and gauge field $A$ have no components along $\bR^{1,3}$. Whether a solution of the form \eqref{eq:10d_ansatz} exists then reduces to questions about the geometry of $X$ and tensors/gauge fields living on it. Thanks to this, in what follows, we abuse notation somewhat and think of $g$, $\varphi$, $H$ and $A$ as restricted to $X$.

The gauge field $A$ is a connection on a vector bundle $V$ over $X$ with curvature $F$. The nowhere-vanishing spinor $\epsilon$ defines an $\SU 3$ structure on $X$, equivalent to a non-degenerate real two-form $\omega$ and a complex three-form $\Omega$ satisfying
\begin{equation}
\omega\wedge\Omega=0,\qquad\frac{\ii}{\Vert\Omega\Vert^{2}}\Omega\wedge\bar{\Omega}=\frac{1}{3!}\omega\wedge\omega\wedge\omega.
\end{equation}
In particular, $\omega$ defines a hermitian metric and $\Omega$ defines an almost complex structure on $X$. In our conventions, given in Appendix \ref{app:spinors}, these are given by
\begin{equation}
\omega_{mn}=-\ii\epsilon^{\dagger}\gamma_{mn}\gamma_{*}\epsilon,\qquad\Omega_{mnp}=-\ii\epsilon^{\dagger}\gamma_{mnp}(1+\gamma_{*})\epsilon.
\end{equation}

The ten-dimensional supersymmetry conditions \eqref{eq:10_susy} reduce to a system of conditions on the internal manifold $X$. The supersymmetry conditions become
\begin{equation}
D_{m}\epsilon=0,\qquad\feyn D\epsilon=0,\qquad\feyn F\epsilon=0,
\end{equation}
where $\feyn{F}=\tfrac{1}{2}F_{mn}\gamma^{mn}$ and the supersymmetry operators are defined by\footnote{Note that $\feyn{D}\neq \gamma^m D_m$, but instead comes from a certain combination of the gravitino and dilatino variations in \eqref{eq:10_susy}.}
\begin{subequations}\label{eq:susy_operators}
\begin{align}
D_{m}\epsilon & \coloneqq\nabla_{m}\epsilon+\tfrac{1}{8}H_{mnp}\gamma^{np}\epsilon\equiv\nabla_{m}^{-}\epsilon,\\
\feyn D\epsilon & \coloneqq\gamma^{m}\nabla_{m}\epsilon+\tfrac{1}{24}H_{mnp}\gamma^{mnp}-\gamma^{m}\partial_{m}\varphi.
\end{align}
\end{subequations}
These conditions imply that $\epsilon^{\dagger}\epsilon$ is constant on $X$, and so can be set to unity, corresponding to the usual normalisation $\Vert\Omega\Vert^{2}=8$. The remaining conditions from the vanishing of the supersymmetry variations are equivalent to\footnote{In order to draw a parallel with other $G$-structures later on, the supersymmetry conditions \eqref{eq:strominger1} can be rephrased in terms of the intrinsic torsion of the $\SU 3$ structure as
\begin{equation*}
\dd\omega=\dd\varphi\wedge\omega+\ii(H_{\mathbf{6}}^{(2,1)}-H_{\mathbf{6}}^{(1,2)}),\qquad
\dd\Omega=2\,\dd\varphi\wedge\Omega,\qquad
H=\star\ee^{2\varphi}\dd(\ee^{-2\varphi}\omega)=-i(\partial-\Bar{\partial})\varphi\wedge\omega+H_{\mathbf{6}}.
\end{equation*}}
\begin{equation}\label{eq:strominger1}
	\begin{gathered}
		\dd(\ee^{-2\varphi}\omega\wedge\omega)=0,\qquad \dd(\ee^{-2\varphi}\Omega)=0,\\ H=\ii(\bar{\partial}-\partial)\omega,\qquad\Omega\wedge F=\omega\lrcorner F=0.
	\end{gathered}
\end{equation}
The first condition tells us that $X$ is a complex manifold with a holomorphically trivial canonical bundle. The second condition is equivalent to the hermitian metric on $X$ being conformally balanced. The third condition uniquely defines the three-form field strength $H$ as the torsion of the Bismut connection. The final two conditions imply that $V$ is a holomorphic bundle whose curvature is type $(1,1)$ and satisfies the hermitian Yang--Mills equation.

The supersymmetry conditions alone are not sufficient to ensure the background solves the supergravity equations of motion. One also needs to satisfy the Bianchi identity for the three-form flux, given in \eqref{eq:bianchi}. Note that the choice of connection $\nabla^{+}$ is crucial here: as shown in \cite{0908.2927} and discussed in \cite{Martelli:2010jx,Fernandez:2014kwa}, the supersymmetry conditions plus the Bianchi identity imply the heterotic equations of motion if and only if the connection $\nabla^{+}$ on $TX$ that appears in the Bianchi identity is an $\SU 3$ instanton, i.e.~its curvature two-form satisfies $\feyn{R}^{+}\epsilon=0$. Working to $\mathcal{O}(\ap)$, given the supersymmetry conditions and the Bianchi identity, the Hull connection $\nabla^{+}$ is \emph{automatically} an instanton, so there is no further constraint. However, if one chooses to use some \emph{other} connection in the Bianchi identity, the implied equations of motion will not be those of heterotic supergravity unless that connection is an $\SU 3$ instanton. However, to $\mathcal{O}(\ap)$, the Hull connection is singled out among the possible choices by compatibility of supersymmetry with the equations of motion. In particular, the Chern connection on $TX$ is \emph{not} an instanton in general, and so if one uses the Chern connection in the Bianchi identity, one will \emph{not} recover the correct equations of motion.\footnote{If the background has a large-volume limit, or equivalently a smooth $\ap\to0$ limit, the zeroth-order background is necessarily Calabi--Yau. The curvature two-forms of the Hull and Chern connections then differ only at order $\ap$, whereas one needs the curvature to be an instanton only to zeroth order, so one can use either connection~\cite{delaOssa:2014msa}. However, this is not true for non-K\"ahler solutions with torsion, which are the main focus of this paper.}

The above discussion also holds for spacetimes of the form $\bR^{1,p}\times X_{9-p}$. In all cases, the connection in the Bianchi identity must be an instanton to reproduce the correct equations of motion, and the Hull connection automatically satisfies this condition. For this paper, unless noted otherwise, we always take $\nabla^{+}$ to be the Hull connection to ensure that the supersymmetry conditions in \eqref{eq:strominger1} plus the Bianchi identity give a solution to the correct supergravity equations of motion.

For what follows, it will be useful to distinguish between a supersymmetric geometry and a supersymmetric solution. A supersymmetric geometry will solve the supersymmetry conditions in \eqref{eq:strominger1}:
\begin{equation}
\text{Supersymmetric geometry}\,\equiv\,\begin{gathered}\dd(\ee^{-2\varphi}\omega\wedge\omega)=0,\qquad\dd(\ee^{-2\varphi}\Omega)=0,\\
H=\ii(\bar{\partial}-\partial)\omega,\qquad\Omega\wedge F=0.
\end{gathered}
\label{eq:susy_geometry}
\end{equation}
These are equivalent to $D_m \epsilon=\feyn{D}\epsilon=0$ and the bundle $V$ being holomorphic. A supersymmetric solution is a supersymmetric geometry which also satisfies the Bianchi identity \eqref{eq:bianchi} and the hermitian Yang--Mills equation in \eqref{eq:strominger1}:
\begin{equation}
\text{Supersymmetric solution}\,\equiv\,\text{Supersymmetric geometry}+\text{Bianchi}+\text{HYM}.\label{eq:susy_solution}
\end{equation}
A supersymmetric solution is so-called as it automatically solves the supergravity equations of motion.

When $F=H=0$, the heterotic string on $\bR^{1,3}\times X$ admits solutions where $X$ is Calabi--Yau, i.e.~a Ricci-flat K\"ahler threefold. In this case, Yau's proof of the Calabi conjecture guarantees the existence of a solution given an initial K\"ahler metric on a threefold with $c_1(X)=0$. It is currently not known whether there is a similar existence guarantee for heterotic solutions to the Hull--Strominger system. Specifically, given a compact complex threefold with a trivial canonical bundle and a conformally balanced hermitian metric, it is not known which further conditions are needed to ensure a solution to the Hull--Strominger system exists. There are two obvious conditions that must be satisfied. First, since $H$ is gauge invariant, $\dd H$ is exact and so the right-hand side of the Bianchi identity \eqref{eq:bianchi} must be trivial in cohomology when restricted to $X$. This requires that the first Pontryagin classes of the tangent bundle of $X$ and the vector bundle $V$ over it are equal. Second, the connection on the holomorphic bundle $V$ should satisfy the hermitian Yang--Mills equation, which holds if and only if $V$ is (poly)stable~\cite{Donaldson:1985zz,10.1215/S0012-7094-87-05414-7,https://doi.org/10.1002/cpa.3160390714}. It was conjectured that these are actually sufficient for existence~\cite{Yau2010}, though it now seems that further notions of stability are needed~\cite{Ashmore:2019rkx,Garcia-Fernandez:2023vah}.

\section{Anomaly flow}\label{sec:anomaly_flow}

Known compact solutions to the Hull--Strominger system are few and far between, with the major difficulties coming from the presence of non-K\"ahler geometry in the form of the conformally balanced condition and the non-trivial Bianchi identity for the flux.

\emph{Anomaly flow} as originally defined is a geometric flow for a hermitian metric on a complex threefold~\cite{1508.03315}. Up to a subtlety with the choice of connection, the fixed points of the flow satisfy the Bianchi identity or ``anomaly cancellation condition'' of heterotic string theory, hence the name. In terms of an $\SU 3$ structure on $X$, the flow as originally defined is given by\footnote{The notation used in \cite{1508.03315} is somewhat non-standard compared with the description of the Hull--Strominger system in the physics literature. Our notation is related to that of \cite{1508.03315} (denoted as subscript AF) by
\begin{equation*}
\ee^{-2\varphi}=\Vert\Omega_{\text{AF}}\Vert,\qquad\Omega=\sqrt{8}\frac{\Omega_{\text{AF}}}{\Vert\Omega_{\text{AF}}\Vert},\qquad\omega=\omega_{\text{AF}},\qquad\alpha'=-2\,\alpha'_{\text{AF}},\qquad t=\tfrac{1}{2}t_{\text{AF}}.
\end{equation*}
}
\begin{subequations}\label{eq:flow}
\begin{align}
	\partial_{t}\left(\ee^{-2\varphi}\omega\wedge\omega\right) & =\dd H-\frac{\ap}{4}(\tr F\wedge F-\tr R^{\text{Ch}}\wedge R^{\text{Ch}}),\\
\partial_{t}\left(\ee^{-2\varphi}\Omega\right) & =0,\\
h^{-1}\partial_{t}h & =-\omega\lrcorner F,
\end{align}
\end{subequations}
where the three-form $H$ is fixed to $H=\ii(\bar{\partial}-\partial)\omega$, $R^{\text{Ch}}$ is the curvature of the Chern connection on $TX$, and the gauge field $A$ which defines the curvature $F$ is determined by a hermitian metric $h$ on the fibres of $V$ via $A=h^{-1}\partial h$.\footnote{The flow for $h$ is known as Donaldson heat flow~\cite{10.1215/S0012-7094-87-05414-7}. This is gauge equivalent to Yang--Mills flow~\cite{atiyahbott}, where the connection evolves via $\partial_{t}A=-\dd_{A}^{\dagger}F$, with the right-hand side given by the negative of the gradient of the Yang--Mills functional. (See \cite{donaldson1997geometry} for a summary.)} As written, the fixed points of the flow satisfy
\begin{equation}
\dd H-\frac{\ap}{4}(\tr F\wedge F-\tr R^{\text{Ch}}\wedge R^{\text{Ch}})=0,\qquad\omega\lrcorner F=0,\label{eq:bianchi_chern}
\end{equation}
which corresponds to the Bianchi identity for heterotic supergravity (albeit with the Chern connection rather than the Hull connection) and the curvature $F$ being primitive.

In more detail, consider a complex threefold $X$ with a trivial canonical bundle and a vector bundle $V$ such that $p_1(X)=p_1(V)$. Choose an initial nowhere-vanishing holomorphic $(3,0)$-form $\Omega$, a scalar dilaton field $\varphi$ and a hermitian form $\omega$ which satisfy
\begin{equation}
\dd(\ee^{-2\varphi}\Omega)=0,\qquad\dd(\ee^{-2\varphi}\omega\wedge\omega)=0,\label{eq:susy}
\end{equation}
so that $\ee^{-2\varphi}\Omega$ is holomorphic and the hermitian metric is conformally balanced.\footnote{Note that there are methods for constructing conformally balanced metrics, so this initial data is not particularly constraining. See, for example, \cite{Tosatti:2013yla}.} The flow in \eqref{eq:flow} preserves both of these conditions. The three-form $\ee^{-2\varphi}\Omega$ does not flow at all, so if it was initially closed, it remains closed along the flow. The four-form $\ee^{-2\varphi}\omega\wedge\omega$ flows by a closed four-form, and so if it was closed at the start of the flow, it remains closed along it. In addition, given an initial hermitian metric $h$ on the bundle $V$, the corresponding curvature $F$ is holomorphic, $\Omega\wedge F=0$.
Since the flow for the gauge field $A$ is defined in terms of $h$, the flow also preserves holomorphicity of $F$. Thus, given initial data which solves \eqref{eq:susy}, the endpoint of the flow in \eqref{eq:flow}, if it exists, is a solution to the Bianchi identity and the hermitian Yang--Mills equation. Thus, anomaly flow seems to provide solutions to the Hull--Strominger system within the conformally balanced class of the initial hermitian metric. Given this, we should think of the anomaly flow as generalised Ricci flow where the initial data is a supersymmetric geometry and the flow itself preserves supersymmetry.

Note that, since $\ee^{-2\varphi}\Omega$ does not flow, the associated complex structure (which can be defined using only $\Omega$~\cite{Hitchin:2000jd}) is invariant along the flow. If $\Omega$ and $\omega$ are to continue to define an $\SU3$ structure along the flow (as they should), this means that $\omega\wedge\omega$ should remain a $(2,2)$-form. Looking at the first of the flow equations in \eqref{eq:flow}, $\dd H=2\ii\partial\bar{\partial}\omega$ and $\tr F\wedge F$ are manifestly $(2,2)$-forms, and, thanks to the properties of the Chern connection, the final term $\tr R^{\text{Ch}}\wedge R^{\text{Ch}}$ is also type $(2,2)$ and so there is no issue.

As shown in \cite{1610.02739}, thanks to the properties of hermitian forms on $n$-folds, given an initial conformally balanced metric, the flow of the four-form in \eqref{eq:flow} implies a flow for the hermitian metric and the dilaton:
\begin{equation}
\partial_{t}g_{mn}=\tfrac{1}{4}\ee^{2\varphi}\omega_{m}{}^{p_{1}}\omega^{p_{2}p_{3}}\mathcal{B}^{\text{Ch}}_{np_{1}p_{2}p_{3}},\qquad\partial_{t}\varphi=\tfrac{1}{32}\ee^{2\varphi}\omega^{m_{1}m_{2}}\omega^{m_{3}m_{4}}\mathcal{B}^{\text{Ch}}_{m_{1}m_2m_3 m_{4}},\label{eq:flow_in_H}
\end{equation}
where
\begin{equation}
\mathcal{B}^{\text{Ch}}=2\ii\partial\bar{\partial}\omega-\frac{\ap}{4}(\tr F\wedge F-\tr R^{\text{Ch}}\wedge R^{\text{Ch}})
\end{equation}
is simply the Bianchi identity in \eqref{eq:bianchi_chern} with $H$ replaced by $\ii(\bar{\partial}-\partial)\omega$. The flow for the metric can then be written as a modified Ricci flow of the form
\begin{equation}
\partial_{t}g_{mn}=-\ee^{2\varphi}R_{mn}+\dots
\end{equation}
where the other terms are built from the ``torsion'' or three-form flux $H$, and further correction terms appearing at $\mathcal{O}(\ap)$ which are quadratic in the curvature. As is the case for standard Ricci flow, one might expect that the flow for both the metric and dilaton is simply the equations of motion for each field. However, due to the use of the Chern connection in \eqref{eq:bianchi_chern}, there is generally no clean relation between the anomaly flow as currently formulated and the heterotic supergravity equations of motion. In particular, fixed points of the flow do not correspond to solutions of heterotic supergravity to $\mathcal{O}(\ap)$. As we will see in the next section, this can be fixed by using the Hull connection in the Bianchi identity.

\subsection{Equations of motion}\label{eq:eom}

As we show in Appendix \ref{app:flow_equation}, without $\ap$ corrections, the flow of the metric \eqref{eq:flow_in_H} can be rewritten as
\begin{equation}
\partial_{t}g_{mn}=-\ee^{2\varphi}\left(R_{mn}+2\nabla_{m}\nabla_{n}\varphi-\tfrac{1}{4}H_{mpq}H_{n}{}^{pq}\right).\label{eq:zeroth_metric}
\end{equation}
The right-hand side here is simply the equation of motion for the metric in heterotic supergravity to zeroth order in $\ap$. Since the $\mathcal{O}(\ap)$ term in the anomaly condition \eqref{eq:bianchi} is essential for non-trivial $H$-flux when $X$ is compact, we need to understand how the flow is altered by the inclusion of the leading $\ap$ corrections.

Using the flow in \eqref{eq:flow} with $R^{\text{Ch}}$ given by the curvature of the Chern connection, the link with heterotic supergravity breaks down at $\mathcal{O}(\ap)$: for example, one finds additional terms in $\partial_{t}g_{mn}$ which prevent it from being interpreted as the supergravity equation of motion for the metric to $\mathcal{O}(\ap)$. This is simply reflecting the fact that if one uses an arbitrary connection in the anomaly condition \eqref{eq:bianchi}, the supergravity equations of motion and the anomaly condition are \emph{not} compatible with $N=1$ supersymmetry at $\mathcal{O}(\ap)$. Instead, supersymmetry requires that the connection is chosen to be the Hull connection $\nabla^{+}$. If one does this and takes the curvature in the anomaly flow to be defined by the Hull connection, one recovers the link with the supergravity equations of motion and its interpretation as a modified Ricci flow.

To emphasise this point, suppose that one defines anomaly flow with an arbitrary connection $\tilde{\nabla}$ in the Bianchi identity:
\begin{equation}
\partial_{t}(\ee^{-2\varphi}\omega\wedge\omega)=2\ii\partial\bar{\partial}\omega-\frac{\ap}{4}(\tr F\wedge F-\tr\tilde{R}\wedge\tilde{R}).
\end{equation}
The flow of the metric can be extracted from this equation by contracting with the hermitian form, giving
\begin{equation}
\partial_{t}g_{mn}=\frac{1}{24}\ee^{2\varphi}(\omega\wedge\omega)_{m}{}^{p_{1}p_{2}p_{3}}\left(2\ii\partial\bar{\partial}\omega-\frac{\ap}{4}(\tr F\wedge F-\tr\tilde{R}\wedge\tilde{R})\right)_{np_{1}p_{2}p_{3}}.\label{eq:metric_flow}
\end{equation}
Assuming a supersymmetric geometry, one has
\begin{equation}
-\frac{1}{24}(\omega\wedge\omega)_{m}{}^{p_{1}p_{2}p_{3}}(2\ii\partial\bar{\partial}\omega)_{np_{1}p_{2}p_{3}}=R_{mn}+2\nabla_{m}\nabla_{n}\varphi-\frac{1}{4}H_{mpq}H_{n}{}^{pq},
\end{equation}
which we recognise as the equation of motion for the metric restricted to $X$. So, to zeroth order in $\ap$, the flow equation for the metric is given by \eqref{eq:zeroth_metric}, independent of the choice of connection $\tilde{\nabla}$.

What about the $\mathcal{O}(\ap)$ correction in \eqref{eq:metric_flow}? Using the $\SU 3$ invariant spinor $\epsilon$, one can show that
\begin{equation}
\frac{1}{24}(\omega\wedge\omega)_{m}{}^{p_{1}p_{2}p_{3}}(\tr\tilde{R}\wedge\tilde{R})_{np_{1}p_{2}p_{3}}=-\tr\tilde{R}_{mp}\tilde{R}_{n}{}^{p}-\tr\epsilon^{\dagger}\tilde{R}_{mp}\gamma_{n}{}^{p}\tilde{\feyn R}\epsilon,
\end{equation}
with an analogous expression for the $\tr F\wedge F$ term. The calculations in Appendix \ref{app:flow_equation} then show that the flow equation can be written as
\begin{equation}
\partial_{t}g_{mn}=-\ee^{2\varphi}\text{eom}[g]_{mn}+\frac{\ap}{4}\ee^{2\varphi}\left(\tr\epsilon^{\dagger}F_{mp}\gamma_{n}{}^{p}\feyn F\epsilon-\epsilon^{\dagger}\tilde{R}_{mp}\gamma_{n}{}^{p}\tilde{\feyn R}\epsilon\right)+\mathcal{O}(\ap^{2}),\label{eq:extra}
\end{equation}
where $\text{eom}[g]_{mn}$ includes the $\mathcal{O}(\ap)$ terms from \eqref{eq:metric_eom}. At a fixed point of the flow, the gauge field satisfies hermitian Yang--Mills and so the $F$-dependent term vanishes due to $\feyn{F}\epsilon=0$. The metric then solves its own heterotic supergravity equation of motion up to the extra term quadratic in $\tilde{R}$. Since we want a stationary point of the flow to correspond to an honest supergravity solution to $\mathcal{O}(\ap)$, we must insist that the $\tilde{R}$-dependent term either vanishes at the fixed point or is at least $\mathcal{O}(\ap^{2})$ and can be dropped to this order in the $\ap$ expansion. Contracting the $\tilde{R}$-dependent term with $g^{mn}$, one notes that it takes the form $\tr\epsilon^{\dagger}\feyn{\tilde{R}}\feyn{\tilde{R}}\epsilon=\tr|\feyn{\tilde{R}}\epsilon|^{2}$. Since this is positive definite, it vanishes if and only if $\feyn{\tilde{R}}\epsilon=0$. The extra term in \eqref{eq:extra} then vanishes to $\mathcal{O}(\ap)$ for $\feyn{\tilde{R}}\epsilon=0+\mathcal{O}(\ap)$. This requires that $\tilde{\nabla}$ is an $\SU 3$ instanton to \emph{zeroth order} in $\ap$, which singles out the Hull connection $\nabla^{+}$ as the unique choice. As we mentioned earlier, the Chern connection on $TX$ is \emph{not} an instanton in general~\cite{Martelli:2010jx}. Thus, if one takes $\tilde{\nabla}$ to be the Chern connection, the extra term in \eqref{eq:extra} will not be second order in $\ap$ at a fixed point and the metric will not solve its own equation of motion.

We have argued that one must use the Hull connection $\nabla^{+}$ in the Bianchi identity, and the definition of anomaly flow, if one wants fixed points of the flow to correspond to $\mathcal{O}(\ap)$ solutions to the heterotic supergravity equations of motion. We now see how this affects the evolution equations for the metric and dilaton.

\subsection{Anomaly flow with the Hull connection}

For a supersymmetric geometry satisfying \eqref{eq:susy_geometry} and taking the connection in the Bianchi identity to be the Hull connection $\nabla^+$, the flow equation in \eqref{eq:extra} for the metric to $\mathcal{O}(\ap)$ takes the form\footnote{See Appendix \ref{app:flow_equation} for this calculation.}
\begin{equation}
\partial_{t}g_{mn}=-\ee^{2\varphi}\text{eom}[g]_{mn}+\frac{\ap}{4}\ee^{2\varphi}\left(\tr\epsilon^{\dagger}F_{mp}\gamma_{n}{}^{p}\feyn F\epsilon-\epsilon^{\dagger}R_{mp}^{+}\gamma_{n}{}^{p}\feyn{R}^{+}\epsilon\right)+\mathcal{O}(\ap^{2}),\label{eq:flow_metric}
\end{equation}
where the first term is the heterotic equation of the motion for the metric correct to $\mathcal{O}(\ap)$ (given by the restriction of \eqref{eq:metric_eom} to the threefold $X$) and the ``extra'' terms are quadratic in $F$ and $R^{+}$. At this point one might worry that at fixed points, one does not satisfy the supergravity equation of motion, $\text{eom}[g]_{mn}=0$, but instead some corrected equation with extra quadratic curvature terms. However, this is not the case, and it is again the choice of the Hull connection that is crucial. First, the term quadratic in the curvature $R^{+}$ is actually proportional to $\dd H$ thanks to the identity
\begin{equation}
R_{mn}^{+}{}^{pq}\gamma^{mn}\epsilon=-\tfrac{1}{2}\dd H_{mn}{}^{pq}\gamma^{mn}\epsilon,
\end{equation}
which is valid for a supersymmetric geometry.
At a fixed point of the flow, by definition, one satisfies the Bianchi identity for $H$, and so $\dd H\sim\mathcal{O}(\ap)$. The extra term quadratic in $R^{+}$ is then $\mathcal{O}(\ap^{2})$ and so should be dropped to the order of $\ap$ we are working at. Second, given the flow for the gauge connection, we have $\feyn{F}\epsilon=0$ at a fixed point, and so the extra term quadratic in $F$ also vanishes. Thus, at a fixed point one finds
\begin{equation}
\partial_{t}g_{mn}=0\equiv-\ee^{2\varphi}\text{eom}[g]_{mn}+\mathcal{O}(\ap^{2}).
\end{equation}
In other words, the metric at the fixed point automatically solves its own heterotic supergravity equation of motion to $\mathcal{O}(\ap)$.

One sees a similar structure in the flow equation for the dilaton upon choosing the Hull connection in the anomaly condition:
\begin{equation}\label{eq:flow_dilaton}
\partial_{t}\varphi=-\ee^{2\varphi}\text{eom}[\varphi]+\frac{\ap}{16}\ee^{2\varphi}\left(\tr\epsilon^{\dagger}\feyn{F}\feyn{F}\epsilon-\tr\epsilon^{\dagger}\feyn{R}^{+}\feyn{R}^{+}\epsilon\right)+\mathcal{O}(\ap^{2}).
\end{equation}
For the same reasons as above, we see that at fixed points of the flow, the dilaton obeys its own supergravity equation of motion to $\mathcal{O}(\ap)$.\footnote{Note that when written in terms of the equations of motion, it is not obvious that the flow equations for the metric and dilaton preserve the warped volume measure $\ee^{-4\varphi}\vol$, while it follows automatically from $\partial_t(\ee^{-2\varphi}\Omega)=0$. The resolution to this apparent discrepancy is that for a supersymmetric geometry, there are extra relations between $\text{eom}[g]_{mn}$ and $\text{eom}[\varphi]$ which ensure $\partial_t\varphi = \tfrac{1}{8} g^{mn}\partial_t g_{mn}$. These are manifest when the flows are written in terms of the Bianchi identity, or can be computed explicitly, as we do in Appendix \ref{app:flow_equation}.}

Having found a direct link between the heterotic supergravity equations of motion and anomaly flow with the Hull connection, from here onwards when we refer to anomaly flow on a complex three-fold we always mean
\begin{subequations}\label{eq:flow_Hull}
\begin{align}
	\partial_{t}\left(\ee^{-2\varphi}\omega\wedge\omega\right) & =\mathcal{B},\\
\partial_{t}\left(\ee^{-2\varphi}\Omega\right) & =0,\\
h^{-1}\partial_{t}h & =-\omega\lrcorner F,
\end{align}
\end{subequations}
where $\dd H= 2\ii\partial\bar{\partial}\omega$ and $\mathcal{B}$ is the Bianchi identity with the curvature defined by the Hull connection:
\begin{equation}\label{eq:B_def}
\mathcal{B}=\dd H -\frac{\ap}{4}(\tr F\wedge F-\tr R^{+}\wedge R^{+}).
\end{equation}

\subsection{The \texorpdfstring{$\ap$}{alpha prime} expansion}

Before moving to the next section, we pause to discuss the importance of the $\ap$ expansion. As originally formulated, the anomaly flow is treated as a closed system that is exact in $\ap$. This was one of the reasons for choosing the Chern connection in the Bianchi identity for $H$ -- the Chern connection is always of complex type $(2,2)$, which agrees with the complex type of both $\partial\bar\partial\omega$ and $\tr F\wedge F$, and so the condition that the Bianchi identity is satisfied is not over constrained. As we will see in Section \ref{sec:diverse}, if one uses the Hull connection, it generally produces a non-$(2,2)$ component, and so there is no solution to the Bianchi identity \emph{if one assumes the system is exact in $\ap$}. However, this is not the case. The heterotic action, the equations of motion, the supersymmetry conditions and the Bianchi identity itself are correct only to $\mathcal{O}(\ap)$, with all of these receiving corrections at higher orders~\cite{Bergshoeff:1988nn,Bergshoeff:1989de}. For example, if the first-order Bianchi identity is satisfied, $\feyn{R}^+\epsilon=0+\mathcal{O}(\ap)$. Looking at the flow equation for the metric in \eqref{eq:flow_metric}, including the correction term leads to a new $\mathcal{O}(\ap^2)$ term, which one then wants to absorb in the metric equation of motion. However, the supersymmetry variations should then be adjusted in order to maintain compatibility with the action and equations of motion. This process is then expected to continue to all orders in $\ap$.\footnote{See also \cite{1407.7542} for an argument that supersymmetry plus the heterotic Lichnerowicz identity implies corrections to all orders in $\ap$.} With this in mind, one should be consistent and work only to order $\ap$ for both the equations we are trying to solve and the solutions to those equations.

\section{Anomaly flow from a functional}\label{sec:functional}

The aim of this section is to find a functional description of anomaly flow \eqref{eq:flow_Hull}. The impetus behind this is to find a dimension-agnostic formulation of the flow, which can then generalised to seven- and eight-manifolds with $G$-structures. Along the way, we will see that anomaly flow is intimately related to a certain heterotic Lichnerowicz identity that relates the bosonic action to squares of supersymmetry operators.

To begin, let us briefly recall Perelman's description of standard Ricci flow as a \emph{gradient flow}~\cite{math/0211159}. Given the functional
\begin{equation}
\mathcal{S}[g,\varphi]=\int_{X}\ee^{-2\varphi}\vol\left(R+4(\nabla\varphi)^{2}\right),
\end{equation}
which is simply the string-frame action for the metric and dilaton restricted to $X$, its variation is
\begin{equation}
\begin{aligned}
\delta \mathcal{S}[g,\varphi]&=\int_{X}\ee^{-2\varphi}\vol\Bigl[-\delta g_{mn}(R^{mn}+2\nabla^{m}\nabla^{n}\varphi)\\
&\eqspace\phantom{\int_{X}\ee^{-2\varphi}\vol[}+\left(\tfrac{1}{2}g^{mn}\delta g_{mn}-2\delta\varphi\right)\left(R+4\nabla^{2}\varphi-4(\nabla\varphi)^{2}\right)\Bigr].
\end{aligned}
\end{equation}
The variation $\tfrac{1}{2}g^{mn}\delta g_{mn}-2\delta\varphi$ vanishes identically if and only if the variations are correlated so that the warped volume measure $\ee^{-2\varphi}\vol$ is held fixed. Defining a new measure $\dd m=\ee^{-2\varphi}\vol$ and the dilaton by $\varphi=\tfrac{1}{2}\log(\text{vol}/\dd m)$, the symmetric tensor $-(R^{mn}+2\nabla^{m}\nabla^{n}\varphi)$ is the gradient of the functional
\begin{equation}
\mathcal{S}[g]=\int_{X}\dd m\left(R+4(\nabla\varphi)^{2}\right),
\end{equation}
where the functional derivative is defined as
\begin{equation}
    \delta \mathcal{S}[g] = \int_X \dd m\, \frac{\delta \mathcal{S}[g]}{\delta g_{mn}} \delta g_{mn}.
\end{equation}
The evolution of the functional is according to
\begin{equation}\label{eq:functional_time}
\frac{\dd}{\dd t}\mathcal{S}[g]=-\int_{X}\dd m\,(R^{mn}+2\nabla^{m}\nabla^{n}\varphi)\partial_t g_{mn}.
\end{equation}
Given a choice of measure $\dd m$, Perelman then considered the gradient flow
\begin{equation}\label{eq:Ricci}
\partial_{t}g_{mn}=\frac{\delta \mathcal{S}[g]}{\delta g_{pq}}g_{mp}g_{nq}\equiv-(R_{mn}+2\nabla_{m}\nabla_{n}\varphi)
\end{equation}
of $\mathcal{S}[g]$, with the flow of $\varphi$ defined implicitly from the invariance of $\ee^{-2\varphi}\vol$.  Substituting this into \eqref{eq:functional_time}, the functional itself evolves as
\begin{equation}
\frac{\dd}{\dd t}\mathcal{S}[g]=\int_{X}\dd m\,| \text{Ric}+2\op{Hess}\varphi|^2\geq 0 ,
\end{equation}
so that $\mathcal{S}[g]$ is monotonically increasing under the flow. When the flow 
 \eqref{eq:Ricci} exists, it is simply Ricci flow up to time-dependent ($\varphi$-dependent) diffeomorphism.\footnote{This is a particular case of the ``DeTurck trick''.} Thus, a choice of measure $\dd m$ is akin to a choice of gauge, with the resulting flows regarded as the same. 
 
\subsection{A rewriting of anomaly flow}

As written, the anomaly flow equations in \eqref{eq:flow_Hull} are written in terms of the forms $\omega$ and $\Omega$ which define an $\SU 3$ structure on $X$. Recall that one can equivalently encode an $\SU 3$ structure using a metric $g$ on $X$ and a spinor $\epsilon$ transforming as a singlet under $\SU 3$. The dictionary between these two is
\begin{equation}
\omega_{mn}=-\ii\epsilon^{\dagger}\gamma_{mn}\gamma_{*}\epsilon,\qquad\Omega_{mnp}=-\ii\epsilon^{\dagger}\gamma_{mnp}(1+\gamma_{*})\epsilon,
\end{equation}
where the vielbein $e_{m}{}^{a}$ for the metric is used to convert the flat gamma matrices to coordinate indices, $\gamma_{m}=e_{m}{}^{a}\gamma_{a}$. The flow equations in \eqref{eq:flow_Hull} can then be rewritten as evolution equations for the dilaton, vielbein and the spinor as
\begin{subequations}\label{eq:anomaly_flow_equations}
\begin{align}
\partial_{t}\varphi & =-\tfrac{1}{4}\ee^{2\varphi}\epsilon^{\dagger}\feyn{\mathcal{B}}\epsilon,\label{eq:flow1}\\
\partial_{t}e_{m}{}^{a} & =-\tfrac{1}{4}\ee^{2\varphi}\tfrac{1}{3!}\epsilon^{\dagger}\gamma_{m}{}^{n_{1}n_{2}n_{3}}\epsilon\,\mathcal{B}^{a}{}_{n_{1}n_{2}n_{3}},\\
\partial_{t}\epsilon & =0,\label{eq:flow3}
\end{align}
\end{subequations}
where $\mathcal{B}$ is the four-form defined in \eqref{eq:B_def}. Note that the variations of the vielbein and dilaton imply that the warped volume measure $\ee^{-4\varphi} \vol$ is invariant under the flow. 

Having recast the anomaly flow in this way, it is straightforward to see that the right-hand side of the flow equations can be obtained as the field equations of the functional
\begin{equation}\label{eq:6d_functional}
I=\int_{X}\ee^{-2\varphi}\vol\epsilon^{\dagger}\feyn{\mathcal{B}}\epsilon=\int_{X}\ee^{-2\varphi}\omega\wedge\mathcal{B},
\end{equation}
where one uses the identity $\epsilon^{\dagger}\gamma_{(4)}\epsilon=\star\omega$ to obtain the second equality. More precisely, $I$ is treated as a pseudo-action where $\mathcal{B}$ is a background field and so is not itself varied. (We come back to this issue in Section \ref{sec:gradient}, where we give a functional that does not have this constraint.) We then define a functional derivative with respect to the fields $(\varphi,e,\epsilon)$ as\footnote{Note that there is an ambiguity in the variation with respect to $\epsilon$, as it is a variation in the space of Majorana spinors with unit norm $\epsilon^{\dagger}\epsilon=1$, so $\epsilon^{\dagger}\delta\epsilon=0$. This ambiguity is fixed by imposing $\epsilon^{\dagger}\frac{\delta I}{\delta \epsilon}=0$.}
\begin{equation}
    \delta I=\int_X\ee^{-2\varphi}\vol\left(\frac{\delta I}{\delta\varphi}\delta\varphi+\frac{\delta I}{\delta e_m{}^a}\delta e_{m}{}^{a}+\left(\frac{\delta I}{\delta \epsilon}\right)^{\dagger}\delta\epsilon+\delta\epsilon^{\dagger}\frac{\delta I}{\delta \epsilon}\right),
\end{equation}
where the integral is weighted by $\ee^{-2\varphi}\vol$. The functional derivatives of $I$ are easily found to be
\begin{subequations}\label{eq:variation}
\begin{align}
\frac{\delta I}{\delta\varphi}&=-2\epsilon^{\dagger}\feyn{\mathcal{B}}\epsilon,\label{eq:variation1}\\
\frac{\delta I}{\delta e_m{}^a}&=-\tfrac{1}{3!}\epsilon^{\dagger}\gamma^{m n_{1}\dots n_{3}}\epsilon\,\mathcal{B}_{a n_{1}\dots n_{3}}+\epsilon^{\dagger}\feyn{\mathcal{B}}\epsilon \,e^{m}{}_{a},\\
\frac{\delta I}{\delta\epsilon}&=\feyn{\mathcal{B}}\epsilon-(\epsilon^{\dagger}\feyn{\mathcal{B}}\epsilon)\epsilon.\label{eq:variation3}
\end{align}
\end{subequations}
We then define the flow equations as
\begin{subequations}\label{eq:flow_I}
\begin{align}
    \partial_{t}\varphi&=\frac{1}{8}\ee^{2\varphi}\frac{\delta I}{\delta\varphi}, \label{eq:flow1_I}\\
    \partial_{t}e_{m}{}^{a}&=\frac{1}{4}\ee^{2\varphi}\left(\frac{\delta I}{\delta e_n{}^b}g_{mn}\eta^{ab}+\frac{1}{2}\frac{\delta I}{\delta \varphi}e_{m}{}^{a}\right),\\
    \partial_{t}\epsilon&=-\frac{1}{4}\ee^{2\varphi}\frac{\delta I}{\delta \epsilon}\label{eq:flow3_I},
\end{align}
\end{subequations}
with the flow for gauge connection given by the flow of the fibre metric in \eqref{eq:flow_Hull}. The particular combination of variations that appears here is correlated with how one defines the heterotic equations of motion as variations of the string-frame action -- see Appendix \eqref{app:eom}. One then imposes that $\mathcal{B}$ is the heterotic Bianchi identity, with supersymmetry fixing the three-form $H$ as a function of the other fields via $H=\ii(\bar{\partial}-\partial)\omega$.

Using the above equations, one can obtain the flow of $\omega\wedge\omega$ and $\Omega$ as
\begin{subequations}\label{eq:general_B_flow}
\begin{align}
\partial_t \left(\ee^{-2\varphi}\omega\wedge\omega\right) &= \mathcal{B},\\
\partial_{t}\left(\ee^{-2\varphi}\Omega\right)&=-\frac{3\ii}{4}\omega\wedge (\Omega\lrcorner\mathcal{B}).
\end{align}
\end{subequations}
When $\mathcal{B}$ is a $(2,2)$-form -- at least to $\mathcal{O}(\ap)$ -- one has $\Omega\lrcorner\mathcal{B}=0$ and so $\ee^{-2\varphi}\Omega$ does not flow. Looking back to the variations of the functional, from \eqref{eq:variation3} we also see that $\feyn{\mathcal{B}}\epsilon=(\epsilon^{\dagger}\feyn{\mathcal{B}}\epsilon)\epsilon$ and so the spinor does not flow either. The flow equations for $(\varphi,e,\epsilon)$ then exactly match those of the anomaly flow \eqref{eq:anomaly_flow_equations}, with the flow of the four-form and three-form in \eqref{eq:general_B_flow} reducing to \eqref{eq:flow}.

Given this observation, one might wonder whether there is a straightforward way to see that the metric and the dilaton solve their own supergravity equations of motion at the fixed points of the flow. We showed this in the previous section in a somewhat roundabout way using various ad hoc identities that were specific to $\SU3$ structures. As we will see, there is a more general and direct path to this result via supersymmetry.

\subsection{A Lichnerowicz identity in heterotic supergravity}

As shown in \cite{1407.7542,1702.01156}, a generalisation of the Lichnerowicz formula leads to an expression for the ten-dimensional bosonic action $S_{10}$ in a ``BPS-squared'' form as
\begin{equation}
\tfrac{1}{4}S_{10}=\int_{M_{10}}\ee^{-2\varphi}\vol_{10}\left[|\feyn D\varepsilon|^{2}-|D\varepsilon|^{2}+\frac{\ap}{16}\bigl(\tr\bar\varepsilon\feyn F\feyn F\varepsilon-\tr\bar\varepsilon\feyn R^{+}\feyn R^{+}\varepsilon\bigr)\right],\label{eq:10d_BPS}
\end{equation}
where the ten-dimensional supersymmetry parameter $\varepsilon$ is assumed to have unit norm. The differential operators $D$ and $\feyn D$ are simply those that appear in the gravitino and modified dilatino variations:
\begin{subequations}
\begin{align}
\delta\Psi_{M} & =D_{M}\varepsilon\equiv(\nabla_{M}+\tfrac{1}{8}H_{MNP}\Gamma^{NP})\varepsilon,\label{eq:susy_op1}\\
\Gamma^{M}\delta\Psi_{M}-\delta\lambda & =\feyn D\varepsilon\equiv(\feyn{\nabla}+\tfrac{1}{4}\feyn H-\feyn{\partial}\varphi)\varepsilon.\label{eq:susy_op2}
\end{align}
\end{subequations}
Note that $\feyn D$ is not defined as the usual contraction with a gamma matrix: $\feyn D\neq\Gamma^{M}D_{M}$. The equality in \eqref{eq:10d_BPS} holds after integration by parts and taking vanishing boundary conditions for the fields, and also requires that $H$ satisfies the Bianchi identity. From \eqref{eq:10d_BPS}, it is then straightforward to see that supersymmetric solutions -- those satisfying $\feyn D\varepsilon=D\varepsilon=\feyn F\varepsilon=0$ (with $\feyn R^+\varepsilon=0$ following automatically from the definition of the Hull connection) -- have vanishing action, and, upon varying \eqref{eq:10d_BPS}, that supersymmetric solutions solve the equations of motion provided the Bianchi identity for $H$ holds.

As in Section \ref{sec:heterotic_supergravity}, consider a background where the metric is that of four-dimensional Minkowski times a six-manifold $X$. The Majorana--Weyl supersymmetry parameter $\varepsilon$ has a corresponding decomposition in terms of a constant Weyl spinor $\eta_\pm$ on Minkowski space and a nowhere-vanishing spinor $\epsilon$ on $X$:
\begin{equation}
\varepsilon=\eta_{+}\otimes\epsilon_{+}+\eta_{-}\otimes\epsilon_{-},
\end{equation}
where $\epsilon_{\pm}$ are the chiral/anti-chiral parts of the Majorana spinor $\epsilon$. There is then a six-dimensional version of the Lichnerowicz-like identity which relates the bosonic action $S_{\text{B}}$ restricted to $X$ to squares of supersymmetry operators:
\begin{equation}\label{eq:Lichnerowicz}
\tfrac{1}{4}S_{\text{B}}=\int_{X}\ee^{-2\varphi}\vol\left[|\feyn D\epsilon|^{2}-|D\epsilon|^{2}+\frac{\ap}{16}\bigl(\tr\epsilon^{\dagger}\feyn F\feyn F\epsilon-    \tr\epsilon^{\dagger}\feyn{R}^{+}\feyn{R}^{+}\epsilon\bigr)+\tfrac{1}{4}\epsilon^{\dagger}\feyn{\mathcal{B}}\epsilon\right],
\end{equation}
where $\mathcal{B}$ is the Bianchi identity defined in \eqref{eq:B_def}. This gives the analogue of \eqref{eq:10d_BPS} restricted to $X$ \emph{without} assuming that the Bianchi identity holds.

Let us pause to see where this identity comes from. For ease of presentation, we focus on the six-dimensional version, though an identical version holds in ten dimensions, with only the zeroth-order $\ap$ terms. The standard Lichnerowicz (or Lichnerowicz--Weitzenböck) formula~\cite{Lichnerowicz} identifies a particular quadratic combination of differential operators acting on spinors, built from the Levi-Civita derivative $\nabla_{m}$ and the Dirac operator $\feyn{\nabla}=\gamma^{m}\nabla_{m}$, which is tensorial. The difference of squares of the operators acts on a spinor $\epsilon$ as multiplication by the Ricci scalar:
\begin{equation}
    (\nabla^{m}\nabla_{m}-\feyn{\nabla}^{2})\epsilon=\tfrac{1}{4}R\,\epsilon.
\end{equation}
Remarkably, this identity admits a generalisation to torsional backgrounds. Consider the following operators
\begin{subequations}
\begin{align}
D_{m}\epsilon&=\nabla_{m}\epsilon+\alpha_{1}\,\tfrac{1}{2!}H_{mn_{1}n_{2}}\gamma^{n_{1}n_{2}}\epsilon,\\
\feyn{D}\epsilon&=\gamma^{m}\nabla_{m}\epsilon+\alpha_{2}\,\tfrac{1}{3!}H_{m_{1}\dots m_{3}}\gamma^{m_{1}\dots m_{3}}\epsilon+\alpha_{3}\,\nabla_{m}\varphi\,\gamma^{m}\epsilon,
\end{align}
\end{subequations}
parametrised by three real constants $\alpha_{1}$, $\alpha_{2}$ and $\alpha_{3}$. Notice that the Dirac operator $\feyn{D}$ is no longer given by $\gamma^m D_m$. The difference of their squares takes the following form
\begin{equation}
\begin{split}
(D^{m}D_{m}-\feyn{D}^{2})\epsilon&=\tfrac{1}{4}(R-12(\alpha_{1}^{2}-\tfrac{1}{3}\alpha_{2}^{2})H^{2}-4\alpha_{3}^{2}(\nabla\varphi)^{2}-4\alpha_{3}\nabla^{m}\nabla_{m}\varphi)\epsilon\\
&\eqspace-\alpha_{2}\,\tfrac{1}{4!}\dd H_{m_{1}\dots m_{4}}\gamma^{m_{1}\dots m_{4}}\epsilon\\
&\eqspace+\tfrac{1}{2}(\alpha_{1}-\alpha_{2})(\star\dd \star H)_{n_{1}n_{2}}\gamma^{n_{1}n_{2}}\epsilon\\
&\eqspace+\tfrac{1}{4}(\alpha_{1}^{2}-\alpha_{2}^{2})H_{m_{1}m_{2}}{}^{n}H_{m_{3}m_{4}n}\gamma^{m_{1}\dots m_{4}}\epsilon\\
&\eqspace+(\alpha_{1}-\alpha_{2})H_{m_{1}m_{2}}{}^{n}\gamma^{m_{1}m_{2}}\nabla_{n}\epsilon\\
&\eqspace-2\alpha_{3}\,\nabla^{m}\varphi\,(\nabla_{m}\epsilon+\alpha_{2}\,\tfrac{1}{2!}H_{mn_{1}n_{2}}\gamma^{n_{1}n_{2}}\epsilon),
\end{split}
\label{eq:difference_squares}
\end{equation}
where $H^{2}=\tfrac{1}{3!}H_{m_{1}\dots m_{3}}H^{m_{1}\dots m_{3}}$ and $(\nabla\varphi)^{2}=\nabla_{m}\varphi\nabla^{m}\varphi$. For arbitrary choices of the $\alpha_i$, this combination is not tensorial: in addition to scalars multiplying $\epsilon$, it contains two- and four-forms acting via gamma matrices, as well as derivatives $\nabla\epsilon$. Many of these terms can be cancelled by imposing $\alpha_{1}=\alpha_{2}$, leaving only two non-tensorial pieces: the four-form $\dd H$ and the term involving $\nabla\varphi$. Taking $H$ to be closed and $\alpha_{3}=0$, these terms disappear and one recovers Bismut's generalisation of the Lichnerowicz formula~\cite{Bismut:1989mzu}.

The inclusion of the dilaton relies on the following observation: with $\alpha_1=\alpha_2$, the non-tensorial term proportional to $\alpha_{3}$  can be written as $\nabla^{m}\varphi\,D_{m}\epsilon$. Hence, while \eqref{eq:difference_squares} is not tensorial as written, it can be made tensorial by integration by parts.\footnote{A necessary condition is the existence of a normalisable spinor $\epsilon$ on $X$ that can be set to $\epsilon^{\dagger}\epsilon=1$.} Taking into account the conformal factor $\ee^{-2\varphi}$ appearing in the heterotic action, the combination $\ee^{-2\varphi}(|D\epsilon|^{2}-|\feyn{D}\epsilon|^{2})$ can be recast in the form
\begin{equation}
\begin{split}
\ee^{-2\varphi}((D^{m}\epsilon)^{\dagger}D_{m}\epsilon-(\feyn{D}\epsilon)^{\dagger}\feyn{D}\epsilon)&=-\ee^{-2\varphi}\epsilon^{\dagger}(D^{m}D_{m}-\feyn{D}^{2})\epsilon\\
&\eqspace+2\,\ee^{-2\varphi}\nabla^{m}\varphi\,\epsilon^{\dagger}D_{m}\epsilon-2(\alpha_{3}+1)\ee^{-2\varphi}\nabla^{m}\varphi\,\epsilon^{\dagger}\gamma_{m}\feyn{D}\epsilon\\
&\eqspace+\nabla^{m}(\ee^{-2\varphi}\epsilon^{\dagger}(D_{m}-\gamma_{m}\feyn{D})\epsilon)\\
&=-\tfrac{1}{4}\ee^{-2\varphi}(R-8\,\alpha_{1}^{2}H^{2}-4\alpha_{3}(\alpha_{3}+2)(\nabla\varphi)^{2})\\
&\eqspace+\alpha_{1}\,\ee^{-2\varphi}\tfrac{1}{4!}\dd H_{m_{1}\dots m_{4}}\epsilon^{\dagger}\gamma^{m_{1}\dots m_{4}}\epsilon\\
&\eqspace+2\,\ee^{-2\varphi}(\alpha_{3}+1)\nabla^{m}\varphi\,\epsilon^{\dagger}(D_{m}-\gamma_{m}\feyn{D})\epsilon\\
&\eqspace+\nabla^{m}(\ee^{-2\varphi}\epsilon^{\dagger}(D_{m}-\gamma_{m}\feyn{D}+\alpha_{3}\nabla_{m}\varphi)\epsilon).\\
\end{split}
\end{equation}
Remarkably, setting $\alpha_{3}=-1$ is enough to set most of the non-tensorial terms to zero. For this choice, $\ii\feyn{D}$ is self-adjoint with respect to the inner product $\langle\epsilon_{1},\epsilon_{2}\rangle=\int_{X}\ee^{-2\varphi}\vol\epsilon^{\dagger}_{1}\epsilon_{2}$. The last free parameter can be set to $\alpha_{1}=1/4$ in order to give the correct kinetic term for $H$.\footnote{The choice $\alpha_{1}=-1/4$ is also allowed and amounts to a trivial change of sign $H\to-H$.} Integrating this identity over $X$ gives the zeroth-order terms in the Lichnerowicz identity \eqref{eq:Lichnerowicz}:
\begin{equation}
\tfrac{1}{4}\int_{X}\ee^{-2\varphi}\vol\Bigl(R+4(\nabla\varphi)^{2}-\tfrac{1}{2}H^{2}\Bigr)=\int_{X}\ee^{-2\varphi}\vol\Bigl(|\feyn{D}\epsilon|^{2}-|D\epsilon|^{2}+\tfrac{1}{4}\epsilon^{\dagger}\cancel{\dd H}\epsilon\Bigr),
\end{equation}
where $D$ and $\feyn{D}$ match the supersymmetry operators of heterotic supergravity \eqref{eq:susy_operators}.

It is relatively easy to see that the $\ap$ corrections to this identity must be of the form given in \eqref{eq:Lichnerowicz}. For example, the extra $\tr |\feyn{F}\epsilon|^2$ term can be expanded using gamma matrix algebra, giving a $\tr \epsilon^\dagger \cancel{F\wedge F}\epsilon$ term and a $\tr |F|^2$ term. The first of these cancels the $\tr {F\wedge F}$ term in the Bianchi identity, while the second reproduces the kinetic term for the gauge field in the bosonic action. The $R^+$ terms follow similarly.

\subsection{Varying \texorpdfstring{$I$}{I}}

Observe that the final term on the right-hand side of \eqref{eq:Lichnerowicz} is precisely the functional $I$ in \eqref{eq:6d_functional} whose field equations reproduce the anomaly flow. Morally, it appears that variations of $I$ should be related to variations of $S_\text{B}$ up to terms that will vanish for supersymmetry geometries. The precise relation between the two is somewhat delicate. In particular, we thought of $I$ as a functional of $(\varphi,e,\epsilon)$ with $\mathcal{B}$ understood to be a background field and no explicit variation with respect to $B$. Related to this is the fact that for anomaly flow, the Bianchi identity holds only at fixed points. Both of these issues need further comments.

For ease of presentation, we first consider the case of $\ap=0$ so that the above identity reads
\begin{equation}\label{eq:Lichnerowicz_alpha_zero}
\tfrac{1}{4}S_{\text{B}}|_{\ap=0}=\int_{X}\ee^{-2\varphi}\vol\Bigl(|\feyn D\epsilon|^{2}-|D\epsilon|^{2}+\tfrac{1}{4}\epsilon^{\dagger}\cancel{\dd H}\epsilon\Bigr).
\end{equation}
Recall that the bosonic action $S_\text{B}$ is usually thought of as a functional of the supergravity fields $(g,\varphi,B,A)$ (with $A$ playing no role for $\ap\to 0$), and varying the action with respect to these fields gives the heterotic supergravity equations of motion \eqref{eq:bosonic_eom}. Since \eqref{eq:Lichnerowicz_alpha_zero} is an identity, varying the left- and right-hand side with respect to the supergravity fields must give the same result. However, there is a subtlety in the variation with respect to $B$: since it is the three-form flux $H$ that appears in the action, in order for us to vary with respect to $B$, we must first be able to define it as the local potential for $H$. However, as soon as $H$ is defined in terms of a potential, it automatically satisfies the Bianchi identity. The four-form $\dd H$ then vanishes identically and drops out of \eqref{eq:Lichnerowicz_alpha_zero} for any $(g,\varphi,B)$, so the $\dd H$ term is itself never varied. In other words, the supergravity equations of motion follow from varying the right-hand side with respect to $(g,\varphi,B)$, with $\dd H$ thought of as a \emph{background field}. In addition, since \eqref{eq:Lichnerowicz_alpha_zero} holds for any field configuration, the choice of nowhere-vanishing spinor $\epsilon$ is arbitrary. For example, for non-supersymmetric solutions, $\epsilon$ defines a $G$-structure on $X$ but it does not have to solve the supersymmetry conditions, and so does not tell us anything about the supergravity fields! Said differently, since generically $\epsilon$ does not depend on $(g,\varphi,B)$ and the left-hand side of \eqref{eq:Lichnerowicz_alpha_zero} has no $\epsilon$ dependence, the right-hand side of \eqref{eq:Lichnerowicz_alpha_zero} must also be independent of the choice of $\epsilon$. This also means that variation of the right-hand side with respect to the spinor vanishes.

If we think of \eqref{eq:Lichnerowicz_alpha_zero} as a functional of $(g,\varphi,B)$, it does not seem possible to go ``off-shell'' and relax the Bianchi identity as the anomaly flow requires. However, instead of varying in the full space of supergravity fields, we can instead vary in the space of supersymmetric geometries, defined in \eqref{eq:susy_geometry}. For these configurations, since the spinor $\epsilon$ satisfies the Killing spinor equations, $\feyn{D}\epsilon=D \epsilon=0$, $\epsilon$ now contains information about the physical supergravity fields -- for example, specifying $(\varphi,e,\epsilon)$ is sufficient to fix $H$ via \eqref{eq:strominger1}, with its equation of motion following automatically. Thanks to this, for a supersymmetric geometry, we can equivalently vary with respect to $(\varphi,e,\epsilon)$. Furthermore, since \eqref{eq:Lichnerowicz_alpha_zero} is quadratic in the various supersymmetry operators, setting $\feyn{D}\epsilon=D \epsilon=0$ is also sufficient to set the variations of the $|\feyn D\epsilon|^{2}$ and $|D\epsilon|^{2}$ terms to zero, and so these do not contribute to the variation of the right-hand side for supersymmetric geometries. Thus, when we limit ourselves to the space of supersymmetric geometries, the identity \eqref{eq:Lichnerowicz_alpha_zero} becomes
\begin{equation}\label{eq:Lichnerowicz_alpha_zero_susy}
S_{\text{B}}|_{\ap=0}=\int_{X}\ee^{-2\varphi}\vol\epsilon^{\dagger}\cancel{\dd H}\epsilon\equiv I|_{\ap=0}.
\end{equation}
Furthermore, since the anomaly flow preserves the supersymmetry conditions for the $\SU3$ structure, setting $D\epsilon=\feyn{D}\epsilon=0$ at the start of the flow is sufficient to ensure they remain zero along it. From this and the definitions of the equations of motion in \eqref{eq:eom_defs}, it is then clear that starting from a supersymmetric geometry with $\ap=0$, the flow \eqref{eq:flow_I} driven by $I|_{\ap=0}$ actually corresponds to a flow by (minus) the zeroth-order equations of motion for those fields, in agreement with the results of Section \ref{eq:eom}.

Reintroducing the leading $\ap$ corrections in \eqref{eq:Lichnerowicz}, a supersymmetric geometry will satisfy \eqref{eq:Lichnerowicz} with the $D\epsilon$ and $\feyn{D}\epsilon$ terms set to zero:
\begin{equation}\label{eq:Lichnerowicz_susy}
S_{\text{B}}=I+\frac{\ap}{4}\int_{X}\ee^{-2\varphi}\vol\bigl(\tr\epsilon^{\dagger}\feyn F\feyn F\epsilon-    \tr\epsilon^{\dagger}\feyn{R}^{+}\feyn{R}^{+}\epsilon\bigr).
\end{equation}
With the equations of motion defined as in \eqref{eq:eom_defs}, the flow \eqref{eq:anomaly_flow_equations} driven by $I$ now corresponds to a flow by the $\mathcal{O}(\ap)$ equations of motion for those fields, but now with the extra $\ap$ corrections that we found in \eqref{eq:flow_metric} and \eqref{eq:flow_dilaton} thanks to the $\feyn{F}$ and $\feyn{R}^+$ terms present on the right-hand side of \eqref{eq:Lichnerowicz_susy}.

\section{Anomaly flows in diverse dimensions}\label{sec:diverse}

In the previous section, we used the Lichnerowicz identity \eqref{eq:Lichnerowicz} restricted to a six-manifold $X$ in order to relate the Bianchi identity to squares of supersymmetry operators and the bosonic action. Using this and the presence of a spinor defining an $\SU3$ structure, we were able to rephrase the anomaly flow in terms of a functional depending on an invariant spinor and the Bianchi identity. This Lichnerowicz identity also holds for restrictions to seven- or eight-dimensional internal spaces, such as those that admit $\Gx 2$ or $\Spin 7$ structures. It is thus natural to define the anomaly flow on these spaces to be the flow driven by the functional
\begin{equation}
I=\int_{X}\ee^{-2\varphi}\vol\epsilon^{\dagger}\feyn{\mathcal{B}}\epsilon,\label{eq:anomaly_functional}
\end{equation}
where $X$ is a seven- or eight-manifold and $\epsilon$ is the invariant spinor that characterises the $G$-structure on $X$ that is relevant for supersymmetry.  When generating flows, the Bianchi identity $\mathcal{B}$ that appears in the functional is held constant when $I$ is varied; this will prove to be sufficient for writing down flows for $\Gx2$ or $\Spin7$ structures. We come back to this issue in Section \ref{sec:gradient} where we give a functional which does not have this subtlety -- the resulting flows are the same either way.

In the rest of this section, we review anomaly flow for Fu--Yau solutions in six dimensions, and then derive flow equations for the invariant tensors characterising $\Gx2$ or $\Spin7$ structures and analyse the implications of these. In cases where the $G$-structure reduces to $\SU3$, the flow driven by \eqref{eq:anomaly_functional} can be reduced to the usual anomaly flow. Assuming that the structure reduces further to $\SU2$, we find analogues of the Fu--Yau ansatz for three- and four-torus fibrations over K3 surfaces. The endpoints of these flows describe supersymmetric solutions with three- and two-dimensional Minkowski factors. Again, the solutions can preserve either four or eight supercharges ($N=1$ or $N=2$ in four dimensions) depending on the fibration structure chosen. In the case with eight supercharges, one again finds the flow simplifies to a single scalar equation for the dilaton. The existence of these solutions should then follow from the long-time existence of the flow, as in \cite{1610.02740}. We also give a simple example where the external space is AdS$_3$ and the internal seven-manifold is $\text{K3}\times\text{S}^3$.

\subsection{Anomaly flow on \texorpdfstring{T$^2$}{T2} fibrations over K3\label{subsec:fu-yau}}

The case where anomaly flow is best understood is for torsional heterotic backgrounds on threefolds which preserve extended $N=2$ supersymmetry. Heterotic backgrounds with torsion are essentially all variations of a solution first found by dualising an F/M-theory compactification on $\text{K3}\times\text{K3}$~\cite{Dasgupta:1999ss}. Following this, there have been many generalisations in the physics literature~\cite{Becker:2002sx,LopesCardoso:2002vpf,Adams:2006kb,Becker:2008rc,Becker:2009df,Andriot:2009fp,Becker:2009zx,Israel:2011urb,Israel:2023itj}. In the mathematics literature, these have come to be known as Fu--Yau solutions~\cite{hep-th/0604063}. The basic idea behind their use in anomaly flow is that the Fu--Yau ansatz automatically solves all the supersymmetry conditions, leaving only the Bianchi identity to satisfy, and, crucially, the anomaly flow preserves the ansatz. A fixed point of the flow will then solve the supersymmetry conditions and the Bianchi identity.

In detail, the Fu--Yau solutions are non-K\"ahler geometries which preserve either $N=1$ or $N=2$ supersymmetry depending on a choice of fibration structure. The underlying threefold $X$ is a Goldstein--Prokushkin fibration~\cite{Goldstein:2002pg}, i.e.~a holomorphic $\text{T}^{2}$ fibration over a K3 surface such that the total space is complex. The $\SU 2$ structure on the K3 surface is defined by a holomorphic two-form $\Omega_{\text{K3}}$ and a Kähler form $\omega_{\text{K3}}$, which together define a Ricci-flat metric on the base of the fibration. The ansatz for the $\SU 3$ structure on the threefold is then given by\footnote{Our notation follows that of \cite{1403.4298}, apart from using $\mathcal{F}$ for the fibration and reserving $F$ for the Yang--Mills curvature.}
\begin{equation}
\omega=\ee^{2\varphi}\omega_{\text{K3}}+a\tfrac{\ii}{2}\Theta\wedge\bar{\Theta},\qquad\Omega=\ee^{2\varphi}\sqrt{a}\,\Omega_{\text{K3}}\wedge\Theta,\label{eq:ansatz}
\end{equation}
where $a$ is the (constant) area of the $\text{T}^{2}$ fibre. The metric on the threefold is then given by
\begin{equation}
\dd s^{2}(X)=\ee^{2\varphi}\dd s^{2}(\text{K3})+a\,\delta_{IJ}\Theta^{I}\Theta^{J},
\end{equation}
where $\Theta=\Theta^{1}+\ii\Theta^{2}$ and the $\Theta^{I}$ are the globally defined one-forms dual to the Killing vectors of the two-torus.\footnote{More generally, one can allow for non-trivial complex structure on the two-torus via $\Theta = \Theta^1 + \tau \Theta^2$. The case where $\tau$ is constant is straightforward. See \cite{Becker:2009df} for a discussion of when $\tau$ is a holomorphic function on K3.} The fibration is specified by the curvatures $\mathcal{F}^{I}=\dd\Theta^{I}$, which are the pullback of elements of $H^{2}(\text{K3},2\pi\mathbb{Z})$. Importantly, if the $\mathcal{F}^I$ are non-trivial in cohomology, the metric on $X$ is \emph{never} K\"ahler~\cite{Goldstein:2002pg}.

Let us review how the Fu--Yau ansatz solves the supersymmetry conditions. First, the almost complex structure on $X$ is integrable if and only if the complex flux $\mathcal{F}=\mathcal{F}^{1}+\ii \mathcal{F}^{2}$ has no $(0,2)$ component. Given this and the closure of the $\SU 2$ structure forms on the K3 base, it is simple to check that the first two equations in \eqref{eq:strominger1} are satisfied. The last two equations are satisfied by taking $A$ to be a hermitian Yang--Mills connection for the pullback of a stable holomorphic bundle over the K3. The third condition in \eqref{eq:strominger1} then simply fixes $H$ in terms of derivatives of $\omega$. Thus, given a Calabi--Yau metric and a hermitian Yang--Mills connection on the K3, and a choice of complex flux $\mathcal{F}$ on the base with no $(0,2)$ component, the Fu--Yau ansatz \eqref{eq:ansatz} solves the supersymmetry conditions of the Hull--Strominger system. As shown in \cite{1403.4298}, this configuration generically preserves $N=1$ supersymmetry. As discussed earlier, the final condition that one needs to satisfy in order to have a solution to the heterotic equations of motion is the Bianchi identity. The resulting backgrounds generically preserve $N=1$ supersymmetry (four supercharges), however, there is an enhancement to $N=2$ if the $\mathcal{F}^{I}$ are individually type $(1,1)$ and solve the hermitian Yang--Mills equation. One can see this by noting that under $\SU2$ rotations of $(\omega_\text{K3},\re \Omega_{\text{K3}},\im \Omega_{\text{K3}})$, the supersymmetry conditions are invariant apart from the condition that $\mathcal{F}$ has no $(2,0)$ component. If one insists that $\mathcal{F}$ is purely $(1,1)$, this condition is also invariant under the rotations, with the resulting $\SU2$ interpreted as the R-symmetry of a background preserving $N=2$.

Before this, we should say a few words about flux quantisation. It is common lore that a supergravity solution lifts to a string theory solution only if the supergravity fluxes are quantised, i.e.~the integrals of closed $p$-form fluxes over cycles in $X$ must be quantised. For general heterotic backgrounds with flux, this story is somewhat subtle since $H$ is \emph{not} closed but has a non-trivial Bianchi identity. However, for Fu--Yau solutions, one can identify an appropriate set of quantisation conditions~\cite{Melnikov:2012cv,1403.4298}. Decomposing the three-form flux as
\begin{equation}
H=H_{\text{hor}}+H_{I}\wedge\Theta^{I},
\end{equation}
where the horizontal component $H_{\text{hor}}$ lies solely on K3 and $H_{I}$ are two-forms on K3, it is simple to check that the $H_{I}$ are closed. The appropriate quantisation conditions are then
\begin{equation}
H_{I}\in H^{2}(X,2\pi\ap\bZ),
\end{equation}
and so $a$, the area of the $\text{T}^{2}$ fibre, must quantised in units of $\ap$. In particular, this means that $a\sim\mathcal{O}(\ap)$ which should be kept in mind when considering which terms appear at a given order in the $\ap$ expansion.

It was argued in \cite{hep-th/0604063} that the existence of a solution for the $N=2$ case can be reduced to a single Monge--Ampère-like scalar equation for the dilaton:
\begin{equation}
\Delta_{\text{K3}}\ee^{2\varphi}=\text{vol}_{\text{K3}}\,\lrcorner\left(a\,\mathcal{F}_{I}\wedge \mathcal{F}^{I}+\frac{\alpha'}{4}(\tr F\wedge F-\tr R^{\text{Ch}}\wedge R^{\text{Ch}})\right),
\end{equation}
where $\Delta_{\text{K3}}$ is the Laplacian on K3 and the contraction with the inverse volume $\vol_{\text{K3}}$ is equivalent to acting with $\star_{\text{K3}}$. At first sight, this appears to be a simple Poisson equation for the dilaton, and so proving existence of a solution should be trivial. However, there is a hidden dilaton dependence in the $\tr R^{\text{Ch}}\wedge R^{\text{Ch}}$ term of the form $\partial \bar\partial \varphi\wedge \partial \bar\partial \varphi+\dots$ which makes the problem non-linear. Note that \cite{hep-th/0604063} and later works used the Chern connection in the Bianchi identity, which we argued above would not lead to a solution to the supergravity equations of motion. However, following \cite{Melnikov:2017wcf}, one can check that the choice of connection appears in the resulting Monge--Ampère equation for $\varphi$ via the choice of function whose precise form is not important for the existence proof. Thanks to this, the argument of \cite{hep-th/0604063} goes through when one uses $R^+$ instead of $R^{\text{Ch}}$.

For an idea of where this scalar equation comes from, recall that, if one takes $R^+$ to be the Hull connection, only the Bianchi identity remains to be solved. For the $N=2$ solutions, $\dd H$ is four-form which is purely horizontal with respect to the two-torus fibration, and so it is proportional to the volume form on K3. Similarly, as shown in \cite{1403.4298}, the $\tr F\wedge F$ and $\tr R^+\wedge R^+$ terms appearing at $\mathcal{O}(\ap)$ are also horizontal. The Bianchi identity can then be dualised to the single scalar equation above. In the $N=1$ case, one has to use the $\ap$ expansion and remember that the area $a$ of the torus is quantised in units of $\ap$. In a little more detail, one finds that the four-form $\dd H$ generically has a non-horizontal contribution. $a\,\bar{\partial}\mathcal{F}_{(2,0)}\wedge\Bar{\Theta}+\text{c.c.} \subset \dd H $, which is of order $\ap$. Solving the Bianchi identity then constrains the $(2,0)$ part of the complexified curvature $\mathcal{F}$ for the torus fibration to be a constant multiple of $\Omega_{\text{K3}}$ (to first order in $\alpha'$). Allowing for this $(2,0)$ piece, the solution preserves only $N=1$ supersymmetry.

\subsubsection{The \texorpdfstring{$N=2$}{N=2} flow equations}

As shown in \cite{1610.02740}, the anomaly flow preserves the form of the Fu--Yau ansatz. Furthermore, for the $N=2$ class, the anomaly flow reduces to a single flow equation for the dilaton:
\begin{equation}\label{eq:N=2_flow}
\partial_{t}(\ee^{-2\varphi})=\frac{1}{2}\left(\Delta_{\text{K3}}\ee^{2\varphi}
-a\star_{\text{K3}}(\mathcal{F}_{I}\wedge\mathcal{F}^{I})
-\frac{\alpha'}{4}\star_{\text{K3}}(\tr F\wedge F-\tr R^{+}\wedge R^{+})\right)+\mathcal{O}(\ap^{2}).
\end{equation}
In the $N=2$ case, long-time existence and convergence of the anomaly flow have been shown~\cite{1610.02740}, providing an alternative proof of Fu and Yau's original existence proof~\cite{hep-th/0604063}.

Note that the anomaly flow is consistent with the $N=1$ geometry even for arbitrary $\mathcal{F}_{(2,0)}$, with the flow again reducing to a scalar equation for the dilaton. However, it generically requires an extra flow equation for the torus fibration, given by 
\begin{equation}
	\partial_{t}\mathcal{A}=\star_{\text{K3}}\,\text{d}\mathcal{F}_{(2,0)}+\mathcal{O}(\alpha').
\end{equation}
This flow equation preserves $N=1$ supersymmetry, as it satisfies 
\begin{equation}
	\Omega_{\text{K3}}\wedge\partial_{t}\mathcal{A}=\mathcal{O}(\alpha'),\qquad\omega_{\text{K3}}\wedge \partial_{t}\mathcal{A}=\text{d}(-\ii\mathcal{F}_{(2,0)})+\mathcal{O}(\alpha').
\end{equation}
These imply $\partial_{t}(\omega_{\text{K3}}\wedge\mathcal{F})=\partial_{t}(\Omega_{\text{K3}}\wedge\mathcal{F})=\mathcal{O}(\alpha')$, and so the torus curvature $\mathcal{F}$ stays primitive and does not acquire a $(0,2)$ part under the flow.

\subsection{Anomaly flow on \texorpdfstring{$\protect\Gx 2$}{G2} structure manifolds}\label{sec:g2}

Our first new example is the case where $X$ is a seven-manifold admitting a $\Gx2$ structure. Following the conventions of \cite{1702.01156}, a unit-norm $\Gx 2$-invariant spinor $\epsilon$ on a seven-manifold defines a three-form $\phi$ and a four-form $\star\phi$ as
\begin{equation}
\phi=-\ii\epsilon^{\dagger}\gamma_{(3)}\epsilon,\qquad\star\phi=\epsilon^{\dagger}\gamma_{(4)}\epsilon.\label{eq:G2_forms}
\end{equation}
From these, one can construct projectors onto $\Gx 2$ representations -- we give these for three-forms and four-forms in Appendix \ref{app:G-structure}.

\subsubsection{Supersymmetry conditions}

Heterotic supergravity compactified on a manifold with $\Gx 2$ structure admits both Minkowski and AdS$_{3}$ solutions. For a string-frame metric
\begin{equation}
    \dd s^2_{10} = \dd s^2(\text{AdS}_3)+\dd s^2(X),
\end{equation}
and allowing for the NSNS flux to have an external component given by $2\ell^{-1} \vol_{\text{AdS}_3}$, where $\ell$ is the AdS radius, the conditions on the $\Gx2$ structure for $N=1$ supersymmetry in three dimensions are~\cite{Gauntlett:2001ur,Friedrich:2001yp}
\begin{equation}
\dd(\ee^{-2\varphi}\star\phi)=0,\qquad\phi\wedge\dd\phi=-\frac{12}{7\ell}\phi\wedge\star\phi,\qquad\star H=-\ee^{2\varphi}\dd(\ee^{-2\varphi}\phi)-\frac{2}{\ell}\star\phi.\label{eq:G2_susy}
\end{equation}
The Minkowski limit of these equations follows from $\ell\to\infty$.\footnote{The supersymmetry conditions for $\ell\neq0$ come from imposing $D_{m}\epsilon=0$ and $\feyn{D}\epsilon=\frac{\ii}{\ell}\epsilon$.} These equations fix the intrinsic torsion of the $\Gx 2$ structure and the three-form flux to
\begin{subequations}\label{eq:g2_susy_conditions}
\begin{align}
\dd\phi & =\frac{3}{2}\dd\varphi\wedge\phi-\frac{12}{7\ell}\star\phi-\star H_{\rep{27}},\\
\dd\star\phi & =2\dd\varphi\wedge\star\phi,\\
H & =-\frac{2}{7\ell}\phi+\frac{1}{2}\star(\dd\varphi\wedge\phi)+H_{\rep{27}},
\end{align}
\end{subequations}
where the $\rep{27}$ component of $H$ is not fixed by supersymmetry. Note that a possible $\rep{14}$ component of the intrinsic torsion is missing from $\dd\star\phi$, giving what is known as an integrable $\Gx2$ structure.\footnote{Since the $\rep{7}$ component of the intrinsic torsion is proportional to $\dd\varphi$ and hence exact, one can always define a conformally rescaled three-form as $\hat\phi = \ee^{-3\varphi/2}\phi$, such that the resulting $\Gx2$ structure is co-closed, $\dd\hat\star\hat\phi=0$.}  Allowing a non-trivial vector bundle over $X$, the final supersymmetry condition is that its curvature $F$ is a $\Gx 2$ instanton~\cite{Corrigan:1982th,Donaldson:1996kp}. Explicitly, this means
\begin{equation}
F\wedge\star\phi=0,\label{eq:G2_instanton}
\end{equation}
and so $F$ transforms in the $\rep{14}$ representation of $\Gx 2$, with no component in the $\rep{7}$. Thinking back to the Bianchi identity, this means that $\tr F\wedge F$ would have no component in the $\rep{7}$, living only in the $\rep{1}\oplus\rep{27}$. We always assume that $p_1(X)=p_1(V)$, so that solutions to the Bianchi identity are not obstructed. Similarly to the $\SU3$ case, we refer to solutions to \eqref{eq:g2_susy_conditions} as supersymmetric geometries, with supersymmetric solutions also satisfying the $\Gx2$ instanton condition and the Bianchi identity.

\subsubsection{Anomaly flow}

Now consider the flow driven by the functional \eqref{eq:anomaly_functional} where $\epsilon$ is taken to be the unit-norm spinor defining the $\Gx2$ structure on $X$. In this case, the functional $I$ can be simplified and written in terms of the $\Gx2$ three-form as
\begin{equation}
I=\int_{X}\ee^{-2\varphi}\mathcal{B}\wedge\phi.
\end{equation}
The variation of this functional for the vielbein, dilaton and spinor are again given by \eqref{eq:variation}. Relating the variations and the flow equations as in \eqref{eq:flow_I}, the flow is given by
\begin{subequations}\label{eq:G2_flow}
\begin{align}
\partial_{t}\varphi & =-\frac{1}{4}\ee^{2\varphi}\epsilon^{\dagger}\feyn{\mathcal{B}}\epsilon,\label{eq:dilaton}\\
\partial_{t}e_{m}{}^{a} & =-\frac{1}{4}\ee^{2\varphi}\frac{1}{3!}\epsilon^{\dagger}\gamma_{m}{}^{n_{1}n_{2}n_{3}}\epsilon\,\mathcal{B}^{a}{}_{n_{1}n_{2}n_{3}},\label{eq:g2_vielbein}\\
\partial_{t}\epsilon & =-\frac{1}{4}\ee^{2\varphi}\frac{1}{3!}\mathcal{B}_{mn_{1}n_{2}n_{3}}\phi^{n_{1}n_{2}n_{3}}\ii\gamma^{m}\epsilon,
\end{align}
\end{subequations}
where we have used $\gamma^{abcd}\epsilon=\star\phi^{abcd}\epsilon-4\ii\phi^{[abc}\gamma^{d]}\epsilon$ for the $\Gx2$ structure~\cite{hep-th/0303127}.
Note that the spinor does not flow if $\mathcal{B}$ has no component transforming the $\rep7$ representation of $\Gx2$. Since the flow equations again come from the functional \eqref{eq:anomaly_functional}, for a supersymmetric geometry, they correspond to the heterotic supergravity equations of motion by construction.
It is then straightforward to derive the flow of the $\Gx2$ three- and four-form:\footnote{More generally, if one replaces the $-1/4$ coefficients in \eqref{eq:G2_flow} with constants $c_i$, the flow of the three- and four-form are
\begin{equation*}
\begin{aligned}
\partial_t \phi &= \ee^{2\varphi} \bigl(12 c_2 \pi_{\rep1} + (8c_3-6c_2)\pi_{\rep7} - 2c_2 \pi_{\rep{27}}\bigr) \star \mathcal{B},\\
\partial_t \star\phi &= \ee^{2\varphi}\bigl(16 c_2 \Pi_{\rep1} + (8c_3-6c_2)\Pi_{\rep{7}} + 2c_2 \Pi_{\rep{27}}\bigr) \mathcal{B}.
\end{aligned}
\end{equation*}
}
\begin{equation}
\partial_{t}\left(\ee^{-2\varphi}\star\phi\right)=-\tfrac{1}{2}\mathcal{B},\qquad\partial_{t}(\ee^{-2\varphi}\phi)=\tfrac{1}{2}(\pi_{\rep 1}-\pi_{\rep 7}+\pi_{\rep{27}})\star\mathcal{B}.\label{eq:g2_flow}
\end{equation}
From these, it is simple to show that the warped volume $\ee^{-4\varphi}\vol$ is again invariant along the flow. Since the $\Gx 2$ structure also defines a Riemannian metric, the flow of the metric on $X$ can be obtained as
\begin{equation}
\partial_{t}g_{mn}=\ee^{2\varphi}\phi_{m}{}^{p_{1}p_{2}}\left(\tfrac{1}{2}\pi_{\rep{27}}\star\mathcal{B}-\tfrac{1}{3}\pi_{\rep 1}\star\mathcal{B}\right)_{np_{1}p_{2}},
\end{equation}
which is consistent with the flow of the vielbein in \eqref{eq:g2_vielbein}. Furthermore, from \eqref{eq:g2_vielbein} and \eqref{eq:g2_flow}, one can check that the flow obeys
\begin{equation}
    \partial_t\star\phi=\star\left(\tfrac{4}{3} \pi_{\rep{1}} + \pi_{\rep{7}}-\pi_{\rep{27}}\right)\partial_t \phi,
\end{equation}
which agrees with the usual deformations of a $\Gx2$ structure~\cite{2003math......5124B,joyce2007riemannian,Becker:2014rea}

The analysis of this case is somewhat subtle compared to the anomaly flow for the Hull--Strominger system. The reason is that the flow does not preserve all of the supersymmetry conditions. The flow of $\ee^{-2\varphi}\star\phi$ is by a closed four-form, so if one starts with a supersymmetric geometry, it will continue to satisfy the first equation in \eqref{eq:G2_susy} along the flow. The $\phi\wedge\dd\phi=0$ condition is more complicated as it is non-linear: one can write this condition as $\pi_{\rep{1}} \dd \phi = \text{constant}$, where the projector depends on the $\Gx2$ structure.\footnote{This should be compared to the open problem of whether the Laplacian flow of a co-closed $\Gx2$ structure remains co-closed~\cite{2018arXiv181110505G}, since the co-closure condition is again non-linear in $\phi$.} However, a somewhat laborious calculation shows that the second equation in \eqref{eq:G2_susy} is generically not preserved unless $\mathcal{B}$ lies only in the $\rep{1}$ representation (or if $\mathcal{B}=\dd H$ with $H_{\rep{27}}=0$). Moreover, the flow for the $\Gx2$ structure should be supplemented by a flow for the connection on the gauge bundle over $X$. Similar to the $\SU3$ case, since $F\wedge \star\phi=0$ depends on a choice of $\Gx2$ structure, one cannot fix $F$ at the start of the flow.

We can also see this from the intrinsic torsion of the $\Gx2$ structure. Without imposing the second supersymmetry condition, one finds
\begin{equation}
\dd \phi=-\left(\tfrac{2}{\ell}+\tfrac{1}{7}H_{\rep{1}}\right)\star\phi+\tfrac{3}{2}\text{d}\varphi\wedge\phi-\star H_{\mathbf{27}},\qquad
H=\tfrac{1}{7}H_{\mathbf{1}}\phi+\tfrac{1}{2}\star(\text{d}\varphi\wedge\phi)+H_{\mathbf{27}},
\end{equation}
with $\dd\star\phi$ unchanged from \eqref{eq:g2_susy_conditions}. Equivalently, the supersymmetry operators $D$ and $\feyn{D}$ act on the $\Gx{2}$ invariant spinor $\epsilon$ as
\begin{equation}
D_{m}\epsilon=\tfrac{\ii}{8}\left(H_{\mathbf{1}}+\tfrac{2}{\ell}\right)\gamma_{m}\epsilon,\qquad\feyn{D}\epsilon=\tfrac{3\ii}{8}\left(H_{\mathbf{1}}+\tfrac{2}{\ell}\right)\epsilon+\tfrac{\ii}{\ell}\epsilon.
\end{equation}
Using this and the integrability condition \eqref{eq:integrability_eom_metric_6D}, which expresses the metric equation of motion in terms of the supersymmetry operators and the Bianchi identity, the $\Gx2$ flow for the metric can be written as
\begin{equation}
\begin{split}
\partial_{t}g_{mn}&=-\tfrac{1}{12}\ee^{2\varphi}\star\phi_{(m}{}^{p_{1}\dots p_{3}}\text{d}H_{n)p_{1}\dots p_{3}}\\
&=-\ee^{2\varphi}\text{eom}[g]_{mn}+\tfrac{1}{8}\ee^{2\varphi}\left(H_{\mathbf{1}}+\tfrac{2}{\ell}\right)\Bigl[\left(\tfrac{1}{7}H_{\mathbf{1}}+\tfrac{6}{\ell}\right)g_{mn}-2 H_{\mathbf{27}}{}_{mn}\Bigr],
\end{split}
\end{equation}
where we have taken $\ap=0$ for simplicity of presentation. For a supersymmetric geometry, the singlet $H_{\rep{1}}$ is fixed to the constant $-2/\ell$, however, generically the $\Gx2$ flow does not preserve this condition and so breaks supersymmetry along the flow.\footnote{Despite this, the resulting flow evolves by Ricci flow to leading order and is no more than quadratic in the torsion, and so may still be ``reasonable'' in the sense of \cite{Chen_2018,2018arXiv181110505G}.}

To avoid this complication, we consider examples where the $\Gx2$ structure is further reduced to an $\SU3$ or $\SU2$ structure. Our guiding requirement is that the ansatz is compatible with the resulting flow equations and that $\phi\wedge\dd\phi=-\frac{12}{7\ell}\phi\wedge\star\phi$ is automatically satisfied, which is then sufficient to ensure supersymmetry is preserved along the flow. In addition, the form of the ansatz should either allow the gauge connection to be fixed at the start of the flow (as in the $N=2$ Fu--Yau case where $F$ is hermitian Yang--Mills with respect to $\omega_{\text{K3}}$) or suggest a natural flow that ensures $F$ is a $\Gx2$ instanton at fixed points.

\subsubsection{Reduction on a circle}\label{sec:g2_circle}

Consider the special case of a Minkowski background where the $\Gx 2$ structure on $X$ can be reduced further to $\SU 3$, corresponding to $N=2$ supersymmetry in three dimensions. The metric on $X$ can be written as a $\Uni1$ fibration over a six-manifold $X_6$
\begin{equation}\label{eq:metric_g2_s1}
\dd s^{2}(X)=\dd s^{2}(X_{6})+\Theta^{2},
\end{equation}
where $\Theta$ is the real one-form dual to the Killing vector generating the $\Uni 1$ isometry of the circle, and the curvature of the fibration is defined as $\mathcal{F}=\dd\Theta$. The $\Gx 2$ structure forms can be written as
\begin{equation}\label{eq:g2_su3}
\phi=\omega\wedge\Theta+\re\Omega,\qquad\star\phi=-\tfrac{1}{2}\omega\wedge\omega-\im\Omega\wedge\Theta,
\end{equation}
where $(\omega,\Omega)$ define a canonical $\SU 3$ structure on $X_{6}$. We further assume that the vector bundle is the pullback of a bundle over $X_6$ and that the dilaton is independent of the coordinate along the circle.

It is simple to see how this ansatz can preserve at least $N=1$ supersymmetry. First, consider the second of the $\Gx2$ supersymmetry conditions in \eqref{eq:G2_susy}. One can check that this is satisfied automatically provided: 1) $\dd\Omega$ is type $(3,1)$; 2) $\mathcal{F}$ is type $(1,1)$; 3) $\omega\lrcorner \mathcal{F} =0$. In other words, the $\SU3$ structure on $X_6$ defines a complex structure, the $\Uni1$ fibration is holomorphic with respect to this complex structure, and the associated curvature $\mathcal{F}$ satisfies the hermitian Yang--Mills equation on the base. Next is the closure of $\ee^{-2\varphi} \star\phi$: with the previous conditions and the above decomposition, this reduces to the closure of $\ee^{-2\varphi}\omega\wedge \omega$ and $\ee^{-2\varphi}\Omega$, which we recognise as the first two supersymmetry conditions for a Hull--Strominger system on $X_6$. Finally, the requirement that the curvature $F$ is a $\Gx2$ instanton \eqref{eq:G2_instanton}, simplifies to $F\wedge\im\Omega=0$ and $F\wedge\omega\wedge\omega=0$, which require that $F$ is hermitian Yang--Mills on $X_6$. These conditions are sufficient to solve the $N=1$ supersymmetry conditions, with the associated torsion given by
\begin{equation}
    H=-\star \ee^{2\varphi}\dd(\ee^{-2\varphi}\phi)=\star_{6}\ee^{2\varphi}\dd(\ee^{-2\varphi}\omega)-\mathcal{F}\wedge\Theta,
\end{equation}
where $\star_6$ is the Hodge star on $X_6$. Note that $\dd H$ is horizontal with respect to the fibration and is type (2,2). Given the above conditions on the fibration, the gauge bundle and $(\omega,\Omega)$, it is then easy to see how the $N=1$ conditions continue to hold if we rotate $(\re\Omega,\im\Omega)$ in the ansatz \eqref{eq:g2_su3} for the $\Gx2$ structure by constant $\SO2 \simeq \Uni1$ rotations. This implies that the background actually preserves an enhanced $N=2$ supersymmetry (four real supercharges) with an associated $\Uni1$ R-symmetry in three dimensions.

Assuming $\mathcal{B}$ is a horizontal four-form, the anomaly flow equations for the $\Gx 2$ structure forms in \eqref{eq:g2_flow} reduce to
\begin{equation}
\partial_{t}\left(\ee^{-2\varphi}\omega\wedge\omega\right)=\mathcal{B},\qquad\partial_{t}\left(\ee^{-2\varphi}\Omega\right)=0,
\end{equation}
which are precisely the anomaly flow equations for the Hull--Strominger system. However, this not the end of the story. In showing that a reduction on a circle solves the $\Gx2$ supersymmetry conditions, it was essential that the curvature $\mathcal{F}$ was type $(1,1)$ and primitive. Given a choice for $\mathcal{F}$ at the start of the flow, the first of these conditions is clearly preserved, since the complex structure on $X_6$ is unchanged. The second condition, that $\mathcal{F}$ is primitive, is not preserved since $\omega$ itself flows. In other words, $\mathcal{F}$ should solve hermitian Yang--Mills on $X_6$ to preserve supersymmetry, but the definition of this condition depends on $\omega$ which is changing along the flow. One might hope that this can be solved by allowing $\mathcal{A}$ to flow in order to compensate the flow of $\omega$, with $\omega\wedge\omega\wedge\mathcal{F}=0$ reached at fixed points.
However, the simplest way to deal with this problem is to take $\mathcal{F}=0$, so that \eqref{eq:metric_g2_s1} is a metric on a direct product $X_6\times \text{S}^1$ and the $\Gx2$ anomaly flow reduces to the standard $\SU3$ anomaly flow on $X_6$.

\subsection{Anomaly flow on \texorpdfstring{$\text{K3}\times\text{S}^{3}$}{K3 x S3}}

As a special case of the previous subsection, consider a seven-manifold given by a round three-sphere fibred over a K3 surface. It is convenient to describe the K3 factor with a triplet of closed hermitian forms $j_{I}$ characterising its hyper-Kähler structure, which we are free to normalise as\footnote{These can be written as $\Omega_{\text{K3}}=j_1+\ii\, j_2$ and $\omega_{\text{K3}}=j_3$.}
\begin{equation}
\tfrac{1}{2}j_{I}\wedge j_{J}=\delta_{IJ}\vol_{\text{K3}},
\end{equation}
where $\vol_{\text{K3}}$ is the K3 volume form. The round $\text{S}^{3}$ admits a triplet of Maurer--Cartan forms $\sigma_{I}$, dual to the left-invariant vector fields on $\SU 2$. These satisfy the usual identity
\begin{equation}
\dd\sigma_{I}+\tfrac{1}{2}\epsilon_{IJK}\sigma^{J}\wedge\sigma^{K}=0.
\end{equation}
The ansatz for the seven-dimensional metric is a direct product
\begin{equation}
\dd s^{2}(X)=\ee^{2\varphi}\dd s^{2}(\text{K3})+\tfrac{1}{4}\ell^{2}\dd s^{2}(\text{S}^{3}),
\end{equation}
with the AdS radius determining the radius of the three-sphere. The $\Gx 2$ forms can then be obtained by decomposing 
a $\Gx2$ invariant spinor in a manner consistent with this ansatz. The result of this is
\begin{subequations}
\begin{align}
\phi&=\tfrac{1}{2}\ee^{2\varphi}\ell j_{I}\wedge\sigma^I-\tfrac{1}{8}\ell^3 \vol_{\text{S}^{3}},\\
\star\phi&=\tfrac{1}{8}\ee^{2\varphi}\ell^2\epsilon_{IJK}j^{I}\wedge\sigma^J\wedge\sigma^K-\ee^{4\varphi}\vol_{\text{K3}},
\end{align}
\end{subequations}
with $\vol_{\text{S}^{3}}=\sigma_{1}\wedge\sigma_{2}\wedge\sigma_{3}$. It is then simple to check that these automatically satisfy the $N=1$ supersymmetry conditions in \eqref{eq:G2_susy}, with the final equation fixing the three-form flux $H$ in terms of the $\Gx 2$ structure as
\begin{equation}
H=\star_{\text{K3}}\dd\ee^{2\varphi}+\tfrac{1}{4}\ell^{2}\vol_{\text{S}^{3}},
\end{equation}
where $\star_{\text{K3}}$ is the Hodge star associated to $\dd s^{2}(\text{K3})$. Notice that the exterior derivative of $H$ is a top-form on the K3 base,
\begin{equation}
\dd H=\Delta_{\text{K3}}\ee^{2\varphi}\vol_{\text{K3}},
\end{equation}
where $\Delta_{\text{K3}}$ is the scalar Laplacian on K3.

\subsubsection{Anomaly flow}

The curvature of the Hull connection (which appears in the anomaly condition) takes the form
\begin{equation}
R^{+}=\begin{pmatrix}2\sigma^{I}\wedge\sigma_{J} & 0\\
0 & R_{\text{K3}}^{+}
\end{pmatrix},
\end{equation}
where $R_{\text{K3}}^{+}$ is a two-form on K3. One way to satisfy instanton condition is to take $F$ to be a primitive $(1,1)$-form on the K3 base, so that $F_{\text{K3}}\wedge j_{I}=0$. With this choice, the Bianchi identity reduces to a four-form with legs only on the K3:
\begin{equation}
\mathcal{B}=\Delta_{\text{K3}}\ee^{2\varphi}\vol_{\text{K3}}-\frac{\ap}{4}\left(\tr F_{\text{K3}}\wedge F_{\text{K3}}-\tr R_{\text{K3}}^{+}\wedge R_{\text{K3}}^{+}\right).
\end{equation}
This is an equation for a single functional degree of freedom, which can be taken to be the dilaton. The anomaly flow \eqref{eq:g2_flow} for the $\Gx 2$ structure then reduces to a flow for the dilaton alone, much like the anomaly flow for $N=2$ Fu--Yau ansatz. Explicitly, one finds
\begin{equation}
\partial_{t}\ee^{2\varphi}=\tfrac{1}{2}\Delta_{\text{K3}}\ee^{2\varphi}-\frac{\ap}{8}\vol_{\text{K3}}\lrcorner\left(\tr F_{\text{K3}}\wedge F_{\text{K3}}-\tr R_{\text{K3}}^{+}\wedge R_{\text{K3}}^{+}\right).
\end{equation}
This is remarkably similar to the flow equation \eqref{eq:N=2_flow} for the six-dimensional $N=2$ Fu--Yau ansatz, where now there is no contribution from the fibration as our ansatz is a direct product. One would again expect this flow to have long-time existence provided a solution to the Bianchi identity is not topologically obstructed.

\subsection{Anomaly flow on \texorpdfstring{$\text{T}^{3}$}{T3} fibrations over K3}

Following \cite{Melnikov:2017wcf}, there should be supersymmetric solutions to heterotic supergravity with three-dimensional Minkowski space times a seven-manifold $X$, where $X$ is a $\text{T}^{3}$ fibration over a K3 surface. Depending on the fibration structure, these solutions will preserve either $N=2$ or $N=4$ supersymmetry in three dimensions, corresponding to four or eight supercharges respectively, giving a seven-dimensional version of the Fu--Yau ansatz discussed in Section \ref{subsec:fu-yau}.

The metric on $X$ is taken to be
\begin{equation}\label{eq:T3_metric}
\dd s^{2}(X)=\ee^{2\varphi}\dd s^{2}(\text{K3})+a\,\delta_{IJ}\Theta^{I}\Theta^{J},
\end{equation}
where the $\Theta^{I}$ are the globally defined one-forms dual to the Killing vectors of the three-torus fibres, and $a^{3/2}$ is the (constant) volume of the $\text{T}^{3}$. The two-forms $\mathcal{F}^{I}$ that define the fibration satisfy $\mathcal{F}^{I}=\dd\Theta^{I}$ which are the pullback of elements of $H^{2}(\text{K3},2\pi\mathbb{Z})$. There is then a $\Gx 2$ structure on $X$ defined by
\begin{subequations}\label{eq:g2_T3_decomposition}
\begin{align}\phi & =a^{1/2}\ee^{2\varphi}j_{I}\wedge\Theta^{I}-\tfrac{1}{6}a^{3/2}\epsilon_{IJK}\Theta^{I}\wedge\Theta^{J}\wedge\Theta^{K},\\
\star\phi & =\tfrac{1}{2}a\,\ee^{2\varphi}\epsilon_{IJK}j^{I}\wedge\Theta^{J}\wedge\Theta^{K}-\ee^{4\varphi}\vol_{\text{K3}}.
\end{align}
\end{subequations}

\subsubsection{Supersymmetry conditions}

Recalling from \eqref{eq:G2_susy} the supersymmetry conditions for a heterotic Minkowski background with a $\Gx 2$ structure, one finds
\begin{subequations}
\begin{align}
	0=\dd(\ee^{-2\varphi}\star\phi) & \equiv a\,\epsilon_{IJK}j^{I}\wedge \mathcal{F}^{J}\wedge\Theta^{K},\\
0=\phi\wedge\dd\phi & \equiv-\tfrac{1}{3}a^{2}\ee^{2\varphi}j_{I}\wedge \mathcal{F}^{I}\wedge\epsilon_{JKL}\Theta^{J}\wedge\Theta^{K}\wedge\Theta^{L},
\end{align}
\end{subequations}
which correspond to the following conditions on the two-forms defining the fibration:
\begin{equation}
\epsilon_{IJK}j^{J}\wedge\mathcal{F}^{K}=0,\qquad \mathcal{F}^{I}\wedge j_{I}=0.\label{eq:T3_susy}
\end{equation}
The supersymmetry conditions and $\Gx2$ then fix the three-form flux $H$ to
\begin{equation}
\begin{aligned}H & =-\star\ee^{2\varphi}\dd(\ee^{-2\varphi}\phi)\\
 & =\star_{\text{K3}}\dd\ee^{2\varphi}+a\,\delta_{IJ}(\star_{\text{K3}}\mathcal{F}^{I})\wedge\Theta^{J}.
\end{aligned}
\end{equation}

If the $\mathcal{F}^{I}$ are chosen to be orthogonal to all three Kähler forms on K3, $\mathcal{F}^{I}\wedge j_{J}=0$, the resulting solution will preserve $N=4$ supersymmetry. One can see this as the supersymmetry conditions will be satisfied for any $\SO 3\simeq\SU 2$ rotation of the $j_{I}$. The R-symmetry group of the solution is then $\SU 2$, corresponding to eight supercharges or $N=4$ supersymmetry in three dimensions. In this case, the exterior derivative of $H$ is a top-form form on K3, given by
\begin{equation}
\dd H=\Delta_{\text{K3}}\ee^{2\varphi}\vol_{\text{K3}}-a\,\delta_{IJ}\mathcal{F}^{I}\wedge \mathcal{F}^{J}.
\end{equation}
In the generic case, the forms $\mathcal{F}^{I}$ do not have to be orthogonal to the hyper-Kähler structure, and instead can have components proportional to the Kähler forms on K3. Decomposing the $\mathcal{F}^{I}$ as
\begin{equation}
\mathcal{F}^{I}=f^{I}+\tfrac{1}{2}\lambda^{IJ}j_{J},
\end{equation}
where $\lambda_{IJ}=\star_{\text{K3}}(\mathcal{F}_{I}\wedge j_{J})$ and the $f_I$ are type $(1,1)$ and orthogonal to the K\"ahler forms, the supersymmetry conditions \eqref{eq:T3_susy} for the fibration translate to the condition that the matrix of functions $\lambda_{IJ}$ is symmetric and traceless, leaving six (functional) degrees of freedom. Extended $N=4$ supersymmetry then corresponds to $\lambda_{IJ}=0$. With this decomposition, the exterior derivative of $H$ is
\begin{equation}
\dd H=\left(\Delta_{\text{K3}}\ee^{2\varphi}+\tfrac{1}{2}a\,\lambda_{IJ}\lambda^{IJ}\right)\vol_{\text{K3}}-a\,f^{I}\wedge f_{I}+a\,\dd\lambda_{IJ}\wedge j^{I}\wedge\Theta^{J}.
\end{equation}
In particular, $\dd H$ is no longer automatically horizontal (it does not lie solely on K3).

For the gauge sector, supersymmetry requires the curvature $F$ to be a $\Gx 2$ instanton, as in \eqref{eq:G2_instanton}. Since an arbitrary two-form transforms in the $\rep 7\oplus\rep{14}$ representations of $\Gx 2$, this amounts to the vanishing of the $\rep 7$ component of $F$. Given the fibration structure, this is satisfied if $F$ is orthogonal to the triplet of Kähler forms on K3:
\begin{equation}\label{eq:g2_triplet}
F\wedge j_{I}=0,
\end{equation}
which also implies that $F$ is horizontal. This condition is equivalent to $F$ being type $(1,1)$ and primitive, i.e.~it solves hermitian Yang--Mills, so that the gauge bundle is the pullback of a stable holomorphic bundle over K3, with the term $\tr F\wedge F$ appearing in the Bianchi identity then a top-form on K3.

The curvature terms in the Bianchi identity should be computed using the Hull connection $\nabla^{+}$. The connection symbols $\Sigma^{+}$ for the metric \eqref{eq:T3_metric}, in the local frame $(\dd x^{\mu},\sqrt{a}\,\Theta^{I})$, are
\begin{equation}
	\Sigma^{+}=\begin{pmatrix}(\Sigma^{+}_{\text{K3}})^{\mu}{}_{\nu}-\tfrac{a}{2}\lambda_{KL}(j^{K})^{\mu}{}_{\nu}\Theta^{L} & \sqrt{a}\,(f_{J})_{\rho}{}^{\mu}\dd x^{\rho}\\
		-\sqrt{a}\,(f_{J})_{\rho\nu}\dd x^{\rho} & 0
	\end{pmatrix},
\end{equation}
where $x^{\mu}$ denote coordinates on K3, and $\Sigma^{+}_{\text{K3}}$ is the Hull connection on K3 for the metric $\ee^{2\varphi}\dd s^{2}(\text{K3})$ with torsion $\star_{\text{K3}}\dd\ee^{2\varphi}$. For a solution with $N=4$ supersymmetry -- where $\mathcal{F}^{I}\wedge j_{J}=0$ -- the connection $\Sigma^{+}$ is automatically a horizontal one-form. More generally, note that many terms in the above expression come with prefactors of $a$, which by flux quantisation is again quantised in units of $\ap$. Working to $\mathcal{O}(\ap)$, these terms should be neglected in the Bianchi identity as they would lead to $\mathcal{O}(\ap^{2})$ corrections. The remaining contribution is
\begin{equation}
	\tr R^{+}\wedge R^{+}=\tr R^{+}_{\text{K3}}\wedge R^{+}_{\text{K3}}+\mathcal{O}(\ap),
\end{equation}
where $R^{+}_{\text{K3}}$ is the curvature of the Hull connection $\Sigma^{+}_{\text{K3}}$ on K3. Putting this all together, the Bianchi identity then reads
\begin{equation}
	\mathcal{B}\equiv\dd H-\frac{\ap}{4}(\tr F\wedge F-\tr R^{+}\wedge R^{+})=b\vol_{\text{K3}}+a\,\dd\lambda_{IJ}\wedge j^{I}\wedge\Theta^{K}+\mathcal{O}(\ap^{2}),\label{eq:T3_Bianchi}
\end{equation}
where $b$ is given by
\begin{equation}
    b=\Delta_{\text{K3}}\ee^{2\varphi}+\tfrac{1}{2}a\,\lambda_{IJ}\lambda^{IJ}-\star_{\text{K3}}\left(a\,f^{I}\wedge f_{I}+\tfrac{1}{4}\ap(\tr F\wedge F-\tr R^{+}_{\text{K3}}\wedge R^{+}_{\text{K3}})\right).\label{eq:G2_b_def}
\end{equation}

Consider the special case of $\lambda_{IJ}=0$. With the above ansatz, the supersymmetry conditions are automatically satisfied. The only equation left to solve to ensure a solution to the heterotic equations of motion is the vanishing of the Bianchi identity, which reduces to a scalar equation on K3, $b=0$. As was the case for the Fu--Yau ansatz, this is not simply an equation of the form $\Delta_{\text{K3}}\ee^{2\varphi}=\text{sources}$, since $R^{+}_{\text{K3}}$ is the curvature of the Hull connection for $\ee^{2\varphi}\dd s^{2}(\text{K3})$ and so is a function of the dilaton $\varphi$. However, comparing with the analysis of \cite{hep-th/0604063}, we see that $b=0$ is of the same form as the scalar equation for which they proved existence of a solution. Thus, at least for $\lambda_{IJ}=0$, these $\text{T}^{3}$-fibred heterotic flux solutions are guaranteed to exist.

\subsubsection{Anomaly flow}

The anomaly flow for this ansatz follows from the flow of the $\Gx 2$ three- and form-form, given in \eqref{eq:g2_flow}, and the decomposition of $\phi$ and $\star\phi$ in \eqref{eq:g2_T3_decomposition}. One then projects the expression for the Bianchi identity onto the $\rep 1$, $\rep 7$ and $\rep{27}$ representations which appear in the $\Gx 2$ anomaly flow. A straightforward, if tedious, calculation gives
\begin{subequations}
	\begin{align}
		\Pi_{\rep 1}\mathcal{B}
		& =\tfrac{1}{7}b\vol_{\text{K3}}-\tfrac{1}{14}a\,\ee^{-2\varphi}b\,\epsilon_{IJK}j^{I}\wedge\Theta^{J}\wedge\Theta^{K}+\mathcal{O}(\ap^{2}),\\
		\Pi_{\rep 7}\mathcal{B} & =0+\mathcal{O}(\ap^{2}),\\
		\Pi_{\rep{27}}\mathcal{B} & =\tfrac{6}{7}b\vol_{\text{K3}}+\tfrac{1}{14}a\,\ee^{-2\varphi}b\,\epsilon_{IJK}j^{I}\wedge\Theta^{J}\wedge\Theta^{K}+a\,\dd\lambda_{IJ}\wedge j^{I}\wedge\Theta^{J}+\mathcal{O}(\ap^{2}).
	\end{align}
\end{subequations}
The $t$-derivatives of the $\Gx 2$ structure along the flow are then
\begin{subequations}
	\begin{align}
	\partial_{t}(\ee^{-2\varphi}\star\phi) & =a\,\epsilon_{IJK}j^{I}\wedge\partial_{t}\mathcal{A}^{J}\wedge\Theta^{K}-(\partial_{t}\ee^{2\varphi})\vol_{\text{K3}},\\
	\begin{split}\partial_{t}(\ee^{-2\varphi}\phi) & =a^{1/2}j_{I}\wedge\partial_{t}\mathcal{A}^{I}-\tfrac{1}{2}a^{3/2}\ee^{-2\varphi}\,\epsilon_{IJK}\partial_{t}\mathcal{A}^{I}\wedge\Theta^{J}\wedge\Theta^{K}\\
		& \eqspace-a^{3/2}(\partial_{t}\ee^{-2\varphi})\epsilon_{IJK}\Theta^{I}\wedge\Theta^{J}\wedge\Theta^{K},
	\end{split}
\end{align}
\end{subequations}
which contains terms $\partial_{t}\mathcal{A}^{I}$ which depend on the flow of the connection forms for the $\text{T}^{3}$ fibration. Finally, the flow of the dilaton and the connection forms is
\begin{subequations}
\begin{align}
	\partial_{t}\ee^{2\varphi} & =\tfrac{1}{2}b,\\
	\partial_{t}\mathcal{A}_{I} & =\tfrac{1}{2}\star_{\text{K3}}(\dd\lambda_{IJ}\wedge j^{J}).
\end{align}
\end{subequations}

Using the $\SU 2$ algebra obeyed by the hyper-K\"ahler forms, $j_{I}{}^{\mu}{}_{\rho}j_{J}{}^{\rho}{}_{\nu}=-\delta_{IJ}\delta_{\nu}^{\mu}-\epsilon_{IJK}j^{K}{}^{\mu}{}_{\nu}$, one can show that the flow for the connection forms obeys
\begin{equation}
	j^{I}\wedge\partial_{t}\mathcal{A}_{I}=0+\mathcal{O}(\ap),\qquad\epsilon_{IJK}j^{J}\wedge\partial_{t}\mathcal{A}^{K}=-\tfrac{1}{2}\dd(\lambda_{IJ}j^{J})+\mathcal{O}(\ap).
\end{equation}
Compared with the conditions for the torus fibration in \eqref{eq:T3_susy}, we see that the flow preserves supersymmetry to the required order in the $\ap$ expansion. Furthermore, if one starts with $\lambda_{IJ}=0$ corresponding to extended supersymmetry, there is no flow for the fibration connections and so the flow will preserve the full $N=4$ supersymmetry.

Again, one might wonder when the system collapses to a single scalar equation, as was the case for the six-dimensional Fu--Yau ansatz. From the above equations, setting $\lambda_{IJ}=\text{constant}$ implies $\partial_{t}\mathcal{A}_{I}=0$. The $\Gx 2$ flow then reduces to a single equation
\begin{equation}
	\partial_{t}\ee^{2\varphi}=\tfrac{1}{2}b,
\end{equation}
with $b$ given by \eqref{eq:G2_b_def}.\footnote{The difference between the $N=2$ and $N=4$ cases is an additional constant source term in $b$ of the form $\lambda_{IJ}\lambda^{IJ}$.} Stationary points of the system then correspond to $b=0$. As was the case for $N=2$ solutions in six dimensions~\cite{1610.02740}, we expect one can prove long-time existence and convergence for the $N=4$ flow with $\lambda_{IJ}=0$, giving an alternative existence proof for these solutions.

\subsection{Anomaly flow on \texorpdfstring{$\Spin7$}{Spin(7)} structure manifolds}\label{sec:spin7}

As a final example, we point out that one can also write down a generalisation of anomaly flow on eight-manifolds with $\Spin 7$ structures. As in the $\Gx 2$ case, the flow generically does not preserve supersymmetry, however there is again a reduction to an $\SU 3$ structure for which supersymmetry is preserved, giving $\SU 3$ anomaly flow. In addition, there is again a Fu--Yau-like ansatz of a four-torus bundle over a K3 surface where the flow equations reduce to a single scalar equation for the dilaton. The analysis is very similar to the case examined in the previous subsection, so we will be brief.

Let $X$ be an eight-manifold admitting a $\Spin 7$ structure defined by an invariant unit-norm spinor $\epsilon$. The spinor defines 
a self-dual four-form $\Phi$ as
\begin{equation}
\Phi=\epsilon^{\dagger}\gamma_{(4)}\epsilon,
\end{equation}
which can be used to construct projectors of $p$-forms onto $\Spin 7$ representations as in Appendix \ref{app:G-structure}. Heterotic supergravity compactified on $\bR^{1,1}\times X$ preserves minimal supersymmetry in two dimensions (one supercharge) if the $\Spin 7$ structure satisfies \cite{Gauntlett:2001ur,Ivanov:2001ma}
\begin{equation}
\Phi\wedge\star\dd\Phi=12\star\dd\varphi,\qquad H=\star\ee^{2\varphi}\dd(\ee^{-2\varphi}\Phi),\label{eq:spin7_susy}
\end{equation}
and the curvature of the vector bundle over $X$ is a $\Spin 7$ instanton~\cite{Corrigan:1982th,Donaldson:1996kp}. These fix the intrinsic torsion of the $\Spin 7$ structure as
\begin{subequations}
    \begin{align}
\dd\Phi & =\frac{12}{7}\dd\varphi\wedge\Phi-\star H_{\rep{48}},\\
H & =-\frac{2}{7}\star(\dd\varphi\wedge\Phi)+H_{\rep{48}},
\end{align}
\end{subequations}
where $H_{\rep{48}}$ is not fixed by supersymmetry, and require $F$ to transform only in the $\rep{21}$ representation with no component in the $\rep 7$.

We assume the flow is again driven by the functional \eqref{eq:anomaly_functional}, which can be written as
\begin{equation}
I=\int_{X}\ee^{-2\varphi}\mathcal{B}\wedge\Phi.
\end{equation}
The flow of the vielbein, dilaton and spinor are then given by
\begin{subequations}\label{eq:spin7_flow_fields}
    \begin{align}
\partial_{t}\varphi & =-\frac{1}{4}\frac{1}{4!}\ee^{2\varphi}\Phi^{m_{1}\dots}\mathcal{B}_{m_{1}\dots},\\
\partial_{t}e_{m}{}^{a} & =-\frac{1}{4}\frac{1}{3!}\ee^{2\varphi}\Phi_{m}{}^{n_{1}n_{2}n_{3}}\mathcal{B}^{a}{}_{n_{1}n_{2}n_{3}},\\
\partial_{t}\epsilon & =-\frac{1}{4}\frac{1}{4!}\ee^{2\varphi}\Phi^{m_{1}m_{2}m_{3}}{}_{n}\gamma^{m_{4}n}\epsilon\,\mathcal{B}_{m_{1}\dots m_{4}},
\end{align}
\end{subequations}
with which one can show that the four-form flows as\footnote{If one replaces the $-1/4$ coefficients in \eqref{eq:spin7_flow_fields} with constants $c_{i}$, the flow of the four-form can be written as
\[
\partial_{t}(\ee^{-2\varphi}\Phi)=\tfrac{1}{3}\left(84(c_{2}-c_{1})\Pi_{\rep 1}+48(c_{3}-c_{2})\Pi_{\rep 7}+12c_{2}\Pi_{\rep{35}}\right)\mathcal{B}.
\]
}
\begin{equation}
\partial_{t}(\ee^{-2\varphi}\Phi)=-\tfrac{1}{2}(\mathcal{B}-\star\mathcal{B}).
\end{equation}
Notice that the flow is by an anti-self-dual four-form and so it corresponds to deforming by an element of $\Omega^4_{\rep{35}}$. Following \cite{2003mathSpiro}, deformations in the $\rep{7}$ or $\rep{27}$ do not change the $\Spin{7}$ metric, while those in the singlet simply cause a conformal rescaling, so the flow of the four-form by an element of the $\rep{35}$ is in some sense the most interesting kind of $\Spin{7}$ deformation.

If the metric on $X$ has the form of a two-torus fibration over a six-manifold $X_{6}$ with a corresponding reduction of the $\Spin 7$ structure to $\SU 3$, the four-form can be written in terms of $\omega$ and $\Omega$ supported on $X_6$ as
\begin{equation}
\Phi=\omega\wedge\Theta^{1}\wedge\Theta^{2}-\tfrac{1}{2}\omega\wedge\omega+\tfrac{1}{2}\Omega\wedge(\Theta^{2}+\ii\Theta^{1})+\tfrac{1}{2}\bar{\Omega}\wedge(\Theta^{2}-\ii\Theta^{1}),
\end{equation}
where the $\Theta^{I}$ are dual to the Killing vectors of the two-torus fibres. Assuming the $\SU 3$ structure on $X_{6}$ solves the Hull--Strominger system, the Bianchi identity $\mathcal{B}$ is horizontal with respect to the fibration, and the two-torus fibration is hermitian Yang--Mills, the $\Spin 7$ supersymmetry condition \eqref{eq:spin7_susy} is automatically solved with the $\Spin 7$ flow reducing to
\begin{equation}
\partial_{t}(\ee^{-2\varphi}\omega\wedge\omega)=\mathcal{B},\qquad\partial_{t}(\ee^{-2\varphi}\Omega)=0,
\end{equation}
which, together with a flow for the gauge bundle, are exactly the evolution equations for anomaly flow. As was the case for a circle fibration over six-manifold is Section \ref{sec:g2_circle}, it is not possible to fix $\dd\Theta^I$ once and for all at the start of the flow and still preserve supersymmetry, since they should be hermitian Yang--Mills with respect to $\omega$, which itself flows. Again, one might try to define a flow for the $\mathcal{A}^{I}$
or simply take $\dd\Theta^I=0$ so that the metric is a product of a $X_6$ with a two-torus. In this latter case, the $\Spin{7}$ flow reduces to $\SU3$ anomaly flow on the six-manifold.

If $X$ is a holomorphic $\text{T}^{4}$ fibration over a K3 surface, with the curvatures individually solving hermitian Yang--Mills on the base, the background will preserve eight supercharges and the above flow reduces to a single scalar flow equation for the dilaton:
\begin{equation}
\partial_t \ee^{2\varphi}=\frac{1}{2}\Bigl(\Delta_{\text{K3}}\ee^{2\varphi}-a\,\star_{\text{K3}}(f^{I}\wedge f_{I})-\tfrac{1}{4}\ap\star_{\text{K3}}(\tr F\wedge F-\tr R^{+}_{\text{K3}}\wedge R^{+}_{\text{K3}})\Bigr),
\end{equation}
where $a^{2}$ is the volume of the $\text{T}^{4}$ fibres, the $f^{I}=\dd\Theta^I$ are four two-form curvatures which are hermitian Yang--Mills on K3, and $R^{+}_{\text{K3}}$ is the curvature of the Hull connection on K3. Unlike above, the $f^{I}$ can be fixed as part of the initial data as the K\"ahler forms on K3 do not flow. As in the previous section, this equation is of the same form as that analysed in \cite{1610.02740}, and so these solutions are guaranteed to exist. More generally, one should be able to relax the requirement that the $f_{I}$ are individually hermitian Yang--Mills, with a corresponding reduction in the amount of supersymmetry preserved.

\section{Anomaly flow as a gradient flow}\label{sec:gradient}

In the previous sections, we introduced a functional $I$ which was intimately related to the bosonic action and squares of supersymmetry operators via a heterotic Lichnerowicz identity. Using this functional, we saw that the evolution equations for anomaly flow could be written in terms of variations of $I$, which then allowed us to generalise and extend anomaly flow to seven- and eight-manifolds. Though this procedure does indeed lead to interesting flows, it is not entirely satisfying since the Bianchi identity $\mathcal{B}$ was treated a background field for the purposes of variations. This is not how one usually thinks of a geometric flow as the gradient flow of a functional, as we reviewed for Ricci flow in Section \ref{sec:functional}. In this final section, we remedy this by introducing a functional whose gradient flow does indeed lead to anomaly flow, with its generalisation to seven- and eight-manifolds reproducing the $\Gx 2$ and $\Spin 7$ flows we gave in Sections \ref{sec:g2} and \ref{sec:spin7}.

The trick is to note that for all of the examples considered in the previous section, even if the flow does not preserve all of the supersymmetry conditions, it always preserves the relation between the three-form flux $H$ and a derivative of the invariant form that defines the $G$-structure. For simplicity of presentation, let us focus on the case where $X$ is a six-manifold with an $\SU3$ structure that defines a supersymmetric geometry. The three-form flux $H$ is then given by
\begin{equation}
H=\star\ee^{2\varphi}\dd(\ee^{-2\varphi}\omega).\label{eq:H_def}
\end{equation}
The curious point about these heterotic backgrounds is that the equation of motion for $H$ follows automatically from this expression, while the Bianchi identity for $H$ is non-trivial. If we introduce a dual three-form flux $\tilde{H}$ via
\begin{equation}
\tilde{H}=\ee^{-2\varphi}\star H,
\end{equation}
the equation of motion for $\tilde{H}$ is then the Bianchi identity for $H$, and vice versa. Since the equation of motion for $H$ is automatically satisfied given \eqref{eq:H_def}, we have $\dd\tilde{H}=0$ which allows us to introduce a dual two-form potential $\tilde{B}$ via $\tilde{H}=\dd\tilde{B}$. We can identify this potential as
\begin{equation}
\tilde{B}=-\ee^{-2\varphi}\omega,
\end{equation}
up to an exact term.

Now consider a new functional $\mathcal{I}$ given by
\begin{equation}
\mathcal{I}[\varphi,e,\epsilon,\tilde{B}]=\int_{X}\bigl(\tilde{B}+\ee^{-2\varphi}\omega\bigr)\wedge\mathcal{B},\label{eq:calI_def}
\end{equation}
which is a functional of $(\varphi,e,\epsilon)$ and a two-form potential $\tilde{B}$. When varying this, one finds
\begin{equation}
\begin{aligned}\delta\mathcal{I} & =\int_{X}\left(\delta\tilde{B}\wedge\mathcal{B}+\delta\bigl(\ee^{-2\varphi}\omega\bigr)\wedge\mathcal{B}+\bigl(\tilde{B}+\ee^{-2\varphi}\omega\bigr)\wedge\delta\mathcal{B}\right)\\
 & \equiv\int_{X}\ee^{-2\varphi}\vol\left(\frac{\delta\mathcal{I}}{\delta\varphi}\delta\varphi+\frac{\delta\mathcal{I}}{\delta e_{m}{}^{a}}\delta e_{m}{}^{a}+\left(\frac{\delta\mathcal{I}}{\delta\epsilon}\right)^{\dagger}\delta\epsilon+\delta\epsilon^{\dagger}\frac{\delta\mathcal{I}}{\delta\epsilon}+\frac{\delta\mathcal{I}}{\delta\tilde{B}_{mn}}\delta\tilde{B}_{mn}\right).
\end{aligned}
\end{equation}
The variation of the Bianchi identity, $\delta\mathcal{B}$, will be complicated and contain variations of $(\varphi,e,\tilde{B})$. However, if one varies and then restricts to the case where the torsion condition \eqref{eq:H_def} is satisfied, the two-form $\tilde{B}+\ee^{-2\varphi}\omega$ that appears in front of the variation will vanish.\footnote{Since $\mathcal{B}$ is assumed to be trivial in cohomology, it is sufficient that $\tilde{B}+\ee^{-2\varphi}\omega$ is zero up to an exact term, which is precisely the gauge freedom in defining $\tilde{B}$.} In this case, the $\delta\mathcal{B}$ term will not contribute at all to the resulting variations of $\mathcal{I}$, and the above simplifies to
\begin{equation}\label{eq:B_var}
\delta\mathcal{I}=\int_{X}\delta\tilde{B}\wedge\mathcal{B}+\int_{X}\delta\bigl(\ee^{-2\varphi}\omega\bigr)\wedge\mathcal{B}.
\end{equation}
The second term on the right-hand side is exactly the variation of the functional $I$ in \eqref{eq:6d_functional} when one treats $\mathcal{B}$ as a background field. The variations of $\mathcal{I}$ with respect to $(\varphi,e,\epsilon)$ then exactly reproduce the variations of $I$ given in \eqref{eq:variation}:
\begin{equation}
\frac{\delta\mathcal{I}}{\delta\varphi}=\left.\frac{\delta I}{\delta\varphi}\right|_{\text{\ensuremath{\mathcal{B}} bg field}},\qquad\frac{\delta\mathcal{I}}{\delta e_{m}{}^{a}}=\left.\frac{\delta I}{\delta e_{m}{}^{a}}\right|_{\text{\ensuremath{\mathcal{B}} bg field}},\qquad\frac{\delta\mathcal{I}}{\delta\epsilon}=\left.\frac{\delta I}{\delta\epsilon}\right|_{\text{\ensuremath{\mathcal{B}} bg field}}.
\end{equation}
Defining the flow equations as in \eqref{eq:flow_I}, it follows that anomaly flow is the gradient flow of the functional $\mathcal{I}$ without assuming that $\mathcal{B}$ is a background field.

What about the flow of the seemingly new degree of freedom $\tilde{B}$? One can compute this from the variation
\begin{equation}
\frac{\delta\mathcal{I}}{\delta\tilde{B}}=\ee^{2\varphi}(\star\mathcal{B}+\Delta),
\end{equation}
where $\Delta =\star\dd\left(H-\ee^{2\varphi}\star\dd(\ee^{-2\varphi}\omega)\right) + \ap(\dots)$, and we have not written the $\ap$ terms.\footnote{These can be computed using 
\[\delta R^{+}_{m_{1}m_{2}n_{1}n_{2}}=-\nabla_{[m_{1}}\delta H_{m_{2}]n_{1}n_{2}}-H_{[m_{1}|n_{1}}{}^{p}\delta H_{|m_{2}]n_{2}p},\]
but they are not particularly enlightening, so we do not give them explicitly.}
Defining the flow of $\tilde{B}$ as $\partial_{t}\tilde{B}=-\frac{1}{2}\ee^{-2\varphi}\frac{\delta I}{\delta\tilde{B}}$, and noting that anomaly flow fixes the flow of the hermitian form as $\partial_{t}(\ee^{-2\varphi}\omega)=\tfrac{1}{2}\star\mathcal{B}$, one finds
\begin{equation}
\partial_{t}(\tilde{B}+\ee^{-2\varphi}\omega)=-\tfrac{1}{2}\Delta.
\end{equation}
Crucially, for $\tilde{B}=-\ee^{2\varphi}\omega$, $\Delta$ vanishes. Thus, if one fixes $\tilde{B}=-\ee^{2\varphi}\omega$ at the start of the flow, the combination $\tilde{B}+\ee^{-2\varphi}\omega$ does not flow.
The condition $\tilde{B}=-\ee^{-2\varphi}\omega$ is then preserved by the flow, ensuring that the variation of $\mathcal{B}$ never contributes to \eqref{eq:B_var}.
In summary, anomaly flow can be understood as the gradient flow of the functional $\mathcal{I}$, defined in \eqref{eq:calI_def}, where the initial data is a supersymmetric geometry, and in particular satisfies \eqref{eq:H_def} with this condition preserved along the flow. 

Everything we have said generalises readily to seven- and eight-manifolds. For example, the functionals are then given by
\begin{subequations}
\begin{align}
\mathcal{I}_{\Gx 2} & =\int_{X}\bigl(\tilde{B}+\ee^{-2\varphi}\phi\bigr)\wedge\mathcal{B},\\
\mathcal{I}_{\Spin 7} & =\int_{X}\bigl(\tilde{B}+\ee^{-2\varphi}\Phi\bigr)\wedge\mathcal{B},
\end{align}
\end{subequations}
where $\tilde{B}$ are now three- and four-form potentials respectively. In these cases, as we have emphasised in previous sections, anomaly flow does not preserve all of the supersymmetry conditions for the backgrounds. However, it turns out it does preserve the relation between the torsion $H$ and the derivative of an invariant form, which is all that is needed for the flow driven by $\mathcal{I}$ to reduce to the relevant anomaly flow. For example, the $\Gx 2$ flow in \eqref{eq:G2_flow} preserves the closure of $\ee^{-2\varphi}\star\phi$. This is equivalent to the defining spinor being parallel with respect to $\nabla^{-}$, with its torsion given by $H=-\star\ee^{2\varphi}\dd(\ee^{-2\varphi}\phi)$~\cite{Friedrich:2001nh,0908.2927}. For the $\Spin 7$ flow in \eqref{eq:spin7_flow_fields}, the torsion can be defined via $H=\star\ee^{2\varphi}\dd(\ee^{-2\varphi}\Phi)$ without any further conditions on $\Phi$~\cite{Ivanov:2001ma,0908.2927}.

One might wonder if the new $\tilde{B}\wedge\mathcal{B}$ term can be motivated from heterotic supergravity alone. In fact, it is exactly the term that one must add to the bosonic action after dualising $H$. Upon dualising, the kinetic term for $H$ becomes a kinetic term for $\tilde{H}$, which can then be varied by assuming $\tilde{H}=\dd \tilde{B}$. However, the equation of motion for $\tilde{H}$ should reproduce the Bianchi identity for $H$. In order for this to happen, one has to add a Chern--Simons term to the bosonic action of the form
\begin{equation}
    \int_X \tilde{B}\wedge \mathcal{B}.
\end{equation}
This is exactly the new term that appears in $\mathcal{I}$. Compared with the Lichnerowicz identity in \eqref{eq:Lichnerowicz}, this amounts to adding $\tfrac{1}{4}\int_X \tilde{B}\wedge\mathcal{B}$ to both sides of the identity! Thus, we see that the formulation of anomaly flow in terms of the functional $\mathcal{I}$ follows from rewriting the heterotic Lichnerowicz identity in terms of the dual three-form flux $\tilde{H}$.

\section{Discussion}\label{sec:discussion}

In this paper, we investigated the relationship between supersymmetry and geometric flows in heterotic supergravity, which are driven by the Bianchi identity for the three-form flux $H$. We demonstrated how these flows can be obtained from a functional that appears in a heterotic Lichnerowicz identity for the bosonic action. On a complex threefold defining a supersymmetric geometry, this reproduced the anomaly flow. Crucially, we used the appropriate connection in the Bianchi identity to guarantee that the flow's fixed points satisfy the supergravity equations of motion. We then extended anomaly flow to seven- and eight-dimensional manifolds with G$_2$ or Spin(7) structures. The resulting flows do not generically preserve supersymmetry, but they are still modified Ricci flows. In cases where the structure group reduces further to $\SU3$ or $\SU2$, we showed that the flows reduced to anomaly flow or gave generalisations of anomaly flow on a Goldstein--Prokushkin fibration. Finally, we proposed a related functional whose gradient flow agrees with these new anomaly flows. There are a number of directions for future work, some of which we comment on here.

\subsubsection*{Connections with spinor flows}
It would be interesting to understand the relation (if any) between flows driven by a Bianchi identity and various \emph{spinor flows} that have been discussed in the literature. In the case without flux~\cite{2012arXiv1207.3529A}, spinor flow was introduced as the gradient flow of the ``energy'' functional $\int_X \vol |\nabla\epsilon|^2$. The flow has short-time existence with fixed points that are covariantly constant spinors. The case with flux was examined in \cite{2021arXiv211200814C}, where a flow was introduced whose fixed points are spinors that are constant with respect to a covariant derivative but now with flux contributions. In addition, the flow for the flux enforces both its equation of motion and a Bianchi identity. In our case, the Lichnerowicz identity \eqref{eq:Lichnerowicz} gives a natural starting point for examining these flows in the heterotic context. In the case without flux, the supersymmetry operators reduce to the Levi-Civita connection $\nabla$, so that the $|D\epsilon|^2$ term reproduces the above energy functional (up to a normalisation factor of the dilaton). In the case with flux, one has both $|D\epsilon|^2$ and $|\feyn{D}\epsilon|^2$ terms in the identity. However, at least for an $\SU3$ structure, one can set $\feyn{D}\epsilon=0$ by imposing the supersymmetry conditions coming from the heterotic superpotential~\cite{Gurrieri:2004dt,Benmachiche:2008ma,delaOssa:2015maa,McOrist:2016cfl,Ashmore:2018ybe} -- these amount to the conditions on $\Omega$ and $H$ in the Hull--Strominger system. The remaining conditions are the Bianchi identity for $H$, hermitian Yang--Mills for $F$ and that the hermitian metric is conformally balanced. One might hope that the spinor flow within this restricted space of solutions has an interesting physical interpretation and is better behaved than ``vanilla'' spinor flow. We hope to comment on this in future work.

\subsubsection*{$N=1$ supergravity and generalised geometry}

The Lichnerowicz identity can be used to make a precise connection between the geometry of $X$ and the language of four-dimensional $N=1$ supergravity. After compactifying on $X$, the supergravity potential $\mathcal{V}$ can be written in terms of a superpotential $\mathcal{W}$ and a K\"ahler potential $\mathcal{K}$ as
\begin{equation}
\mathcal{V}=\ee^{\mathcal{K}}\left(g^{\alpha\bar{\beta}}\delta_{\alpha}\mathcal{W}\,\delta_{\bar{\beta}}\bar{\mathcal{W}}-3|\mathcal{W}|^{2}\right)+\tfrac{1}{2}(\re f)^{\mathcal{AB}}\mathcal{P}_{\mathcal{A}}\mathcal{P}_{\mathcal{B}},
\end{equation}
where $g_{\alpha\bar{\beta}}$ is a K\"ahler metric and $\delta_{\alpha}\mathcal{W}=\partial_{\alpha}\mathcal{W}-(\partial_{\alpha}\mathcal{K})\mathcal{W}$ are covariant variations, both on the space of chiral fields (labelled by $\alpha$), $\mathcal{P}_{\mathcal{A}}$ are moment maps for the action of symmetries on the fields, and $(\re f)_{\mathcal{AB}}$ is a positive-definite invariant metric on the Lie algebra of the symmetry group (interpreted as a metric on the space of gauge multiplets). In fact, the supergravity potential $\mathcal{V}$ is given by the (negative of the) bosonic action restricted to the internal manifold $X$. For an $\SU 3$ compactification, one can see this story explicitly using the expression for the four-dimensional superpotential in terms of the geometry of $X$~\cite{Gurrieri:2004dt,Benmachiche:2008ma,delaOssa:2015maa,McOrist:2016cfl,Ashmore:2018ybe}. Using this, one can show that imposing the F-term constraints, equivalent to $\delta_{\alpha}\mathcal{W}=0$, sets $\feyn D\epsilon=0$. In terms of the $\SU 3$ structure, this imposes $\dd(\ee^{-2\varphi}\Omega)=0$ and $H=\ii(\bar{\partial}-\partial)\omega$, but it is not sufficient to recover the conformally balanced condition. The remaining Killing spinor equation simplifies to
\begin{equation}
D_{m}\epsilon=\ii Q_{m}\gamma_{*}\epsilon,\qquad Q=\ii(\partial-\Bar{\partial})\varphi-\ii(W_{4}^{(1,0)}-W_{4}^{(0,1)}),
\end{equation}
where the $W_{4}$ torsion class is unfixed and appears in $\omega\wedge\dd\omega=W_{4}\wedge\omega\wedge\omega$~\cite{Grana:2004bg}. The Lichnerowicz identity then reduces to
\begin{equation}
S_{\text{B}}=\int_{X}\ee^{-2\varphi}\vol\left[-4|W_{4}-\dd\varphi|^{2}+\epsilon^{\dagger}\cancel{\dd H}\epsilon\right].
\end{equation}
Since we have imposed the superpotential conditions, one might imagine that this is equivalent to the square of the moment map, since the potential should be
\begin{equation}
-S_{\text{B}}=\mathcal{V}\equiv\tfrac{1}{2}(\re f)^{\mathcal{AB}}\mathcal{P}_{\mathcal{A}}\mathcal{P}_{\mathcal{B}},
\end{equation}
with the minimum of the potential corresponding to the ``D-term'' condition $\mathcal{P}_{\mathcal{A}}=0$. In fact, this exactly agrees with the picture in generalised geometry~\cite{1910.04795,Ashmore:2019rkx} (at least after solving the Bianchi identity). After imposing the superpotential conditions, one finds that supersymmetry requires the vanishing of the moment map
\begin{equation}
\mu(\lambda)=\int_{X}\lambda\wedge\dd(\ee^{-2\varphi}\omega\wedge\omega)=2\int_{X}\ee^{-2\varphi}\lambda\wedge(W_{4}-\dd\varphi)\wedge\omega\wedge\omega,
\end{equation}
for all one-forms $\lambda$. The one-form $\lambda$ is a component of a ``generalised vector'', parametrising an element of the algebra $\mathfrak{gdiff}$ of diffeomorphisms and gauge transformations on $X$, so that $\mu$ takes values in $\det T^{*}X\otimes\mathfrak{gdiff}^{*}$. We can then view the above component of $\mu$ as a vector density:
\begin{equation}
\mu=2\,\ee^{-2\varphi}\vol\star\left((W_{4}-\dd\varphi)\wedge\omega\wedge\omega\right)^{\sharp}.
\end{equation}
The metric $\re f$ is then simply the ``generalised metric'' $G$, with the potential given by the Yau functional~\cite{https://doi.org/10.1002/cpa.3160310304} associated to the moment map:
\begin{equation}
\begin{split}\mathcal{V} & =\int_{X}\vol_{G}^{-1}G^{\mathcal{AB}}\mu_{\mathcal{A}}\mu_{\mathcal{B}}=\int_{X}(\ee^{-2\varphi}\vol)^{-1}g_{mn}\mu^{m}\mu^{n}\\
 & \propto\int_{X}\ee^{-2\varphi}\vol|W_{4}-\dd\varphi|^{2}.
\end{split}
\end{equation}
We see that this expression has exactly the right form to give the potential due to the remaining D-term. It is then simple to include the leading-order $\ap$ corrections in this argument using the moment map given in \cite{Ashmore:2019rkx}.

\subsubsection*{$\ap$ corrections and the existence of string vacua}
For the Hull--Strominger system, the anomaly flow is known to have short-time existence when the length scale set by the curvature is large compared with $\sqrt{\ap}$~\cite{1508.03315}. Furthermore, there are ``stability'' results which show that, for $\ap=0$, the flow is stable around Calabi--Yau metrics~\cite{2005.05670}. From a physical point of view, it would be interesting to understand the inverse of this problem, namely what kind of $\ap$ corrections can be included without breaking known existence theorems? Without fluxes, there is a worldsheet argument along these lines which shows that starting from a Calabi--Yau solution, adding $\ap$ corrections does not spoil the existence of a solution to the spacetime equations of motion (instead leading to a non-Ricci-flat K\"ahler metric)~\cite{Nemeschansky:1986yx}. This has now been extended to $(0,2)$ sigma models with a smooth $\ap\to0$ limit (which notably does not include the T$^2$-fibred backgrounds discussed in this paper).\footnote{Spacetime analogues of these results have appeared for Calabi--Yau~\cite{Becker:2015wga} and exceptional holonomy backgrounds~\cite{Becker:2014rea}.} With this in mind, one might hope that if long-time existence and convergence can be shown for a given ansatz at first order in $\ap$, these properties persist when small, higher-order $\ap$ corrections are included.

\subsubsection*{Numerical approximations to non-K\"ahler metrics}
A more practical question is whether anomaly flow, or any of the geometric flows we have described, are amenable to numerical simulation. There is now a long history of studying Ricci flow numerically for a variety of Riemannian and Lorentzian manifolds~\cite{Headrick:2006ti, Doran:2007zn, Adam:2011dn, Holzegel_2007, Garfinkle:2003an, Holzegel:2007ud, DeBiasio:2022nsd},\footnote{See \cite{wiseman_book} for a nice review.} so one might wonder whether the flows we have discussed can be used to construct heterotic flux backgrounds numerically. We saw that in certain cases, the system reduced to a single flow equation for the dilaton, which was equivalent to solving the Bianchi identity. These are also the systems which have long-term existence and convergence. This is important as it gives one confidence that if a numerical algorithm appears to converge to a solution, it is likely to be the solution that is known to exist, rather than a ``non-solution'' that is numerically close to solving the equations of motion. In another direction, rather than solving these systems by flowing, one might simply solve the equation for the dilaton using some other method, such as by using a neural network to parametrise $\varphi$ and then minimising a loss function which encodes the Bianchi identity. This approach has been used recently for finding numerical Calabi--Yau metrics~\cite{Jejjala:2020wcc,Afkhami-Jeddi:2021qkf,Ashmore:2021ohf,Larfors:2021pbb,Larfors:2022nep,Berglund:2022gvm,Gerdes:2022nzr} and hermitian Yang--Mills connections~\cite{Ashmore:2021rlc}, and is likely to be a powerful technique for solving non-linear PDEs in the future.

\subsection*{Acknowledgements}

It is a pleasure to thank André Coimbra, Dan Israël, Jock McOrist, Ilarion Melnikov, Savdeep Sethi and Eirik Eik Svanes for useful discussions. AA is supported in part by NSF Grant No.~PHY2014195 and in part by the Kadanoff Center for Theoretical Physics. AA also acknowledges the support of the European Union’s Horizon 2020 research and innovation program under the Marie Skłodowska-Curie grant agreement No.~838776. RM is supported in part by ERC Grant 787320-QBH Structure and by ERC Grant 772408-Stringlandscape.

\appendix

\section{Conventions}\label{app:conventions}

Throughout this paper, ten-dimensional frame indices are denoted by $(A,B,\ldots)=0,1,\dots,9$. When compactifying on an internal space of dimension $d$, we denote internal frame indices by $(a,b,\ldots)=1,\dots,d$. Likewise, we denote by $(M,N,\dots)$ and $(m,n,\dots)$ the ten- and $d$-dimensional coordinate frame indices. The indices $(I,J,K,L)$ label coordinates on tori or elements of $\SU2$ triplets, with no distinction between raised and lowered indices.

\subsection{Heterotic string theory}\label{app:eom}

The effective action for the bosonic sector of heterotic string theory, correct to $\mathcal{O}(\ap)$, is
\begin{equation}
S_{\text{B}}=\int\ee^{-2\varphi}\vol_{10}\left(R+4(\nabla\varphi)^{2}-\frac{1}{2}H^{2}-\frac{\ap}{4}\left(\tr |F|^2 - \tr |R^+|^2\right)\right),
\end{equation}
where $H^{2}=\tfrac{1}{3!}H^{MNP}H_{MNP}$, $|F|^2 = \tfrac{1}{2!} F_{MN}F^{MN}$ and $\vol_{10}$ is the volume form associated to the string-frame metric $g_{MN}$. Anomaly cancellation requires that $H$ satisfies a non-trivial Bianchi identity, given by
\begin{equation}
    \dd H = \frac{\ap}{4}\left(\tr F\wedge F - \tr R^+\wedge R^+\right).
\end{equation}

In the $\ap\to0$ limit, the bosonic action simplifies to
\begin{equation}
S_{\text{B}}=\int\ee^{-2\varphi}\vol_{10}\Bigl(R+4(\nabla\varphi)^{2}-\tfrac{1}{2}H^{2}\Bigr),
\end{equation}
The variation of this action can be written in terms of the one-loop beta functions for the worldsheet string as
\begin{equation}
\begin{split}\delta S_{\text{B}} & =\int\ee^{-2\varphi}\vol_{10}\Bigl[-\delta g_{MN}\beta^{MN}(g)-\delta B_{MN}\beta^{MN}(B)\\
 & \phantom{=\int\ee^{-2\varphi}\vol_{10}\Bigl[}+\left(\tfrac{1}{2}g^{MN}\delta g_{MN}-2\delta\varphi\right)\left(g^{PQ}\beta_{PQ}(g)-4\beta(\varphi)\right)\Bigr],
\end{split}
\label{eq:variation_action}
\end{equation}
where the beta functions are
\begin{subequations}
    \begin{align}
\beta_{MN}(g) & =R_{MN}+2\nabla_{M}\nabla_{N}\varphi-\tfrac{1}{4}H_{MPQ}H_{N}{}^{PQ},\\
\beta(\varphi) & =-\tfrac{1}{2}\nabla^{2}\varphi+(\nabla\varphi)^{2}-\tfrac{1}{4}H^{2},\\
\beta_{MN}(B) & =-\tfrac{1}{2}\nabla^{P}(\ee^{-2\varphi}H_{PMN}).
\end{align}
\end{subequations}
The renormalisation group flow equations are then given by $\partial_{t}\boldsymbol{\Phi}=-\beta(\boldsymbol{\Phi})$ for $\boldsymbol{\Phi}=(g,\varphi,B)$. A consistent string background must be Weyl invariant, which requires that the beta functions all vanish. The supergravity equations of motion defined in Equations \eqref{eq:bosonic_eom} are then linear combinations of the beta functions:
\begin{subequations}
    \begin{align}
\text{eom}[g]_{MN}&=\beta_{MN}(g),\\
\text{eom}[\varphi]&=\tfrac{1}{4}\left(\beta^{P}{}_{P}(g)-4\beta(\varphi)\right),\\
\text{eom}[B]_{MN}&=-2\beta_{MN}(B).
\end{align}
\end{subequations}
From \eqref{eq:variation_action}, the variations of the bosonic action with respect to $(g,\varphi,B)$ are given by
\begin{subequations}
    \begin{align}
\frac{\delta S_{\text{B}}}{\delta g_{MN}} & =-\ee^{-2\varphi}\left(\text{eom}[g]^{MN}-2g^{MN}\text{eom}[\varphi]\right),\\
\frac{\delta S_{\text{B}}}{\delta\varphi} & =-8\ee^{-2\varphi}\text{eom}[\varphi],\\
\frac{\delta S_{\text{B}}}{\delta B_{MN}} & =\tfrac{1}{2}\ee^{-2\varphi}\text{eom}[B]_{MN}.
\end{align}
\end{subequations}
Said differently, the equations of motion and the variations are related by
\begin{subequations}\label{eq:eom_defs}
    \begin{align}
\text{eom}[g]_{MN} & =-\ee^{2\varphi}\left(\frac{\delta S_{\text{B}}}{\delta g_{PQ}}g_{MP}g_{NQ}+\frac{1}{4}g_{MN}\frac{\delta S_{\text{B}}}{\delta\varphi}\right),\\
\text{eom}[\varphi] & =-\frac{1}{8}\ee^{2\varphi}\frac{\delta S_{\text{B}}}{\delta\varphi},\\
\text{eom}[B]_{MN} & =2\ee^{2\varphi}\frac{\delta S_{\text{B}}}{\delta B_{MN}}.
\end{align}
\end{subequations}

\subsection{Connections and supersymmetry}

In heterotic string theory, there are a number of connections which commonly appear:

\paragraph{Levi-Civita}

The Levi-Civita connection $\nabla$ is the unique metric compatible connection with vanishing torsion. It is defined from derivatives of the metric in the usual way.

\paragraph{Bismut}

The supersymmetry variation of the gravitino is
\begin{equation}
\delta\psi_{M}=\nabla_{M}\varepsilon+\tfrac{1}{8}H_{MNP}\Gamma^{NP}\varepsilon+\mathcal{O}(\ap^{2}),
\end{equation}
which can be written in terms of the Bismut connection $\nabla^-$ as
\begin{equation}
\nabla_{M}^{-}\varepsilon=\nabla_{M}\varepsilon+\tfrac{1}{8}H_{MNP}\Gamma^{NP}\varepsilon.\label{eq:Bismut}
\end{equation}

\paragraph{Hull}

It is not the Bismut connection that appears in the heterotic action or anomaly condition. The connection that appears is
\begin{equation}
\nabla_{M}^{+}\varepsilon=\nabla_{M}\varepsilon-\tfrac{1}{8}H_{MNP}\Gamma^{NP}\varepsilon.\label{eq:Hull}
\end{equation}
This connection is known as the Hull connection.

\paragraph{Chern}

The Chern connection is the unique connection which is hermitian and compatible with the holomorphic structure on $T^{1,0}X$. We denote this connection by $\nabla^{\text{Ch}}$ with connection symbols $\Sigma^{\text{Ch}}$.

\subsection{Spinors and gamma matrices}\label{app:spinors}

\subsubsection*{\texorpdfstring{$\text{Cliff}(1,9)$}{Cliff(1,9)}}

In ten spacetime dimensions, the gamma matrices $\Gamma^{A}$ satisfy the anticommutation relation
\begin{equation}
\lbrace\Gamma_{A},\Gamma_{B}\rbrace=2\,\eta_{AB}\,1_{32},
\end{equation}
where $\eta_{AB}$ is the Minkowski metric with Lorentzian signature $(-,+,\dots,+)$. The Clifford algebra is generated by antisymmetrised products $\Gamma_{A_{1}\dots A_{r}}=\Gamma_{[A_{1}}\dots\Gamma_{A_{r}]}$. The chirality matrix is defined as $\Gamma_{*}=\frac{1}{10!}\varepsilon^{A_{1}\dots A_{10}}\Gamma_{A_{1}\dots A_{10}}$, with $\varepsilon^{01\dots9}=1$. Gamma matrices satisfy the hermitian conjugation property 
\begin{equation}
(\Gamma^{A})^{\dagger}=\Gamma_{0}\,\Gamma^{A}\Gamma_{0},
\end{equation}
and the Dirac conjugate of a ten-dimensional spinor $\varepsilon$ is defined as $\Bar{\varepsilon}=\varepsilon^{\dagger}\,\ii\Gamma^{0}$. The charge conjugation matrix $\Gamma_{\text{C}}$ is the unique (up to a phase) hermitian matrix satisfying the two properties $(\Gamma_{A})^{\transpose}=-\Gamma_{\text{C}}^{-1}\Gamma_{A}\Gamma_{\text{C}}$ and $(\Gamma_{\text{C}})^{\transpose}=-\Gamma_{\text{C}}.$ A Majorana spinor $\varepsilon$ obeys $\varepsilon=\Gamma_{\text{C}}\,\bar{\varepsilon}^{\transpose}$.

Though defined with frame indices, gamma matrices can be converted to coordinate indices using the ten-dimensional vielbein by $\Gamma^{M}=e^{M}{}_{A}\Gamma^{A}$. For an $r$-form $A$, the Clifford product $\feyn{A}$ is defined by
\begin{equation}
\feyn{A}=\frac{1}{r!}A_{M_{1}\dots M_{r}}\,\Gamma^{M_{1}\dots M_{r}}.
\end{equation}

\subsubsection*{\texorpdfstring{$\text{Cliff}(6)$}{Cliff(6)}}
In six Euclidean dimensions, the gamma matrices are hermitian matrices $\gamma^{a}=(\gamma^{a})^{\dagger}$ that satisfy the anticommutation relation
\begin{equation}
\lbrace\gamma_{a},\gamma_{b}\rbrace=2\,\delta_{ab}\,1_{8}.
\end{equation}
The chirality matrix is defined as $
\gamma_{*}=-\frac{\ii}{6!}\,\varepsilon_{a_{1}\dots a_{6}}\gamma^{a_{1}\dots a_{6}}$, and appears in the duality relation
\begin{equation}
\gamma^{a_{1}\dots a_{r}}=-\ii(-1)^{r(r-1)/2}\frac{1}{(6-r)!}\varepsilon^{a_{1}\dots a_{r}b_{1}\dots b_{6-r}}\gamma_{b_{1}\dots b_{6-r}}\gamma_{*}.
\end{equation}
The charge conjugation matrix $\gamma_{\text{C}}$ is defined by $(\gamma^{a})^{\transpose}=-\gamma_{\text{C}}^{-1}\gamma^{a}\gamma_{\text{C}}$ and $(\gamma_{\text{C}})^{\dagger}=(\gamma_{\text{C}})^{\transpose}=\gamma_{\text{C}}$. 
Pseudo-Majorana spinors are defined by the constraint $\epsilon=\gamma_{\text{C}}\epsilon^{*}$.

When compactifying the ten-dimensional theory on an internal space $X$ equipped with a non-vanishing pseudo-Majorana spinor $\epsilon$, spinors admit a decomposition into internal and external components. A ten-dimensional Majorana--Weyl spinor $\varepsilon$ gives a four-dimensional Majorana spinor $\eta$ via
\begin{equation}
\varepsilon=\eta_{+}\otimes\epsilon_{+}+\eta_{-}\otimes\epsilon_{-},
\end{equation}
where $\eta_{\pm}$ (respectively $\epsilon_{\pm}$) are left and right components of $\eta$ (respectively $\epsilon$).

\subsubsection*{\texorpdfstring{$\text{Cliff}(7)$}{Cliff(7)}}

In seven Euclidean dimensions, the gamma matrices are hermitian matrices $\gamma^{a}=(\gamma^{a})^{\dagger}$ that satisfy the anticommutation relation
\begin{equation}
\lbrace\gamma_{a},\gamma_{b}\rbrace=2\,\delta_{ab}\,1_{8},
\end{equation}
where now $(a,b)=1,\dots,7$. Chirality cannot be defined in odd dimensions, and the duality relation satisfied by the gamma matrices is now
\begin{equation}
\gamma^{a_{1}\dots a_{r}}=-\ii(-1)^{r(r-1)/2}\frac{1}{(7-r)!}\varepsilon^{a_{1}\dots a_{r}b_{1}\dots b_{7-r}}\gamma_{b_{1}\dots b_{7-r}}.
\end{equation}
The charge conjugation matrix is defined as in six dimensions. 

\subsubsection*{\texorpdfstring{$\text{Cliff}(8)$}{Cliff(8)}}

In eight Euclidean dimensions, the gamma matrices are hermitian matrices $\gamma^{a}=(\gamma^{a})^{\dagger}$ that satisfy the anticommutation relation
\begin{equation}
\lbrace\gamma_{a},\gamma_{b}\rbrace=2\,\delta_{ab}\,1_{16},
\end{equation}
with $(a,b)=1,\dots,8$. The chirality matrix is defined as $
\gamma_{*}=\frac{1}{8!}\,\varepsilon_{a_{1}\dots a_{8}}\gamma^{a_{1}\dots a_{8}}$, and appears in the duality relation
\begin{equation}
\gamma^{a_{1}\dots a_{r}}=(-1)^{r(r-1)/2}\frac{1}{(8-r)!}\varepsilon^{a_{1}\dots a_{r}b_{1}\dots b_{8-r}}\gamma_{b_{1}\dots b_{8-r}}\gamma_{*}.
\end{equation}
The charge conjugation matrix is defined as in six dimensions.

\subsection{\texorpdfstring{$G$}{G}-structures}\label{app:G-structure}

\subsubsection*{$\protect\Gx 2$ structures}

Following the conventions of \cite{1702.01156}, a $\Gx 2$ structure is defined from an invariant unit-norm spinor $\epsilon$ by
\begin{equation}
\phi=-\ii\epsilon^{\dagger}\gamma_{(3)}\epsilon,\qquad\star\phi=\epsilon^{\dagger}\gamma_{(4)}\epsilon.
\end{equation}
The spaces of $p$-forms on a manifold with a $\Gx 2$ structure admits decompositions into $\Gx 2$ modules. Of particular importance for us are three- and four-forms, which decompose as
\begin{equation}
\begin{aligned}\Omega^{3} & =\Omega_{\rep 1}^{3}\oplus\Omega_{\rep 7}^{3}\oplus\Omega_{\rep{27}}^{3},\\
\Omega^{4} & =\Omega_{\rep 7}^{4}\oplus\Omega_{\rep{14}}^{4},
\end{aligned}
\end{equation}
where the subscript denotes the relevant $\Gx 2$ representation. We define projectors onto these subspaces as $\pi_{\rep d}$ and $\Pi_{\rep d}$ for three- and four-forms respectively. These can be defined explicitly as
\begin{equation}
\begin{aligned}(\pi_{\rep 1})_{m_{1}m_{2}m_{3}}{}^{n_{1}n_{2}n_{3}} & =\tfrac{1}{42}\phi_{m_{1}m_{2}m_{3}}\phi^{n_{1}n_{2}n_{3}},\\
(\pi_{\rep 7})_{m_{1}m_{2}m_{3}}{}^{n_{1}n_{2}n_{3}} & =\tfrac{1}{24}\star\phi_{m_{1}m_{2}m_{3}p}\star\phi^{n_{1}n_{2}n_{3}p},\\
(\pi_{\rep{27}})_{m_{1}m_{2}m_{3}}{}^{n_{1}n_{2}n_{3}} & =\tfrac{1}{7}\phi_{m_{1}m_{2}m_{3}}\phi^{n_{1}n_{2}n_{3}}+\tfrac{1}{8}\star\phi_{m_{1}m_{2}m_{3}p}\star\phi^{n_{1}n_{2}n_{3}p}\\
&\eqspace+\tfrac{3}{2}\star\phi_{[m_{1}m_{2}}{}^{n_{1}n_{2}}\delta_{m_{3}]}^{n_{3}]},\\
(\Pi_{\rep 1})_{m_{1}\dots m_{4}}{}^{n_{1}\dots n_{4}} & =\tfrac{1}{168}\star\phi_{m_{1}\dots m_{4}}\star\phi^{n_{1}\dots n_{4}},\\
(\Pi_{\rep 7})_{m_{1}\dots m_{4}}{}^{n_{1}\dots n_{4}} & =\tfrac{1}{6}\delta_{[m_{1}}^{[n_{1}}\phi_{m_{2}m_{3}m_{4}]}\phi^{n_{2}n_{3}n_{4}]},\\
(\Pi_{\rep{27}})_{m_{1}\dots m_{4}}{}^{n_{1}\dots n_{4}} & =\tfrac{1}{28}\star\phi_{m_{1}\dots m_{4}}\star\phi^{n_{1}\dots n_{4}}+\tfrac{1}{2}\delta_{[m_{1}}^{[n_{1}}\phi_{m_{2}m_{3}m_{4}]}\phi^{n_{2}n_{3}n_{4}]}\\
&\eqspace+3\delta_{[m_{1}}^{[n_{1}}\delta_{m_{2}}^{n_{2}}\star\phi_{m_{3}m_{4}]}{}^{n_{3}n_{4}]},
\end{aligned}
\end{equation}
where these are defined to act as, for example, $(\pi_{\rep 1}\phi)_{m_{1}m_{2}m_{3}}=(\pi_{\rep 1})_{m_{1}m_{2}m_{3}}{}^{n_{1}n_{2}n_{3}}\phi_{n_{1}n_{2}n_{3}}$.

\subsubsection*{Spin(7) structures}

A $\Spin 7$ structure is defined by an invariant spinor $\epsilon$ of unit-norm  as
\begin{equation}
\Phi=\epsilon^{\dagger}\gamma_{(4)}\epsilon.
\end{equation}
Similar to the $\Gx 2$ case, the space of four-forms admits a decomposition into $\Spin 7$ modules as
\begin{equation}
\Omega^{4}=\Omega_{\rep 1}^{4}\oplus\Omega_{\rep 7}^{4}\oplus\Omega_{\rep{27}}^{4}\oplus\Omega_{\rep{35}}^{4},
\end{equation}
where the first three summands have eigenvalue $+1$ under the Hodge star, while the final summand, $\Omega^4_{\rep{35}}$, has eigenvalue $-1$ and so is anti-self-dual. There are a corresponding set of projectors onto these subspaces constructed from the $\Spin 7$ four-form:
\begin{equation}
\begin{aligned}(\Pi_{\rep 1})_{m_{1}\dots m_{4}}{}^{n_{1}\dots n_{4}} & =\tfrac{1}{336}\Phi_{m_{1}\dots m_{4}}\Phi^{n_{1}\dots n_{4}},\\
(\Pi_{\rep 7})_{m_{1}\dots m_{4}}{}^{n_{1}\dots n_{4}} & =-\tfrac{1}{48}\Phi_{[m_{1}m_{2}m_{3}}{}^{[n_{4}}\Phi^{n_{1}n_{2}n_{3}]}{}_{m_{4}]}+\tfrac{1}{48}\Phi_{p[m_{1}m_{2}m_{3}}\Phi^{p[n_{1}n_{2}n_{3}}\delta_{m_{4}]}^{n_{4}]},\\
(\Pi_{\rep{27}})_{m_{1}\dots m_{4}}{}^{n_{1}\dots n_{4}} & =\delta_{[m_{1}}^{n_{1}}\delta_{m_{2}}^{n_{2}}\delta_{m_{3}}^{n_{3}}\delta_{m_{4}]}^{n_{4}}+\tfrac{1}{56}\Phi_{m_{1}\dots m_{4}}\Phi^{n_{1}\dots n_{4}}-\tfrac{1}{16}\Phi_{[m_{1}m_{2}m_{3}}{}^{[n_{4}}\Phi^{n_{1}n_{2}n_{3}]}{}_{m_{4}]}\\
 & \eqspace-\tfrac{5}{48}\Phi_{p[m_{1}m_{2}m_{3}}\Phi^{p[n_{1}n_{2}n_{3}}\delta_{m_{4}]}^{n_{4}]},\\
(\Pi_{\rep{35}})_{m_{1}\dots m_{4}}{}^{n_{1}\dots n_{4}} & =-\tfrac{1}{48}\Phi_{m_{1}\dots m_{4}}\Phi^{n_{1}\dots n_{4}}+\tfrac{1}{12}\Phi_{[m_{1}m_{2}m_{3}}{}^{[n_{4}}\Phi^{n_{1}n_{2}n_{3}]}{}_{m_{4}]}\\
&\eqspace+\tfrac{1}{12}\Phi_{p[m_{1}m_{2}m_{3}}\Phi^{p[n_{1}n_{2}n_{3}}\delta_{m_{4}]}^{n_{4}]}.
\end{aligned}
\end{equation}

\section{Flow equation of the metric}\label{app:flow_equation}

In this appendix, we show explicitly how the flow equation of the metric can be recast in the form \eqref{eq:flow_metric}, where the internal part of the Einstein equation \eqref{eq:metric_eom} appears explicitly. This formulation refines the flow equation derived in \cite{1610.02739}, by linking the anomaly flow to the heterotic equations of motion. The derivation relies on two observations. First, as underlined in \cite{1610.02739}, the flow equation of the metric can be obtained from the flow equations \eqref{eq:flow} of the $\SU3$ structure forms for supersymmetric geometries. Second, when written in terms of the heterotic degrees of freedom, the tensor driving the flow of the metric reduces to the metric equation of motion, up to terms that are $\mathcal{O}(\alpha'^{2})$ at fixed points of the flow.

In the following, we use holomorphic coordinates $(z^{i}$, $\Bar{z}^{\Bar{\imath}})$ instead of working with real coordinates $x^{m}$. Holomorphic indices are denoted by $(i,j,k,l)=1,\dots,3$, while real indices are taken as $(m,n,p,q)=1,\ldots,6$.

\subsection{Integration of the flow equation}
Contracting the flow equation of $\ee^{-2\varphi}\omega\wedge\omega$ with the hermitian form yields
\begin{equation}
2\ee^{-2\varphi}\bigl(-\partial_{t}g_{\bar{\imath}j}-g_{\Bar{\imath}j}\,g^{\Bar{k}l}\partial_{t}g_{\Bar{k}l}+4\,g_{\Bar{\imath}j}\,\partial_{t}\varphi\bigr)=g^{\Bar{k}l}\left(\dd H-\frac{\alpha'}{4}(\tr F\wedge F-\tr R^{+}\wedge R^{+})\right)_{\Bar{k}l\Bar{\imath}j}.
\end{equation}
This flow equation is simplified considerably by noticing that the invariance of $\ee^{-2\varphi}\Omega$ under the flow implies the preservation of the volume form $\ee^{-4\varphi}\vol=-\ee^{-4\varphi}\frac{1}{3!}\omega^{3}$. This translates to the relation
\begin{equation}
\partial_{t}\varphi=\frac{1}{4}g^{\Bar{\imath}j}\partial_{t}g_{\Bar{\imath}j}.
\end{equation}
between the metric and dilaton flows. The  metric flow equation then simplifies to
\begin{equation}
\partial_{t}g_{\Bar{\imath}j}=-\frac{1}{2}\ee^{2\varphi}(\dd H-\frac{\alpha'}{4}(\tr F\wedge F-\tr R^{+}\wedge R^{+}))^{k}{}_{k\Bar{\imath}j}.
\end{equation}
In order to recast this equation as a modified Ricci flow, the contraction $\dd H^{k}{}_{k\Bar{\imath}j}$ is rewritten in \cite{1610.02739} by making explicit the dependence on the Chern connection. Indeed, the Chern curvature tensor has components
\begin{equation}
R^{\text{Ch}}_{\Bar{\imath}_{1}j_{1}\Bar{\imath}_{2}j_{2}}=\partial_{\Bar{\imath}_{1}}\partial_{j_{1}}g_{\Bar{\imath}_{2}j_{2}}-g^{\Bar{k}l}\partial_{\Bar{\imath}_{1}}g_{\Bar{\imath}_{2}l}\,\partial_{j_{1}}g_{j_{2}\Bar{k}},
\end{equation}
so antisymmetrising over the two pairs of indices $\Bar{\imath}_{1},\Bar{\imath}_{2}$ and $j_{1},j_{2}$ gives
\begin{equation}
R^{\text{Ch}}_{\Bar{\imath}_{1}j_{1}\Bar{\imath}_{2}j_{2}}-R^{\text{Ch}}_{\Bar{\imath}_{2}j_{1}\Bar{\imath}_{1}j_{2}}-R^{\text{Ch}}_{\Bar{\imath}_{1}j_{2}\Bar{\imath}_{2}j_{1}}+R^{\text{Ch}}_{\Bar{\imath}_{2}j_{2}\Bar{\imath}_{1}j_{1}}=\frac{1}{2}\dd H_{\Bar{\imath}_{1}\Bar{\imath}_{2}j_{1}j_{2}}-H_{\Bar{\imath}_{1}\Bar{\imath}_{2}}{}^{\Bar{k}}H_{j_{1}j_{2}\Bar{k}}.
\end{equation}
Contraction of this identity with the hermitian metric yields, upon using relations between contractions of the Chern curvature for conformally balanced metrics~\cite{1610.02739},
\begin{equation}
-\frac{1}{2}\dd H^{k}{}_{k\Bar{\imath}j}=R^{\text{Ch}}_{\Bar{\imath}j}+H_{\Bar{\imath}}{}^{\Bar{k}l}H_{j\Bar{k}l},
\end{equation}
where $R^{\text{Ch}}_{\Bar{\imath}j}$ is the contraction of the Chern curvature with the hermitian form as $R^{\text{Ch}}_{\Bar{\imath}j}=g^{\Bar{k}l}R^{\text{Ch}}_{\Bar{k}l\Bar{\imath}j}$.\footnote{In \cite{1610.02739}, this tensor is denoted $-\Tilde{R}_{\Bar{\imath}j}$.} The metric flow equation then simplifies to
\begin{equation}
\partial_{t}g_{\Bar{\imath}j}=\ee^{2\varphi}(R^{\text{Ch}}_{\Bar{\imath}j}+H_{\Bar{\imath}}{}^{\Bar{k}l}H_{j\Bar{k}l})+\frac{\alpha'}{8}\ee^{2\varphi}(\tr F\wedge F-\tr R^{+}\wedge R^{+})^{k}{}_{k\Bar{\imath}j},
\label{eq:anomaly_flow_equation}
\end{equation}
which agrees with the metric flow equation of \cite{1610.02739} (up to change of notation).

\subsection{Recasting as equation of motion}\label{ap:recasting_eom}

In order to recast the flow \eqref{eq:anomaly_flow_equation} as driven by the heterotic equation of motion for the metric, one needs to properly identify the tensor $R^{\text{Ch}}_{\Bar{\imath}j}+H_{\Bar{\imath}}{}^{\Bar{k}l}H_{j\Bar{k}l}$. It is natural to expect that this tensor is related to the Ricci tensor of the internal manifold; indeed, these tensors agree for Kähler manifolds. For a non-Kähler manifold, with torsion given by $H=\ii(\Bar{\partial}-\partial)\omega$, the tensor $R^{\text{Ch}}_{\Bar{\imath}j}$ differs from the Ricci tensor by two-derivative terms constructed from $H$. Once one identifies these terms, one can make the link with the relevant equation of motion of heterotic supergravity.

\subsubsection*{Riemann and Chern curvatures}
On non-Kähler manifolds, the Chern connection $\nabla^{\text{Ch}}$ and the Levi-Civita connection $\nabla$ do not coincide. In holomorphic coordinates, the connection one-forms have components
\begin{equation}
\Sigma_{m}^{\text{Ch}}{}^{a}{}_{b}\,\dd x^{m}=\begin{pmatrix}
\Sigma^{\text{Ch}}_{m}{}^{i}{}_{j}&0\\0&\Sigma^{\text{Ch}}_{m}{}^{\bar{\imath}}{}_{\bar{\jmath}}\\
\end{pmatrix}\dd x^{m},\qquad
\Sigma_{m}{}^{a}{}_{b}\,\dd x^{m}=\begin{pmatrix}
\Sigma_{m}{}^{i}{}_{j}&\Sigma_{m}{}^{i}{}_{\bar{\jmath}}\\\Sigma_{m}{}^{\bar{\imath}}{}_{j}&\Sigma_{m}{}^{\bar{\imath}}{}_{\bar{\jmath}}\\
\end{pmatrix}\dd x^{m}.
\end{equation}
The difference between the two connections can be expressed in terms of the torsion $H$ as
\begin{equation}
\Sigma^{\text{Ch}}_{m}{}^{a}{}_{b}\,\dd x^{m}-\Sigma_{m}{}^{a}{}_{b}\,\dd x^{m}=\begin{pmatrix}
-\frac{1}{2}H_{m}{}^{i}{}_{j}&\frac{1}{2}H_{m}{}^{i}{}_{\bar{\jmath}}\\\frac{1}{2}H_{m}{}^{\bar{\imath}}{}_{j}&-\frac{1}{2}H_{m}{}^{\bar{\imath}}{}_{\bar{\jmath}}\\
\end{pmatrix}\dd x^{m}.
\end{equation}
The difference between the corresponding curvatures $R^{\text{Ch}}_{m_{1}m_{2}}{}^{n_{1}}{}_{n_{2}}$ and $R_{m_{1}m_{2}}{}^{n_{1}}{}_{n_{2}}$ can be worked out accordingly as
\begin{subequations}
\label{eq:Riemann_curvature}
\begin{align}
R^{\text{Ch}}_{\Bar{\imath}_{1}i_{2}}{}^{j_{1}}{}_{j_{2}}-R_{\Bar{\imath}_{1}i_{2}}{}^{j_{1}}{}_{j_{2}}&=-\frac{1}{2}\nabla_{\Bar{\imath}_{1}}H_{i_{2}}{}^{j_{1}}{}_{j_{2}}+\frac{1}{2}\nabla_{i_{2}}H_{\Bar{\imath}_{1}}{}^{j_{1}}{}_{j_{2}}-\frac{3}{4}H_{i_{2}}{}^{j_{1}}{}_{\Bar{k}}H_{\Bar{\imath}_{1}j_{2}}{}^{\Bar{k}}\nonumber
\\
&\eqspace-\frac{1}{4}H_{\Bar{\imath}_{1}}{}^{j_{1}\Bar{k}}H_{i_{2}j_{2}\Bar{k}}+\frac{1}{4}H_{i_{2}}{}^{j_{1}\Bar{k}}H_{\Bar{\imath}_{1}j_{2}\Bar{k}}
\nonumber\\
\begin{split}
	&=-\frac{1}{2}\nabla^{\text{Ch}}_{\Bar{\imath}_{1}}H_{i_{2}}{}^{j_{1}}{}_{j_{2}}+\frac{1}{2}\nabla^{\text{Ch}}_{i_{2}}H_{\Bar{\imath}_{1}}{}^{j_{1}}{}_{j_{2}}-\frac{1}{4}H_{i_{2}}{}^{j_{1}}{}_{\Bar{k}}H_{\Bar{\imath}_{1}j_{2}}{}^{\Bar{k}}\\
	&\eqspace+\frac{1}{4}H_{\Bar{\imath}_{1}}{}^{j_{1}\Bar{k}}H_{i_{2}j_{2}\Bar{k}}-\frac{1}{4}H_{i_{2}}{}^{j_{1}\Bar{k}}H_{\Bar{\imath}_{1}j_{2}\Bar{k}},\\
\end{split}\\
R^{\text{Ch}}_{i_{1}i_{2}}{}^{j_{1}}{}_{j_{2}}-R_{i_{1}i_{2}}{}^{j_{1}}{}_{j_{2}}&=-\nabla_{[i_{1}}H_{i_{2}]}{}^{j_{1}}{}_{j_{2}}+\frac{1}{2}H^{j_{1}}{}_{[i_{1}}{}^{\Bar{k}}H_{i_{2}]j_{2}\Bar{k}}\nonumber\\
&=-\nabla^{\text{Ch}}_{[i_{1}}H_{i_{2}]}{}^{j_{1}}{}_{j_{2}}-\frac{1}{2}H^{j_{1}}{}_{[i_{1}}{}^{\Bar{k}}H_{i_{2}]j_{2}\Bar{k}}-\frac{1}{2}H_{i_{1}i_{2}}{}^{k}H^{j_{1}}{}_{j_{2}k},
\\
R^{\text{Ch}}_{\Bar{\imath}_{1}i_{2}}{}^{\Bar{\jmath}_{1}}{}_{j_{2}}-R_{\Bar{\imath}_{1}i_{2}}{}^{\Bar{\jmath}_{1}}{}_{j_{2}}
&=\frac{1}{2}\nabla_{\Bar{\imath}_{1}}H_{i_{2}}{}^{\Bar{\jmath}_{1}}{}_{j_{2}}-\frac{1}{2}\nabla_{i_{2}}H_{\Bar{\imath}_{1}}{}^{\Bar{\jmath}_{1}}{}_{j_{2}}-\frac{1}{4}H_{\Bar{\imath}_{1}}{}^{\Bar{\jmath}_{1}\Bar{k}}H_{i_{2}j_{2}\Bar{k}}+\frac{1}{4}H_{i_{2}}{}^{\Bar{\jmath}_{1}\Bar{k}}H_{\Bar{\imath}_{1}j_{2}\Bar{k}}\nonumber\\
&=\frac{1}{2}\nabla^{\text{Ch}}_{\Bar{\imath}_{1}}H_{i_{2}}{}^{\Bar{\jmath}_{1}}{}_{j_{2}}-\frac{1}{2}\nabla^{\text{Ch}}_{i_{2}}H_{\Bar{\imath}_{1}}{}^{\Bar{\jmath}_{1}}{}_{j_{2}}-\frac{1}{4}H_{\Bar{\imath}_{1}}{}^{\Bar{\jmath}_{1}\Bar{k}}H_{i_{2}j_{2}\Bar{k}}+\frac{1}{4}H_{i_{2}}{}^{\Bar{\jmath}_{1}\Bar{k}}H_{\Bar{\imath}_{1}j_{2}\Bar{k}},
\\
R^{\text{Ch}}_{i_{1}i_{2}}{}^{\Bar{\jmath}_{1}}{}_{j_{2}}-R_{i_{1}i_{2}}{}^{\Bar{\jmath}_{1}}{}_{j_{2}}&=\nabla_{[i_{1}}H_{i_{2}]}{}^{\Bar{\jmath}_{1}}{}_{j_{2}}\nonumber\\
&=\nabla^{\text{Ch}}_{[i_{1}}H_{i_{2}]}{}^{\Bar{\jmath}_{1}}{}_{j_{2}},\\
R^{\text{Ch}}_{\Bar{\imath}_{1}\Bar{\imath}_{2}}{}^{\Bar{\jmath}_{1}}{}_{j_{2}}-R_{\Bar{\imath}_{1}\Bar{\imath}_{2}}{}^{\Bar{\jmath}_{1}}{}_{j_{2}}&=\nabla_{[\Bar{\imath}_{1}}H_{\Bar{\imath}_{2}]}{}^{\Bar{\jmath}_{1}}{}_{j_{2}}-\frac{1}{2}H_{[\Bar{\imath}_{1}}{}^{\Bar{\jmath}_{1}k}H_{\Bar{\imath}_{2}]j_{2}k}-\frac{1}{2}H_{[\Bar{\imath}_{1}}{}^{\Bar{\jmath}_{1}\Bar{k}}H_{\Bar{\imath}_{2}]j_{2}\Bar{k}}\nonumber\\
\begin{split}
&=\nabla^{\text{Ch}}_{[\Bar{\imath}_{1}}H_{\Bar{\imath}_{2}]}{}^{\Bar{\jmath}_{1}}{}_{j_{2}}-\frac{1}{2}H_{[\Bar{\imath}_{1}}{}^{\Bar{\jmath}_{1}k}H_{\Bar{\imath}_{2}]j_{2}k}-\frac{1}{2}H_{[\Bar{\imath}_{1}}{}^{\Bar{\jmath}_{1}\Bar{k}}H_{\Bar{\imath}_{2}]j_{2}\Bar{k}}\\
&\eqspace+\frac{1}{2}H_{\Bar{\imath}_{1}\Bar{\imath}_{2}}{}^{\Bar{k}}H^{\Bar{\jmath}_{1}}{}_{j_{2}\Bar{k}},
\end{split}
\end{align}
\end{subequations}
with the remaining components obtained by antisymmetry of indices or complex conjugation. Here the only non-vanishing components of the Chern curvature are $R^\text{Ch}_{\Bar{\imath}_{1}i_{2}}{}^{j_{1}}{}_{j_{2}}$.

With non-vanishing torsion, curvature tensors have less symmetry than on a Kähler manifold. As an example, the Chern curvature is no longer symmetric under permutation of indices, but satisfies
\begin{equation}
R^{\text{Ch}}_{\Bar{\imath}_{1}j_{1}\Bar{\imath}_{2}j_{2}}-R^{\text{Ch}}_{\Bar{\imath}_{2}j_{1}\Bar{\imath}_{1}j_{2}}=\nabla^{\text{Ch}}_{j_{1}}H_{\Bar{\imath}_{1}\Bar{\imath}_{2}j_{2}}.
\label{eq:permutation_Chern_curvature}
\end{equation}
It should also be noted that the use of holomorphic coordinates can hide some ambiguities. Notably, the Levi-Civita connection is compatible with the hermitian metric, but not with the complex structure. Thus, special care should be taken when working with the Levi-Civita covariant derivative $\nabla$. In particular, $\nabla$ is not compatible with contraction with the hermitian form. This issue can be avoided by working with the Chern connection (which is compatible with the complex structure), and converting the result back to Levi-Civita at the end.

\subsubsection*{Ricci tensor}
The Ricci tensor can be computed from the Riemann curvature \eqref{eq:Riemann_curvature}. Its components take the form
\begin{subequations}
\label{eq:Ricci_tensor_1}
\begin{align}
R_{ij}&=\frac{1}{2}\nabla^{\text{Ch}}_{\Bar{k}}H^{\Bar{k}}{}_{ij}-\frac{1}{2}\nabla^{\text{Ch}}_{k}H^{k}{}_{ij}+\nabla^{\text{Ch}}_{i}H_{j}+\frac{1}{2}H_{ij}{}^{k}H_{k}+\frac{1}{2}H_{i}{}^{\Bar{k}l}H_{j\Bar{k}l},\\
\begin{split}
	R_{\Bar{\imath}j}&=R^{\text{Ch}}_{k\Bar{\imath}}{}^{k}{}_{j}+\frac{1}{2}\nabla^{\text{Ch}}_{\Bar{k}}H^{\Bar{k}}{}_{\Bar{\imath}j}-\frac{1}{2}\nabla^{\text{Ch}}_{k}H^{k}{}_{\Bar{\imath}j}+\nabla^{\text{Ch}}_{\Bar{\imath}}H_{j}+\frac{1}{2}H_{\Bar{\imath}j}{}^{k}H_{k}\\
	&\eqspace-\frac{1}{2}H_{\Bar{\imath}j}{}^{k}H_{k}-\frac{1}{2}H_{\Bar{\imath}}{}^{\Bar{k}l}H_{j\Bar{k}l}+\frac{1}{4}H_{\Bar{\imath}}{}^{\Bar{k}\Bar{l}}H_{j\Bar{k}\Bar{l}},
\end{split}
\end{align}
\end{subequations}
where $H_{i}=H^{j}{}_{ji}$ and $H_{\Bar{\imath}}=-H^{j}{}_{j\Bar{\imath}}$.

The form $H=\ii(\Bar{\partial}-\partial)\omega$ of the torsion implies several constraints on covariant derivatives $\nabla H$, which simplify \eqref{eq:Ricci_tensor_1}.\footnote{Such constraints can equivalently be extracted from the first Bianchi identity associated to the Chern connection.} As an example, $H$ satisfies $\partial H=\Bar{\partial}H(=i\partial\Bar{\partial}\omega)$, or in components $\partial_{[j_{1}}H_{j_{2}]\Bar{\imath}_{1}\Bar{\imath}_{2}}=\partial_{[\Bar{\imath}_{1}}H_{\Bar{\imath}_{2}]j_{1}j_{2}}$. This equality can be dressed with covariant derivatives to give
\begin{equation}
\nabla^{\text{Ch}}_{j_{1}}H_{j_{2}\Bar{\imath}_{1}\Bar{\imath}_{2}}-\nabla^{\text{Ch}}_{j_{2}}H_{j_{1}\Bar{\imath}_{1}\Bar{\imath}_{2}}-\nabla^{\text{Ch}}_{\Bar{\imath}_{1}}H_{\Bar{\imath}_{2}j_{1}j_{2}}+\nabla^{\text{Ch}}_{\Bar{\imath}_{2}}H_{\Bar{\imath}_{1}j_{1}j_{2}}=0.
\end{equation}
Contraction with the hermitian metric then implies
\begin{equation}
\nabla^{\text{Ch}}_{k}H^{k}{}_{\Bar{\imath}j}+\nabla^{\text{Ch}}_{\Bar{k}}H^{\Bar{k}}{}_{\Bar{\imath}j}=-\nabla^{\text{Ch}}_{\Bar{\imath}}H_{j}+\nabla^{\text{Ch}}_{j}H_{\Bar{\imath}}.
\label{eq:torsion_identity_1}
\end{equation}
The torsion also obeys $\partial_{[i_{1}}H_{i_{2}i_{3}]\Bar{\jmath}}=0$, which can be put in the covariant form
\begin{equation}
\nabla^{\text{Ch}}_{[i_{1}}H_{i_{2}i_{3}]\Bar{\jmath}}=-H_{\Bar{\jmath}k[i_{1}}H_{i_{2}i_{3}]}{}^{k}.
\end{equation}
Contraction of this equation with the hermitian metric implies
\begin{equation}
\nabla^{\text{Ch}}_{k}H^{k}{}_{ij}=\nabla^{\text{Ch}}_{i}H_{j}-\nabla^{\text{Ch}}_{j}H_{i}+H_{ij}{}^{k}H_{k}.
\label{eq:torsion_identity_2}
\end{equation}
As the torsion has no $(3,0)$ component, it also satisfies
\begin{equation}
\nabla^{\text{Ch}}_{\Bar{k}}H^{\Bar{k}}{}_{ij}=0.
\label{eq:torsion_identity_3}
\end{equation}

As the contraction of the Chern curvature $R^{\text{Ch}}_{\Bar{\imath}j}$ which appears in the flow equation \eqref{eq:anomaly_flow_equation} differs from the contraction $R^{\text{Ch}}_{k\Bar{\imath}}{}^{k}{}_{j}$ which appears in the Ricci tensor \eqref{eq:Ricci_tensor_1}, it is also necessary to relate the two of them. This can be done by contracting \eqref{eq:permutation_Chern_curvature} with the hermitian metric, yielding
\begin{equation}
R^{\text{Ch}}_{k\Bar{\imath}}{}^{k}{}_{j}=-R^{\text{Ch}}_{\Bar{\imath}j}+\nabla^{\text{Ch}}_{k}H^{k}{}_{\Bar{\imath}j}.
\label{eq:Chern_tensors}
\end{equation}
Using identities \eqref{eq:torsion_identity_1}, \eqref{eq:torsion_identity_2} and \eqref{eq:torsion_identity_3} for derivatives of the torsion, as well as \eqref{eq:Chern_tensors}, the Ricci tensor can be brought to the form
\begin{subequations}
\label{eq:Ricci_tensor_2}
\begin{align}
R_{ij}&=\frac{1}{2}\nabla^{\text{Ch}}_{i}H_{j}+\frac{1}{2}\nabla^{\text{Ch}}_{j}H_{i}+\frac{1}{2}H_{i}{}^{\Bar{k}l}H_{j\Bar{k}l}
\nonumber\\
&=\frac{1}{2}\nabla_{i}H_{j}+\frac{1}{2}\nabla_{j}H_{i}+\frac{1}{2}H_{i}{}^{\Bar{k}l}H_{j\Bar{k}l},\\
R_{\Bar{\imath}j}&=-R^{\text{Ch}}_{\Bar{\imath}j}+\frac{1}{2}\nabla^{\text{Ch}}_{\Bar{\imath}}H_{j}+\frac{1}{2}\nabla^{\text{Ch}}_{j}H_{\Bar{\imath}}+\frac{1}{2}H_{\Bar{\imath}j}{}^{k}H_{k}-\frac{1}{2}H_{\Bar{\imath}j}{}^{\Bar{k}}H_{\Bar{k}}-\frac{1}{2}H_{\Bar{\imath}}{}^{\Bar{k}l}H_{j\Bar{k}l}+\frac{1}{4}H_{\Bar{\imath}}{}^{\Bar{k}\Bar{l}}H_{j\Bar{k}\Bar{l}}\nonumber\\
&=-R^{\text{Ch}}_{\Bar{\imath}j}+\frac{1}{2}\nabla_{\Bar{\imath}}H_{j}+\frac{1}{2}\nabla_{j}H_{\Bar{\imath}}-\frac{1}{2}H_{\Bar{\imath}}{}^{\Bar{k}l}H_{j\Bar{k}l}+\frac{1}{4}H_{\Bar{\imath}}{}^{\Bar{k}\Bar{l}}H_{j\Bar{k}\Bar{l}}.
\end{align}
\end{subequations}

\subsubsection*{Flow equation}
The expressions \eqref{eq:Riemann_curvature} and \eqref{eq:Ricci_tensor_2} for the Riemann and Ricci tensor are valid for any complex manifold and do not rely on supersymmetry. Assuming a supersymmetric geometry gives the additional equation
\begin{equation}
H_{m}=-2\nabla_{m}\varphi,
\end{equation}
which can be obtained from the conformally balanced condition. Combining this equation with \eqref{eq:Ricci_tensor_2}, we find that for supersymmetric geometries,
\begin{subequations}
\begin{align}
&R_{ij}+2\nabla_{i}\nabla_{j}\varphi-\frac{1}{4}H_{i}{}^{mn}H_{jmn}=0,\\
&R_{\Bar{\imath}j}+2\nabla_{\Bar{\imath}}\nabla_{j}\varphi-\frac{1}{4}H_{\Bar{\imath}}{}^{mn}H_{jmn}=-(R^{\text{Ch}}_{\Bar{\imath}j}+H_{\Bar{\imath}}{}^{\Bar{k}l}H_{j\Bar{k}l}).
\end{align}
\end{subequations}
The tensor driving the flow in \eqref{eq:anomaly_flow_equation} appears explicitly, allowing us rewrite the flow in the form
\begin{equation}
\partial_{t}g_{\Bar{\imath}j}=-\ee^{2\varphi}\left(R_{\Bar{\imath}j}+2\nabla_{\Bar{\imath}}\nabla_{j}\varphi-\tfrac{1}{4}H_{\Bar{\imath}}{}^{mn}H_{jmn}\right)+\frac{\alpha'}{8}\ee^{2\varphi}(\tr F\wedge F-\tr R^{+}\wedge R^{+})^{k}{}_{k\Bar{\imath}j}.
\label{eq:flow_metric_0th_order}
\end{equation}
Similarly, one can show that the metric stays hermitian under the flow, i.e.~$\partial_{t}g_{ij}=0$, in agreement with the vanishing of the non-mixed components of the Einstein equation.

The flow equation \eqref{eq:flow_metric_0th_order} includes the metric equation of motion at zeroth order in $\alpha'$. A similar derivation can be considered for the $\mathcal{O}(\alpha')$ terms in the flow equation in order to obtain the equation of motion correct to first order in $\ap$. Instead of showing this explicitly, it turns out that there is an alternative (and easier) way to obtain the same result, which makes use of supersymmetry.

\subsection{A detour through supersymmetry}

The underlying supersymmetry of the heterotic theory leads to additional structure in the field equations. In particular, somewhat remarkably, the bosonic equations of motion can be recast as integrability conditions of the supersymmetry variations~\cite{0908.2927,Gauntlett:2002sc,Kunitomo:2009mx,Martelli:2010jx}. Consequently, at zeroth order in $\alpha'$, the equations of motion can be built by acting on a non-vanishing Majorana--Weyl spinor $\varepsilon$ with combinations of supersymmetry operators. For example, the Einstein equation can be expressed as
\begin{equation}\begin{split}
&R_{MN}+2\nabla_{M}\nabla_{N}\varphi-\tfrac{1}{4}H_{M}{}^{P_{1}P_{2}}H_{NP_{1}P_{2}}\\
&=\left(\bar{\varepsilon}\,\Gamma_{(M}\feyn{D}D_{N)}\varepsilon-\bar{\varepsilon}\,\Gamma_{(M}D_{N)}\feyn{D}\varepsilon+\tfrac{1}{2}H_{(M}{}^{PQ}\,\bar{\varepsilon}\,\Gamma_{N)}\Gamma_{P}D_{Q}\varepsilon+\text{c.c.}\right)\\
&\eqspace+\tfrac{1}{12}\dd H_{(M}{}^{P_{1}\dots P_{3}}\,\bar{\varepsilon}\Gamma_{N)P_{1}\dots P_{3}}\varepsilon.
\end{split}
\label{eq:integrability_eom_metric_10D}
\end{equation}
For compactifications to four-dimensional Minkowski space, the only non-trivial components of this equation are along the internal space. Bilinears of $\varepsilon$ then reduce to bilinears of the internal spinor $\epsilon$ that appears in a decomposition of $\varepsilon$, and the resulting equation is
\begin{equation}\begin{split}
&R_{mn}+2\nabla_{m}\nabla_{n}\varphi-\tfrac{1}{4}H_{m}{}^{p_{1}p_{2}}H_{np_{1}p_{2}}\\
&=\left(\epsilon^{\dagger}\gamma_{(m}\feyn{D}D_{n)}\epsilon-\epsilon^{\dagger}\gamma_{(m}D_{n)}\feyn{D}\epsilon+\tfrac{1}{2}H_{(m}{}^{pq}\,\epsilon^{\dagger}\gamma_{n)}\gamma_{p}D_{q}\epsilon+\text{c.c.}\right)\\
&\eqspace+\tfrac{1}{12}\dd H_{(m}{}^{p_{1}\dots p_{3}}\,\epsilon^{\dagger}\gamma_{n)p_{1}\dots p_{3}}\epsilon.
\end{split}
\label{eq:integrability_eom_metric_6D}
\end{equation}
For a supersymmetric geometry, both $D_{m}\epsilon$ and $\feyn{D}\epsilon$ vanish, and \eqref{eq:integrability_eom_metric_6D} reduces to
\begin{equation}
R_{mn}+2\nabla_{m}\nabla_{n}\varphi-\tfrac{1}{4}H_{m}{}^{p_{1}p_{2}}H_{np_{1}p_{2}}=-\tfrac{1}{24}\dd H_{(m}{}^{p_{1}\dots p_{3}}\,(\omega\wedge\omega)_{n)p_{1}\dots p_{3}},
\end{equation}
where the last term has been rewritten using $\epsilon^{\dagger}\gamma_{m_{1}\dots m_{4}}\epsilon=(\star\omega)_{m_{1}\dots m_{4}}$. In holomorphic components, the non-mixed components of this equation vanish. For the mixed components, one obtains
\begin{equation}
R_{\Bar{\imath}j}+2\nabla_{\Bar{\imath}}\nabla_{j}\varphi-\tfrac{1}{4}H_{\Bar{\imath}}{}^{m_{1}m_{2}}H_{jm_{1}m_{2}}=\tfrac{1}{2}\dd H^{k}{}_{k\Bar{\imath}j},
\end{equation}
which exactly reproduces the result derived in Appendix \ref{ap:recasting_eom}.

Inclusion of the $\mathcal{O}(\alpha')$ part of the heterotic equation of motion is then straightforward. For the curvature of the gauge field, $F_{mn}$, expanding the product of gamma matrices in the Clifford basis yields
\begin{equation}
\begin{split}
\tr \epsilon^{\dagger}F_{mp}\gamma_{n}{}^{p}\feyn{F}\epsilon&=\tfrac{1}{2}\tr( F_{mp}F_{q_{1}q_{2}})(\epsilon^{\dagger}\gamma_{n}{}^{pq_{1}q_{2}}\epsilon-2\delta^{q_{1}}_{n}\delta^{q_{2}}_{p})\\
&=-\tfrac{1}{24}\tr (F\wedge F)_{m}{}^{p_{1}\dots p_{3}}(\omega\wedge\omega)_{np_{1}\dots p_{3}}-\tr F_{m}{}^{p}F_{np}.
\end{split}
\end{equation}
Using this equation, the contraction of $\tr F\wedge F$ with the hermitian form that appears in \eqref{eq:flow_metric_0th_order} can readily be rewritten as the gauge contribution to the metric equation of motion, with an additional contribution involving the supersymmetry operator $\feyn{F}\epsilon$. The exact same derivation can be followed for $R^{+}$. Taken together, this expresses \eqref{eq:flow_metric_0th_order} in terms of the metric equation of motion complete to first order in $\alpha'$. 

Explicitly, one finds that the equation of motion for the metric \eqref{eq:metric_eom} restricted to $X$
\begin{equation}
\begin{split}
\text{eom}[g]_{mn}&=\left(\epsilon^{\dagger}\gamma_{(m}\feyn{D}D_{n)}\epsilon-\epsilon^{\dagger}\gamma_{(m}D_{n)}\feyn{D}\epsilon+\tfrac{1}{2}H_{(m}{}^{pq}\,\epsilon^{\dagger}\gamma_{n)}\gamma_{p}D_{q}\epsilon+\text{c.c.}\right)\\
&\eqspace+\frac{\ap}{4}\ee^{2\varphi}\left(\tr\epsilon^{\dagger}F_{mp}\gamma_{n}{}^{p}\feyn F\epsilon-\epsilon^{\dagger}R_{mp}^{+}\gamma_{n}{}^{p}\feyn{R}^{+}\epsilon\right)+\tfrac{1}{12}\mathcal{B}_{(m}{}^{p_{1}\dots p_{3}}\,\epsilon^{\dagger}\gamma_{n)p_{1}\dots p_{3}}\epsilon.
\end{split}
\end{equation}
Similarly, dilaton equation of motion \eqref{eq:dilaton_eom} restricted to $X$ can be written as
\begin{equation}
    \text{eom}[\varphi]=\epsilon^{\dagger}(D^{2}-\feyn{D}^{2})\epsilon-2\nabla^{m}\varphi\,\epsilon^{\dagger}D_{m}\epsilon+\frac{\ap}{16}\epsilon^{\dagger}(\tr\feyn{F}\feyn{F}-\tr\feyn{R}^{+}\feyn{R}^{+})\epsilon+\tfrac{1}{4}\epsilon^{\dagger}\feyn{\mathcal{B}}\epsilon.
\end{equation}

\bibliographystyle{utphys}
\bibliography{inspire,extra}

\end{document}